\documentclass[11pt,a4paper]{article}

\pdfoutput=1  %%% If your are submitting a pdflatex (i.e. if you have images in pdf, png or jpg format)

\usepackage{jheppub}  %%% Produce preprints in a form suitable for submission to JHEP. For details on the use of the package, please see the JHEP-author-manual.pdf
\usepackage[utf8]{inputenc} %%% Enables reading accent directly from editor
\graphicspath{{./figures/}} %%% Path to include figures files
\usepackage{caption}
\usepackage{subcaption}

\usepackage{epstopdf}

\title{Homogeneous isotropization and equilibration of a strongly coupled plasma with a critical point}

\author[a]{Renato Critelli,}
\author[b]{Romulo Rougemont,}
\author[a]{and Jorge Noronha}

\affiliation[a]{Instituto de F\'{i}sica, Universidade de S\~{a}o Paulo, Rua do Mat\~{a}o, 1371, Butant\~{a}, CEP 05508-090, S\~{a}o Paulo, SP, Brazil}
\affiliation[b]{International Institute of Physics, Federal University of Rio Grande do Norte,
Campus Universit\'{a}rio - Lagoa Nova, CEP 59078-970, Natal, Rio Grande do Norte, Brazil}

\abstract{We use holography to investigate the process of homogeneous isotropization and thermalization in a strongly coupled $\mathcal{N} = 4$ Super Yang-Mills plasma charged under a $U(1)$ subgroup of the global $SU(4)$ R-symmetry which features a critical point in its phase diagram. Isotropization dynamics at late times is affected by the critical point in agreement with the behavior of the characteristic relaxation time extracted from the analysis of the lowest non-hydrodynamic quasinormal mode in the $SO(3)$ quintuplet (external scalar) channel of the theory. In particular, the isotropization time may decrease or increase as the chemical potential increases depending on whether one is far or close enough to the critical point, respectively. On the other hand, the thermalization time associated with the equilibration of the scalar condensate, which happens only after the system has relaxed to a (nearly) isotropic state, is found to always increase with chemical potential in agreement with the characteristic relaxation time associated to the lowest non-hydrodynamic quasinormal mode in the $SO(3)$ singlet (dilaton) channel. These conclusions about the late dynamics of the system are robust in the sense that they hold for different initial conditions seeding the time evolution of the far-from-equilibrium plasma.}

\emailAdd{renato.critelli@usp.br, rrougemont@iip.ufrn.br, noronha@if.usp.br}
\arxivnumber{1709.03131}

\begin{document}
\maketitle

%%%%%%%%%%%%%%%%%%%%%%%%%
\section{Introduction and summary}
\label{sec:Intro}

With the advent of the holographic gauge/gravity duality \cite{Maldacena:1997re,Gubser:1998bc,Witten:1998qj,Witten:1998zw}, it has become possible to study physical properties of some strongly coupled quantum systems using classical gravity in higher dimensions (for reviews see, e.g., \cite{McGreevy:2009xe,CasalderreySolana:2011us,Adams:2012th}). Concerning strongly correlated quantum fluids, this framework has made it possible the investigation of several aspects of such systems, such as their thermodynamics and hydrodynamics \cite{Kovtun:2004de,Bhattacharyya:2008jc,Baier:2007ix}, quasinormal modes \cite{Horowitz:1999jd,Kovtun:2005ev,Heller:2013fn}, and also the far-from-equilibrium dynamics describing the relaxation of holographic fluids toward thermodynamic equilibrium in many different settings (see e.g. \cite{Chesler:2013lia,vanderSchee:2014qwa} for recent reviews).

One of the main attractive features of holography is its unique ability to deal with the entire evolution of a strongly coupled fluid within a single framework, starting from far-from-equilibrium anisotropic initial states and dynamically evolving them passing through different stages comprising several kinds of characteristic ``relaxation times'' until reaching thermodynamic equilibrium. These different relaxation times characterize, for instance, the onset of applicability of hydrodynamics (known as the ``hydrodynamization time'', which also depends on the specific formulation of hydrodynamics considered), the onset of nearly isotropization of the system (when the longitudinal and transverse pressures in a given flow are approximately equal), the onset of applicability of the equilibrium equation of state in nonconformal plasmas (known as the ``EoSization time'' \cite{Attems:2016ugt}), and the onset of true thermalization (when all the physical observables of the theory have approximately relaxed to their equilibrium values). In fact, one of the main outcomes of holographic investigations of far-from-equilibrium strongly coupled quantum fluids was the conclusion that in some cases the system may hydrodynamize when the fluid is still significantly anisotropic and far-from-equilibrium \cite{Chesler:2009cy,Heller:2011ju,Attems:2017zam,Romatschke:2017vte}. This finding, together with recent calculations \cite{Heller:2013fn,Buchel:2016cbj,Denicol:2016bjh,Heller:2016rtz} that showed that the gradient expansion diverges, have significantly changed our understanding of relativistic hydrodynamics (for a review see \cite{Florkowski:2017olj}).  

The fast expanding fireball produced in ultrarelativistic heavy ion collisions at RHIC \cite{Arsene:2004fa,Adcox:2004mh,Back:2004je,Adams:2005dq} and LHC \cite{Aad:2013xma} is probably the most remarkable example of a dynamical system actually realized in nature featuring a very rich and complex time evolution characterized by both hard (i.e., perturbative) and soft (nonperturbative) physics. Just before the heavy ions collide, the gluon density inside these large nuclei at very high energies is expected to saturate producing the so-called color glass condensate (CGC) \cite{McLerran:2001sr,Iancu:2003xm,Weigert:2005us,Gelis:2010nm}, which has become the starting point for the initial conditions in heavy ion collisions. Right after the collision the medium has an enormous amount of energy concentrated in a very small volume, which starts to rapidly expand passing through different stages. Before 1 fm/c after the collision the system is expected to be a highly dense coherent medium dominated by the dynamics of classical QCD fields called ``glasma'' \cite{Gelis:2012ri}.\footnote{An intermediate stage between the color \emph{glass} condensate and the quark-gluon \emph{plasma}.} As the system keeps expanding, the glasma decoheres towards a new state of QCD matter called the quark-gluon plasma (QGP) \cite{Gyulassy:2004zy,Heinz:2013th,Shuryak:2014zxa}, whose relevant degrees of freedom correspond to deconfined quarks and gluons.

In practice, the QGP produced in heavy ion collisions is well described by viscous hydrodynamics with an equilibrium equation of state and transport coefficients which are compatible with soft physics expectations \cite{Ryu:2015vwa,Bernhard:2016tnd,Pratt:2015zsa,Monnai:2017cbv}, indicating that at this stage the system is strongly coupled, contrary to the scenario just after the collision where weak coupling physics plays a prominent role. As the QGP keeps expanding and cooling down it eventually enters in the crossover region \cite{Aoki:2006we,Borsanyi:2016ksw} and hadronizes, giving place to a hadron gas. Later stages of the temporal evolution of heavy ion collisions include the regions of chemical freeze-out\footnote{When inelastic collisions between the produced hadrons cease and the relative ratio between the different kinds of particles remains fixed.} and the thermal or kinetic freeze-out\footnote{When the average distance between the hadrons is large enough to make the nuclear interaction between them effectively negligible, causing the momentum distribution of the particles to remain fixed.}. After this stage, the produced hadrons are essentially free and the yields of their decays reach the detectors of the experimental apparatus providing a large amount of information about the evolution of the system.

The full dynamical evolution of high energy heavy ion collisions outlined above cannot be completely described by the gauge/gravity duality since the former encompasses both hard and soft physics while the latter can only deal with strongly coupled systems. In fact, it is well-known that the asymptotically free ultraviolet regime of QCD cannot be described by the gauge/gravity duality given that holographic models are usually characterized by strongly coupled ultraviolet fixed points. On the other hand, even though a rigorous top-down construction of a gravity dual of the hydrodynamized and strongly coupled QGP is missing, there are phenomenological bottom-up Einstein-Maxwell-Dilaton (EMD) gauge/gravity constructions which have been shown to successfully describe in practice, not only on a qualitative level but also quantitatively, a plethora of thermodynamical and hydrodynamical observables characterizing the physics of the strongly coupled QGP under many different situations \cite{Gubser:2008ny,Gubser:2008yx,DeWolfe:2010he,DeWolfe:2011ts,Finazzo:2014cna,Rougemont:2015wca,Rougemont:2015ona,Finazzo:2015xwa,Rougemont:2017tlu,Critelli:2017oub,Rougemont:2015oea,Finazzo:2016mhm,Critelli:2016cvq}. Therefore, one striking question which is posed in face of the above considerations is the following: in which cases could one hope, at least in principle, to obtain possible useful insights (or even some quantitatively accurate results) for the far-from-equilibrium dynamics of heavy ion collisions using holographic techniques?

The answer to the question above is still unsettled but one may gauge the applicability of gauge/gravity models to the analysis of heavy ion collisions by looking at some recent results comparing lattice QCD calculations (which are performed in equilibrium) with heavy ion experimental data. This kind of comparison may help to identify under which conditions the expanding fireball would probe, through more stages, a strongly coupled regime of QCD.

For instance, in Ref.\ \cite{Monnai:2017cbv} it was shown that in central heavy ion collisions at RHIC and LHC the experimentally extracted ratio between the pressure and the internal energy density of the QGP, $(p/\epsilon)_{\textrm{exp}}(T_{\textrm{eff}})=0.21\pm 0.10$, is in good agreement with the corresponding lattice QCD estimate, $(p/\epsilon)_{\textrm{lQCD}}(T_{\textrm{eff}})\approx 0.23$, where $T_{\textrm{eff}}$ is half-way between the effective temperatures associated with the collision energies $\sqrt{s}=200$ GeV and $\sqrt{s}=2.76$ TeV (at these high energies, the baryon chemical potential is negligible compared to the temperature of the medium). Moreover, in Ref.\ \cite{Pratt:2015zsa} a state-of-the-art Bayesian analysis was simultaneously applied to several physical observables while varying the free parameters of the model. The results in the space of parameters of the model which match combined heavy ion data measured at RHIC and LHC within the interval $\sqrt{s}=200$ GeV --- $2.76$ TeV were also found to be consistent with results from lattice QCD simulations. These findings show that the QGP produced in these collisions at high energies is not only adequately described by hydrodynamics but also that the (equilibrium) QCD equation of state obtained in lattice simulations may be trustfully used in hydrodynamic simulations of the spacetime evolution of the system at these high energies (i.e., $\sqrt{s}\gtrsim 200$ GeV).

Furthermore, it was also shown in Refs. \cite{Borsanyi:2013hza,Borsanyi:2014ewa} that ratios between higher order susceptibilities of baryon and electric charges calculated in equilibrium on the lattice give a good description of experimentally measured ratios between moments of net-proton and net-electric charge multiplicity distributions for collision energies $\sqrt{s}\ge 39$ GeV; however, for $\sqrt{s}< 39$ GeV the compatibility observed at higher energies between the set of chemical freeze-out parameters extracted from the independent analysis performed in the baryon and electric charge sectors on the lattice is not warranted \cite{Borsanyi:2014ewa}. This suggests that at chemical freeze-out the system produced in heavy ion collisions is less close to equilibrium if the system has a higher chemical potential (corresponding to a lower collision energy). 

Recently, using a phenomenological holographic EMD model which quantitatively reproduces lattice QCD thermodynamics at zero and nonzero baryon chemical potential \cite{Rougemont:2015wca,Rougemont:2017tlu}, it was found that an increase in the baryon chemical potential decreases the shear viscosity times temperature to enthalpy density ratio, $\eta T/(\epsilon+p)$, which gives a measure of the fluidity of the medium \cite{Liao:2009gb}\footnote{At zero density this ratio reduces to the shear viscosity to entropy density ratio, $\eta/s$.}. This indicates that the QGP at finite baryon density remains strongly coupled. From the above considerations, one may expect that the gauge/gravity duality is more likely to produce useful insights for the far-from-equilibrium dynamics of heavy ion collisions when the system is doped with a nonzero chemical potential\footnote{This is so because, as discussed before, there are indications that the baryon dense QGP experiences relevant far-from-equilibrium effects while remaining strongly coupled through more stages than in the case of higher energy collisions (where the chemical potential is negligible). In fact, this may provide a partial explanation for the findings of Ref.\ \cite{Casalderrey-Solana:2016xfq} where a qualitative agreement between the rapidity distribution of baryon charge in a holographic shock wave analysis at finite density and in heavy ion collisions at moderate and low energies was found to disappear at full RHIC or LHC energies (i.e., $\sqrt{s}\gtrsim 200$ GeV).}. This is exactly the scenario we are interested in analyzing in the present work.

The literature regarding out-of-equilibrium holographic dynamics is vast, and even if we restrict ourselves to works that deal with models endowed with a chemical potential there is a substantial amount of papers already available. For non-equilibrium aspects of some strongly coupled plasmas doped with a chemical potential which do not employ numerical relativity and probe the thermalization process using non-local observables (e.g. Wilson's loops), see e.g. Refs.\ \cite{Caceres:2012em,Giordano:2014kya,Caceres:2014pda,Camilo:2014npa}. In the far-from-equilibrium context, there is the study of Ref.\ \cite{Fuini:2015hba} regarding homogeneous equilibration in a charged plasma without a critical point, while Ref.\ \cite{Casalderrey-Solana:2016xfq} considers a shock wave analysis with baryon charge, yet without a critical point. One can also study quantum critical points in the context of holographic quenches, as done in Refs. \cite{Basu:2011ft,Basu:2012gg}. Recent studies about holographic isotropization in the context of Gauss-Bonnet gravity can be found in Ref.\ \cite{Andrade:2016rln}.

One of the main goals of the present work is to assess the impact of a critical point in the far-from-equilibrium dynamics of a relativistic strongly coupled plasma at finite density, which to the best of our knowledge is a question that has not been previously studied in the literature. In fact, this is the simplest problem one may consider that can lead to interesting insights into the strongly coupled dynamics of the QGP near a critical point at finite density, one of the focus of RHIC's Beam Energy Scan (BES) program. Even though the holographic model we consider here is very different from QCD, we believe such a study is important given that there are no other approaches that can be used to perform real-time, far-from-equilibrium calculations at strong coupling near a critical point.

More specifically, we shall investigate here the homogeneous equilibration dynamics of a far-from-equilibrium top-down holographic plasma at finite density known as the 1-R charge black hole (1RCBH) \cite{Gubser:1998jb,Behrndt:1998jd,Kraus:1998hv,Cai:1998ji,Cvetic:1999ne,Cvetic:1999rb}. This model describes a strongly coupled conformal $\mathcal{N}=4$ Super Yang-Mills (SYM) plasma charged under a $U(1)$ subgroup of the global $SU(4)$ R-symmetry and features a critical point (CP) in its phase diagram. Our gravitational setup is different from the one considered in Ref.\ \cite{Fuini:2015hba} where the authors studied the homogeneous equilibration of a strongly coupled SYM plasma with a nonzero charge density without a CP. Indeed, as we are going to see in a moment, the action of the 1RCBH model includes a dilaton field that considerably changes the dynamics of the theory and leads to a CP in its phase diagram.

In order to solve the corresponding numerical relativity problem, in this work we follow the pioneering work of Chesler and Yaffe \cite{Chesler:2008hg} by employing the so-called characteristic formulation of general relativity \cite{Lehner:2001wq}, which is very convenient for asymptotically AdS spacetimes (see, e.g., Ref.\ \cite{Chesler:2013lia} for a review). Alternatively, one may also use the ADM formulation \cite{Arnowitt:1959ah,Arnowitt:1960es} of numerical relativity in asymptotically AdS spacetimes, as done in Refs. \cite{Bantilan:2012vu,Heller:2012je}.

An important remark must be done at this point. As it is well-known, the thermodynamics and the hydrodynamics of the SYM plasma are very different than what is observed in the QGP produced in heavy ion collisions (see, e.g., \cite{Rougemont:2016etk}). Thus, one should not expect to obtain, in general, quantitative insights for the early time dynamics of heavy ion collisions from the study of the far-from-equilibrium dynamics of the 1RCBH model. However, it may be that some qualitative properties derived in the far-from-equilibrium dynamics of this model are robust or ``nearly universal'', as it happens with the shear viscosity to entropy density ratio of holographic fluids where $\eta/s=1/4\pi$ for a broad class of strongly coupled systems \cite{Kovtun:2004de}. In order to look for possible signatures of this kind of ``universality'', one should also consider the far-from-equilibrium dynamics in other holographic models at finite density endowed with a CP.

For instance, the quasinormal mode (QNM) oscillations of the 1RCBH model in the external scalar and vector diffusion channels were found in Ref. \cite{Finazzo:2016psx} to be damped as one increases the $U(1)$ R-charge chemical potential, as long as one is away from criticality. This finding is in qualitative accordance with the observation done in Ref.\ \cite{Rougemont:2015wca} that by increasing the baryon chemical potential far from the CP there is a damping in the QNM oscillations of the external scalar channel of a phenomenological bottom-up EMD model at finite baryon density that quantitatively describes QGP thermodynamics. This damping of the QNM oscillations caused by increasing the chemical potential away from criticality in two very different holographic models may be a signature of robustness or an indication of a general behavior for strongly coupled systems at finite density.

However, close to the CP, the QNM oscillations of the 1RCBH model in the external scalar and vector diffusion channels revert this trend observed at lower densities and increase as the $U(1)$ R-charge chemical potential is enhanced \cite{Finazzo:2016psx}. On the other hand, the behavior of the QNMs of the QCD-like phenomenological EMD model at finite baryon density of Refs.\ \cite{Rougemont:2015wca,Critelli:2017oub} has not yet been investigated close to criticality, a task we postpone for a future work. It would be interesting to check if the same qualitative modification in the QNM oscillations of the 1RCBH model caused by the proximity of the CP is also featured in the phenomenological EMD model, since this could point out to a possible universal effect of the CP on the equilibration dynamics of strongly coupled fluids.

Therefore, even though the 1RCBH model (and the SYM plasma in general) is not a holographic setup suited for direct applications to heavy ion phenomenology, it may potentially contain some general properties displayed by strongly coupled fluids driven out-of-equilibrium. Moreover, one of the main conclusions of the present work will be the distinction of two characteristic equilibration times of the far-from-equilibrium system. Namely, by looking at the imaginary part of the lowest non-hydrodynamic QNMs of the model one may extract an upper bound for some characteristic ``relaxation times'' of the theory, according to the general reasoning first devised in Ref.\ \cite{Horowitz:1999jd}. The aforementioned behavior of the QNMs in the external scalar and vector diffusion channels would, therefore, suggest that the ``equilibration time'' of the 1RCBH model decreases with increasing chemical potential far from the CP, while close to the CP it would instead increase. As we shall see in this work, this is, in fact, the behavior found for the \emph{isotropization} time of the system, which is dominated by the lowest non-hydrodynamic QNM of the external scalar channel of the theory. 

On the other hand, even after (nearly) isotropization is reached, the scalar condensate dual to the bulk dilaton field may still be significantly far from its equilibrium value. Since in the present work the equilibration of the scalar condensate will always be the last equilibration time of the system, we shall associate it with the true thermalization time of the medium. Contrary to the isotropization time, in this model the thermalization time always increase with increasing chemical potential. As we are going to show, this is in consonance with the behavior of the lowest non-hydrodynamic QNM of the dilaton channel. Also, in the near future we intend to extend the present analysis to consider the case of the QCD-like EMD model of Ref. \cite{Critelli:2017oub}, which provided a prediction for the location of the long-sought critical point of the QCD phase diagram in the plane of temperature and baryon chemical potential.

The outline of this work is as follows. In Sec.\ \ref{sec:ModelandEOM} we present the equations of motion for the 1RCBH model assuming a time-dependent and spatially homogeneous anisotropic Ansatz for the bulk fields. This gives a set of nonlinear partial differential equations (PDEs) to be solved numerically. We also discuss the relevant observables of the dual quantum field theory (QFT) that we need to compute to analyze the homogeneous isotropization and thermalization processes in this setup, namely, the one-point functions associated with the expectation value of the boundary stress-energy tensor, $\langle T_{\mu\nu}\rangle$, dual to the bulk metric field $g^{\mu\nu}$, the scalar condensate, $\langle\mathcal{O}_{\phi}\rangle$, dual to the bulk dilaton field $\phi$, and the expectation value of the conserved $U(1)$ R-current, $\langle J_{\mu} \rangle$, dual to the bulk Maxwell field $A^\mu$. The derivation of the general form of these one-point functions via holographic renormalization is presented in Appendix \ref{sec:HoloRen}. In Sec.\ \ref{sec:AsymExp} we perform the near-boundary asymptotic expansion of the bulk fields in order to fix the boundary conditions for the set of PDEs. From this near-boundary analysis, we will be left with a couple of unknown time-dependent ultraviolet coefficients that shall be related with the one-point functions of the dual QFT. In Sec.\ \ref{sec:EqSol}, we briefly review the analytical equilibrium solutions of the 1RCBH model and its thermodynamics. In Sec.\ \ref{sec:FarSol}, we discuss some relevant technical issues related to the numerical time-dependent far-from-equilibrium solutions of the 1RCBH model, such as the choice of the initial conditions, and explain how we numerically solve the set of coupled nonlinear PDEs for the far-from-equilibrium system. Once the numerics is settled, we proceed in Sec.\ \ref{sec:results} to provide the main results of the homogeneous equilibration dynamics of the 1RCBH model. Moreover, in Subsection \ref{sec:MatQNM} we perform the match of the late time behavior of the pressure anisotropy and the scalar condensate with the lowest non-hydrodynamic QNMs of the external scalar and dilaton channels, respectively. While the QNMs of the external scalar channel were obtained in Ref. \cite{Finazzo:2016psx}, we derive in Appendix \ref{sec:QNMs} the QNMs of the dilaton channel. Finally, we close the paper in Sec.\ \ref{sec:outlook} with an outlook of our main results and also point out future perspectives and ongoing investigations.
 
Throughout this paper we work with natural units where $\hbar=c=k_B=1$ and use a mostly plus metric signature. Capital Latin indices $\lbrace M,N, \dots \rbrace$ denote the coordinates $\lbrace t,r,x,y,z \rbrace$ of a five dimensional asymptotically AdS spacetime where the gravity theory is defined, whereas Greek indices $\lbrace \mu,\nu, \dots \rbrace$ represent the coordinates $\lbrace t,x,y,z \rbrace$ of the four dimensional boundary.

%%%%%%%%%%%%%%%%%%%%%%%%%
\section{The holographic model and its equations of motion}
\label{sec:ModelandEOM}

The 1RCBH model \cite{Gubser:1998jb,Behrndt:1998jd,Kraus:1998hv,Cai:1998ji,Cvetic:1999ne,Cvetic:1999rb} first appeared as a solution of the five dimensional $\mathcal{N}=8$ gauged supergravity action \cite{Behrndt:1998jd}, which was later demonstrated to lie within a class of solutions equivalent to near-extremal spinning D3-branes in AdS$_5\times$S$^{5}$ \cite{Cvetic:1999ne}. The Kaluza-Klein compactification of the five sphere S$^{5}$ on the spinning D3-branes solutions leads to the $SU(4)$ R-symmetry and the three independent Cartan subgroups of the R-Symmetry $U(1)_{a}\times U(1)_{b}\times U(1)_{c}$ are associated with three distinct conserved charges $(Q_{a},Q_{b},Q_{c})$ of the black hole background \cite{Behrndt:1998jd}. The general solution is known as the STU model, whilst the 1RCBH model is obtained by considering only one charge, i.e., $Q \equiv Q_{a}$ and $Q_{b}=Q_{c}=0$. A thorough discussion of the matter content of the dual QFT at zero temperature may be found in Ref.\ \cite{DeWolfe:2012uv}. For the thermal plasma at finite density, detailed discussions may be found in Refs.\ \cite{DeWolfe:2011ts,Finazzo:2016psx}. For the sake of completeness, in the present work we briefly review the thermodynamics of the 1RCBH plasma in Sec.\ \ref{sec:thermo}.

The gravitational action of the 1RCBH model is given by \cite{Gubser:1998jb,Behrndt:1998jd,Kraus:1998hv,Cai:1998ji,Cvetic:1999ne,Cvetic:1999rb},
\begin{equation}
\label{eq:action}
S = \frac{1}{2 \kappa_5^2} \int_{\mathcal{M}} d^5x \sqrt{-g} \left[ R - \frac{f(\phi)}{4} F_{MN} F^{MN} - \frac{1}{2} (\partial_{M} \phi)^2 - V(\phi) \right]+S_{GHY}+S_{ct},
\end{equation}
where $\kappa_5^2=8\pi G_5$ with $G_{5}$ being the five dimensional Newton's constant, $F_{MN}=\partial_{M}A_{N}-\partial_{N}A_{M}$, with $A_M$ being the Maxwell gauge field, and
\begin{equation}
S_{GHY} = \frac{1}{\kappa_{5}^2}\int_{\partial\mathcal{M}} d^4 x \sqrt{-\gamma}K
\end{equation}
is the Gibbons-Hawking-York boundary action \cite{York:1972sj,Gibbons:1976ue} needed to provide a well-posed Dirichlet problem. In this term, $\gamma_{\mu\nu}$ denotes the induced metric at the boundary and $K$ represents its extrinsic curvature. The last term in Eq.\ \eqref{eq:action} is the boundary counterterm action needed to remove the divergences of the on-shell action. In Appendix \ref{sec:HoloRen} we give the details regarding the boundary terms and show how to obtain the desired one-point functions of the dual QFT. 

The expressions for the dilaton potential $V(\phi)$ and the Maxwell-Dilaton coupling $f(\phi)$ in the action \eqref{eq:action} which define the top-down construction corresponding to the 1RCBH model are given by
\begin{align}
V(\phi) = -\frac{1}{L^2} \left(8e^{\frac{\phi}{\sqrt{6}}} + 4 e^{-\sqrt{\frac{2}{3}}\phi} \right), \,\,\,
f(\phi) = e^{- 2\sqrt{\frac{2}{3}}\phi},
\end{align}
where $L$ is the AdS radius, which is set to unity henceforth for simplicity. Moreover, by Taylor expanding the dilaton potential in powers of $\phi$ close to the boundary, we obtain
\begin{equation}\label{eq:VTay}
V(\phi) = -12-2\phi^2 +\mathcal{O}(\phi^4),
\end{equation}
which tells us that the mass of the dilaton is given by $m^{2}=-4$. Recalling that the relation between the mass of the dilaton and the scaling dimension of its dual operator in the boundary QFT is given by $m^2=\Delta(\Delta-4)$, one concludes that $\Delta = 2$. Note also that the dilaton field vanishes at the boundary such that the bulk geometry is asymptotically AdS$_5$.

Einstein's equations are obtained from the variation of the EMD action with respect to the metric,
\begin{align}
R_{MN}-\frac{1}{2}g_{MN}R=\kappa_{5}^{2} T_{MN},
\label{eq:ohoho}
\end{align}
where,
\begin{align}
T_{MN} = \frac{1}{\kappa_{5}^{2}}\left[ \frac{1}{2}\partial_M\phi\partial_N\phi+\frac{f(\phi)}{2}F_{MP}F_{N}^{\ P}-\frac{g_{MN}}{2}\left(\frac{1}{2}(\partial_P\phi)^2+V(\phi)+\frac{f(\phi)}{4} F_{PQ} F^{PQ}\right)\right],
\label{eq:ehehe}
\end{align}
is the stress-energy tensor of the matter fields $A_\mu$ and $\phi$. It is usually simpler to work with the ``trace-reversed'' form of Einstein's equations.\footnote{The tensorial algebra using a specific chart in this paper is done with the help of the Mathematica's package ``Riemann Geometry and Tensor Calculus'' (RGTC) \cite{RGTC}.} This may be derived by noting that Eq.\ \eqref{eq:ohoho} implies that
\begin{equation}
R_{MN}= \kappa_{5}^{2} \left(T_{MN}- \frac{1}{3}g_{MN}T^Q_Q \right).
\label{eq:ahaha}
\end{equation}
By substituting Eq.\ \eqref{eq:ehehe} into Eq.\ \eqref{eq:ahaha}, one rewrites Einstein's equations as follows
\begin{align}\label{eq:Einstein}
R_{MN}-\frac{g_{MN}}{3}\left[V(\phi)-\frac{f(\phi)}{4}F_{PQ} F^{PQ}\right] -\frac{1}{2}\partial_M\phi\partial_N\phi-\frac{f(\phi)}{2}F_{MP}F_{N}^{\ P}=0.
\end{align}

On the other hand, Maxwell's equations are obtained by varying the EMD action with respect to the Maxwell field
\begin{equation}\label{eq:MaxwellEq}
\nabla_{M}\left( f(\phi)F^{MN} \right)=0.
\end{equation}
The last equation that one needs to fully specify the dynamics of the model is the Klein-Gordon equation for the dilaton field
\begin{align}\label{eq:KleinGordon}
\nabla^2\phi-\frac{\partial V}{\partial\phi}- \frac{F_{PQ} F^{PQ}}{4}\frac{\partial f}{\partial\phi}=
\frac{1}{\sqrt{-g}}\partial_{M}\left(\sqrt{-g}g^{MN}\partial_N\phi \right)-\partial_\phi V- \frac{F_{PQ} F^{PQ}}{4}\partial_\phi f=0.
\end{align}

To explore the far-from-equilibrium solutions of the model, we adopt the well-known characteristic formulation for asymptotically AdS spacetimes \cite{Chesler:2008hg}. We consider here a time-dependent and spatially homogeneous anisotropic Ansatz for the metric field, which is suited to study homogeneous isotropization dynamics, in which one starts with an anisotropic configuration and an energy density that is conserved as time evolves. We work with generalized infalling Eddington-Finkelstein (EF) coordinates, in terms of which one may write the Ansatz for the line element as follows \cite{Chesler:2008hg},
\begin{equation}\label{lineElement}
ds^2= 2dv\left[dr-A(v,r) dv \right]+\Sigma(v,r)^2\left[e^{B(v,r)}(dx^2+dy^2)+e^{-2B(v,r)}dz^2\right],
\end{equation}
where $v$ represents the EF time, which near the boundary is interpreted as the time of the dual QFT. To see this, recall that the EF time is defined via
\begin{equation}
dv = dt +\sqrt{-\frac{g_{rr}}{g_{tt}}}dr,
\end{equation}
where $g_{rr}$ and $g_{tt}$ are the radial and temporal diagonal components of an asymptotically AdS$_{5}$ metric. Thus, as one goes to the boundary located at $r\rightarrow\infty$, one obtains $v\rightarrow t$. We also remark that in these generalized infalling EF coordinates, infalling radial null geodesics satisfy $v=\textrm{constant}$, while outgoing radial null geodesics satisfy $dr/dv=A(v,r)$ \cite{Chesler:2008hg}\footnote{Note that $A(v,r)$ here is half the corresponding metric function in the convention of Ref.\ \cite{Chesler:2008hg}.}. Furthermore, there is a residual diffeomorphism invariance in Eq.\ \eqref{lineElement} corresponding to the radial shift $r\mapsto r+\lambda(v)$, where $\lambda(v)$ is an arbitrary function of the EF time \cite{Chesler:2013lia}.

With respect to the Maxwell and dilaton fields, the Ans\"{a}tze are
\begin{equation}
A = \Phi(v,r)dv,\qquad \phi = \phi(v,r).
\end{equation}
The $U(1)$ R-charge chemical potential is associated with the boundary value of the Maxwell field $\Phi(v,r)$ in equilibrium, while the dilaton field vanishes at the boundary, as mentioned above.

The resulting equations of motion for the EMD system form a set of coupled nonlinear PDEs,
\begin{subequations}
\begin{align}
4 \Sigma (d_{+}\phi)'+6 \phi'd_{+}\Sigma +6 \Sigma'd_{+}\phi+\Sigma  \partial_\phi f \mathcal{E}^2-2 \Sigma  \partial_{\phi}V &=0, \label{PDE4}\\
\frac{\partial_{\phi}f \phi '}{f}+\frac{3 \Sigma '}{\Sigma } + \frac{\mathcal{E}'}{\mathcal{E}} &=0 \label{PDE6}\\
A''+\frac{1}{12} \left(18 B'd_{+}B-\frac{72 \Sigma'd_{+}\Sigma}{\Sigma^2}+6 \phi'd_{+}\phi-7 f \mathcal{E}^2-2 V\right) &=0, \label{PDE5}\\
(d_{+}\Sigma)' +\frac{2 \Sigma '}{\Sigma }d_{+}\Sigma+\frac{1}{12} \Sigma \left(f \mathcal{E}^2+2 V\right) &=0, \label{PDE2} \\
\Sigma \, (d_{+}B)'+\frac{3 \Sigma'}{2}d_{+}B+\frac{3 d_{+}\Sigma}{2} B' &=0, \label{PDE3} \\
\frac{1}{6} \Sigma  \left(3 \left(B'\right)^2+\left(\phi '\right)^2\right)+\Sigma '' &=0, \label{PDE1} \\
d_{+}(d_{+}\Sigma)-A' \,d_{+}\Sigma + A^2 \Sigma ''+ \frac{\Sigma}{6}\left(3 A^2\left(B'\right)^2 + A^2 \left(\phi'\right)^2 +3(d_{+}B)^2 +(d_{+}\phi)^2 \right) &=0, \label{PDE7}
\end{align}
\end{subequations}
where the prime denotes $\partial_r$, and
\begin{equation}\label{eq:dplus}
d_{+}\equiv \partial_{v}+A(v,r)\partial_r,
\end{equation}
defines the directional derivative along outgoing null vectors. Eq.\ \eqref{PDE4} is the dilaton equation, Eq.\ \eqref{PDE6} is the Maxwell equation, and Eqs.\ \eqref{PDE5}---\eqref{PDE7} are Einstein's equations\footnote{Eqs.\ \eqref{PDE1} and \eqref{PDE7} are constraints, whose derivatives are implied by the dynamical Eqs.\ \eqref{PDE5}---\eqref{PDE3}. Therefore, there is in total five dynamical equations, \eqref{PDE4}---\eqref{PDE3}, for five unknowns, $\{\phi,\Phi,A,\Sigma,B\}$.}. As in Ref.\ \cite{Fuini:2015hba}, we also defined a bulk ``electric field'',
\begin{equation}\label{eq:Efield}
\mathcal{E} \equiv -\Phi'.
\end{equation}
It is important to note that the Maxwell equation \eqref{PDE6} relates $\phi$, $\Sigma$, and $\Phi'$, i.e.
\begin{equation}\label{eq:constraint}
\ln(\Sigma^3\mathcal{E})-2\sqrt{\frac{2}{3}}\phi=\text{constant}.
\end{equation}
We will show in the next section how one can relate the above unknown constant to the $U(1)$ R-charge density $\rho_c$ using results from holographic renormalization. Indeed, Eq.\ \eqref{eq:constraint} expresses the existence of a Gauss charge coming from the Gauss law of classical electromagnetism. 

We explain now the general algorithm to solve the system of PDEs \eqref{PDE4}---\eqref{PDE7}:
\begin{enumerate}
\item Choose an initial profile for $B(v_0,r)$ and $\phi(v_0,r)$, where $v_0$ denotes the initial time;

\item Once $B(v_0,r)$ and $\phi(v_0,r)$ are given, one can solve Eq.\ \eqref{PDE1} to obtain $\Sigma(v_0,r)$. Using the constraint \eqref{eq:constraint} we also determine $\mathcal{E}(v_0,r)$;

\item With $B(v_0,r)$, $\phi(v_0,r)$, $\Sigma(v_0,r)$, and $\mathcal{E}(v_0,r)$ at hand, we proceed to solve Eq.\ \eqref{PDE2} for $d_{+}\Sigma(v_0,r)$;

\item Next we solve Eq.\ \eqref{PDE3} for $d_{+}B(v_0,r)$;

\item Next we solve Eq.\ \eqref{PDE4} for $d_{+}\phi(v_0,r)$;

\item Next we solve Eq.\ \eqref{PDE5} for $A(v_0,r)$;

\item When $B(v_0,r)$, $d_{+}B(v_0,r)$, $\phi(v_0,r)$, $d_{+}\phi(v_0,r)$, and $A(v_0,r)$ are known, it is clear from the definition of $d_{+}$ in Eq.\ \eqref{eq:dplus} that we also have $\partial_v B(v_0,r)$ and $\partial_v \phi(v_0,r)$. With $\left\{B(v_0,r),\partial_v B(v_0,r)\right\}$ and $\left\{\phi(v_0,r),\partial_v \phi(v_0,r)\right\}$ at hand we have now the initial conditions required to evolve in time $B(v_0)$ and $\phi(v_0,r)$ from $v_0$ to $v_0+\Delta v$;

\item Next we repeat the process to obtain the fields at the next instant $v_0+\Delta v$.

\item The constraint \eqref{PDE7} is useful for checking the accuracy of the numerical solutions.
\end{enumerate}

Before we delve into the numerics and solve the system of PDEs \eqref{PDE4}---\eqref{PDE7} following the above algorithm, there are still some technical details that we need to take into account. Some of these details are: the boundary conditions, the equilibrium solutions, and the initial data.

In order to fix the boundary conditions for the EMD fields we first need to perform a near-boundary expansion of the bulk fields. This expansion will reveal which ultraviolet coefficients need to be dynamically fixed. Such analysis is done in Sec.\ \ref{sec:AsymExp}.
The detailed knowledge of the equilibrium solutions, which are discussed in Sec.\ \ref{sec:EqSol}, is very important to determine the final state given some initial geometry. Moreover, it is only possible to predict the equilibrium state because in this homogeneous setup the energy and charge density are constant. Still in Sec.\ \ref{sec:EqSol}, using the formulas derived from holographic renormalization in Appendix \ref{sec:HoloRen}, we provide holographic formulas to calculate the energy density and the pressure coming from $\langle T_{\mu\nu} \rangle$, the charge density that comes from the temporal component of $\langle J^{\mu} \rangle$, and the expectation value of the scalar operator dual to the dilaton field, $\langle\mathcal{O}_{\phi} \rangle$. In Sec.\ \ref{sec:FarSol}, we discuss different initial data and specify the numerical techniques that we shall employ to solve the equations of motion of the 1RCBH model. After that, we will be ready to present in Sec.\ \ref{sec:results} our results for the homogeneous isotropization and thermalization of the 1RCBH plasma, emphasizing in particular the dynamics near the critical point.

%%%%%%%%%%%%%%%%%%%%%%%%%
\section{Near-boundary expansion of the bulk fields}
\label{sec:AsymExp}

In a five dimensional model where the dilaton field has scaling dimension $\Delta=2$ the near-boundary expansions of the bulk fields are given by integer powers of the radial coordinate plus logarithmic terms \cite{Bianchi:2001kw}, i.e.
\begin{align}
A(v,r) & = \frac{1}{2}(r+\lambda(v))^2-\partial_v\lambda(v) + \sum_{n=0,m=0,m<n}^{n,m=\infty}\frac{A_{n,m}(v)}{r^n}\ln^m r, \label{eq:expA}\\
\Sigma(v,r) & = r+\lambda(v) + \sum_{n=0,m=0,m<n}^{n,m=\infty}\frac{\Sigma_{n,m}(v)}{r^n}\ln^m r, \label{eq:expSig}\\
B(v,r) & = \sum_{n=0,m=0,m<n}^{n,m=\infty}\frac{B_{n,m}(v)}{r^n}\ln^m r, \label{eq:expB}\\
\phi(v,r) & = \sum_{n=2,m=0,m<n}^{n,m=\infty}\frac{\phi_{n,m}(v)}{r^n}\ln^m r, \label{eq:expphi}\\
\Phi(v,r) & = \sum_{n=0,m=0,m<n}^{n,m=\infty}\frac{\Phi_{n,m}(v)}{r^n}\ln^m r, \label{eq:expPhi}
\end{align}
where $\lambda(v)$ is the radial shift function associated with the aforementioned residual diffeomorphism invariance of the bulk metric \cite{Chesler:2013lia}. As we shall see in Sec.\ \ref{sec:HoloRen}, the logarithmic terms in the above expansions vanish due to three main facts: the conformal flatness of the boundary, the scaling dimension of the QFT scalar operator dual to the bulk dilaton field, and the conformal symmetry of the system. Therefore, we can already set the logarithmic corrections to zero. We adopt then the following notation
\begin{equation}
\lbrace A_{n},\, \Sigma_{n},\, B_{n},\, \phi_{n},\, \Phi_{n}  \rbrace \equiv \lbrace A_{n,0},\, \Sigma_{n,0},\, B_{n,0},\, \phi_{n,0},\, \Phi_{n,0}  \rbrace.
\end{equation}

Substituting the expansions \eqref{eq:expA}---\eqref{eq:expPhi} into the equations of motion \eqref{PDE4}---\eqref{PDE7}, setting the radial shift function $\lambda(v)$ to zero, and eliminating all possible coefficients in favor of the others, the ultraviolet asymptotic behavior of the EMD fields reads
\begin{align}
A(v,r) &= \frac{r^2}{2}+\frac{H-\phi _2(v){}^2/18}{r^2}-\frac{\phi _2(v) \dot{\phi}_2(v)}{18 r^3}+\mathcal{O}(r^{-4}) \label{eq:nearBdry1} \\
\Sigma(v,r) &= r-\frac{\phi _2(v){}^2}{18 r^3}-\frac{\phi _2(v) \dot{\phi}(v)}{10 r^4} + \mathcal{O}(r^{-5})\\
d_{+}\Sigma(v,r) &= \frac{r^2}{2}+\frac{H+1/36\,\phi _2(v){}^2}{r^2}+\mathcal{O}(r^{-3})\\
B(v,r) &= \frac{B_4(v)}{r^4}+\frac{\dot{B}_4(v)}{r^5}+ \mathcal{O}(r^{-6}) \\
\phi(v,r) &= \frac{\phi _2(v)}{r^2}+\frac{\dot{\phi}_2(v)}{r^3}+\frac{\sqrt{6} \phi _2(v){}^2+9 \ddot{\phi}_2(v)}{12r^4}+  \mathcal{O}(r^{-5})\\
d_{+}\phi(v,r) &= -\frac{\phi _2(v)}{r}+ \mathcal{O}(r^{-2})\\
\Phi(v,r) &= \Phi _0(v)+\frac{\Phi _2(v)}{r^2}+\frac{\sqrt{\frac{2}{3}} \phi _2(v) \Phi _2(v)}{r^4} + \mathcal{O}(r^{-5}). \label{eq:nearBdry5}
\end{align}
where the dot represents the time derivative $\partial_v$.

Furthermore, we find that this near-boundary analysis cannot determine five coefficients: $A_2(v)$ (or, equivalently, the coefficient $H$ defined below), $B_{4}(v)$, $\phi_{2}(v)$, $\Phi_{0}(v)$, and $\Phi_{2}(v)$. With the exception of $\Phi_{0}$, these coefficients are \emph{dynamical}, i.e. we need the bulk solution to determine them. The coefficient $\Phi_{0}(v)$ is fixed by the Dirichlet boundary condition for the Maxwell field which imposes that its boundary value gives the chemical potential associated with the $U(1)$ R-symmetry\footnote{The Dirichlet boundary condition for the metric field $g_{\mu\nu}$ has been already implemented when we imposed an asymptotically AdS$_5$ solution.}
\begin{equation}
\mu = \lim_{v\rightarrow\infty}\Phi_{0}(v).
\end{equation}
The coefficient $H$ is defined by
\begin{equation}
H\equiv A_{2}(v)+\frac{\phi_{2}(v)^{2}}{18}.
\end{equation}
By working out the equations of motion up to $\mathcal{O}(r^{-3})$ in the near-boundary expansions of the bulk fields one concludes that $H$ is a constant. Moreover, as we shall see in a moment, this coefficient $H$ is, up to a numerical factor, the energy of the system which is conserved in this homogeneous setup. On the other hand, the coefficients $B_{4}(v)$ and $\phi_{2}(v)$ are time-dependent quantities related to the pressure anisotropy and the scalar condensate dual to the dilaton field, respectively.

Additionally, the coefficient $\Phi_{2}(v)$ is actually a constant. This can be shown by expanding the equations of motion near the boundary up to $\mathcal{O}(r^{-5})$. Moreover, by exploring relation \eqref{eq:constraint} near the boundary using the expansions \eqref{eq:nearBdry1}---\eqref{eq:nearBdry5}, one finds that,
\begin{equation}\label{eq:Efield2}
\mathcal{E}(v,r) = 2\Phi_{2}\Sigma(v,r)^{-3}e^{2\sqrt{\frac{2}{3}}\phi(v,r)}.
\end{equation}
We shall discuss how the charge density can be related to $\Phi_{2}$ in Sec. \ref{sec:EqSol}.

%%%%%%%%%%%%%%%%%%%%%%%%%
\section{Equilibrium solutions}
\label{sec:EqSol}

Here we briefly review the main features of the equilibrium solution of the 1RCBH plasma and its thermodynamics \cite{DeWolfe:2011ts,Finazzo:2016psx}.

\subsection{Thermodynamics}
\label{sec:thermo}

\begin{figure}[t]
	\centering
	\begin{subfigure}{0.49\textwidth}
		\includegraphics[width=\textwidth]{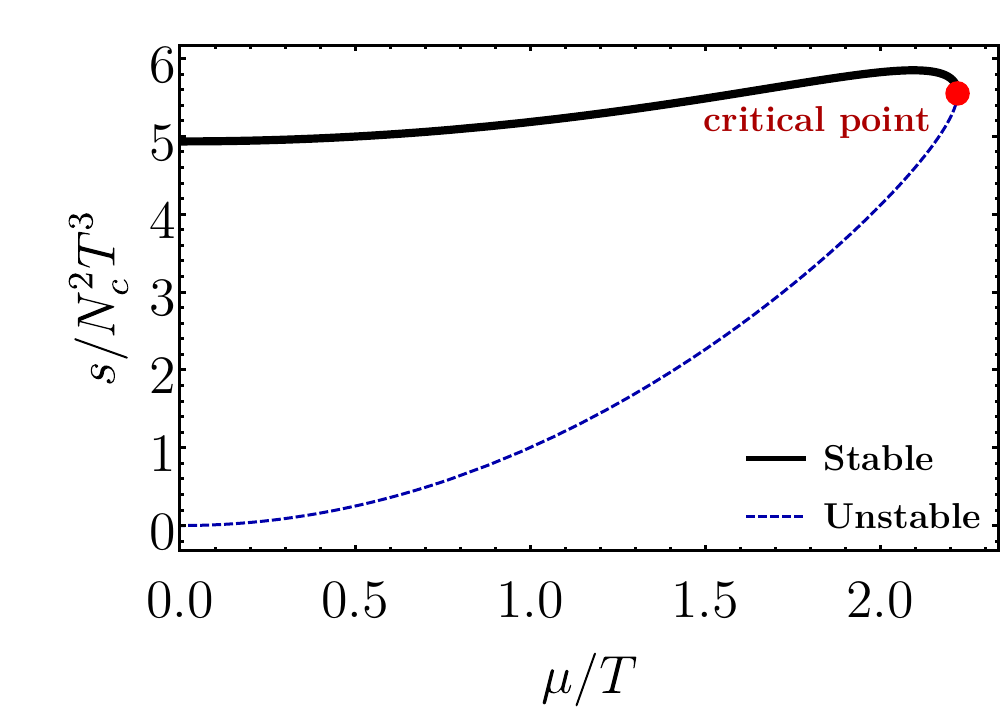}
		\caption{Entropy density}
	\end{subfigure}
	\begin{subfigure}{0.49\textwidth}
		\includegraphics[width=\textwidth]{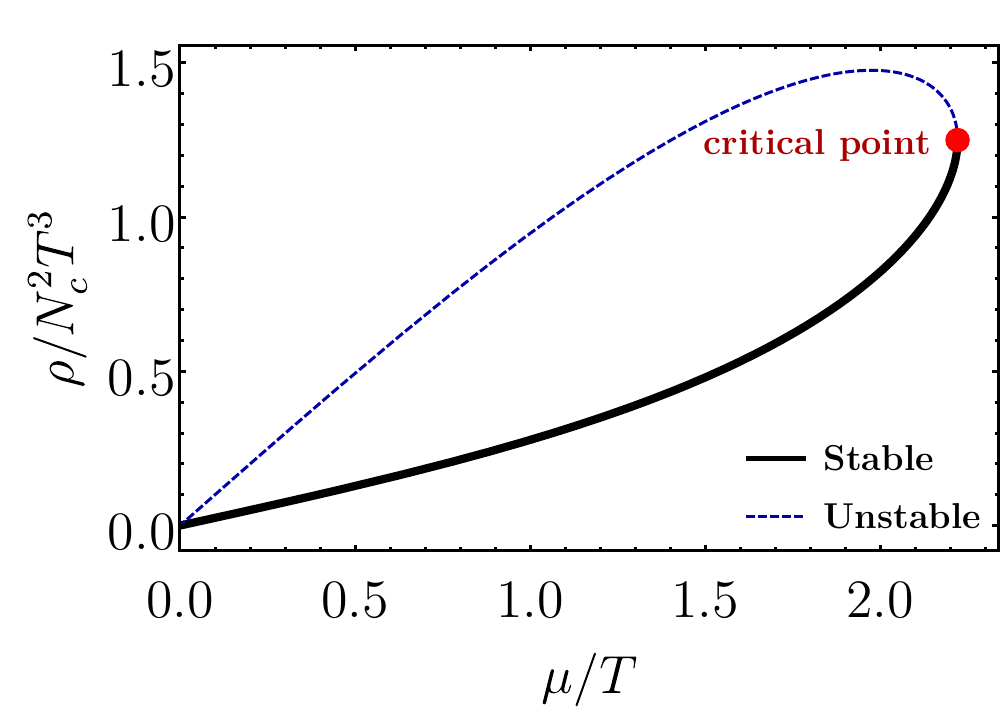}
		\caption{Charge density}
	\end{subfigure}
	\begin{subfigure}{0.49\textwidth}
		\includegraphics[width=\textwidth]{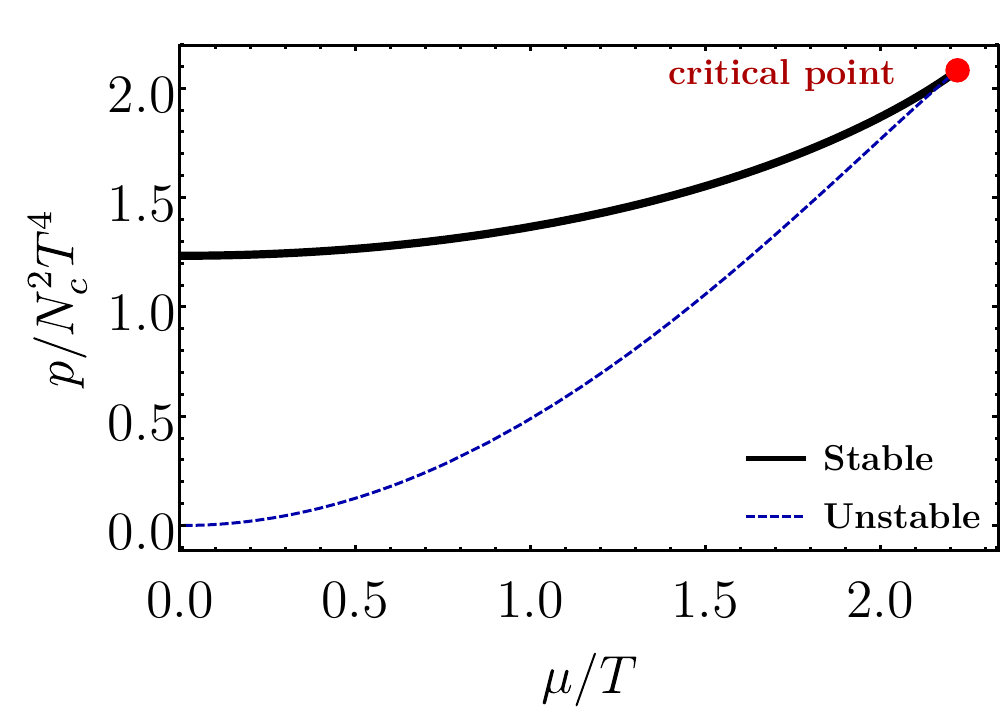}
		\caption{Pressure}
	\end{subfigure}
	\begin{subfigure}{0.49\textwidth}
		\includegraphics[width=\textwidth]{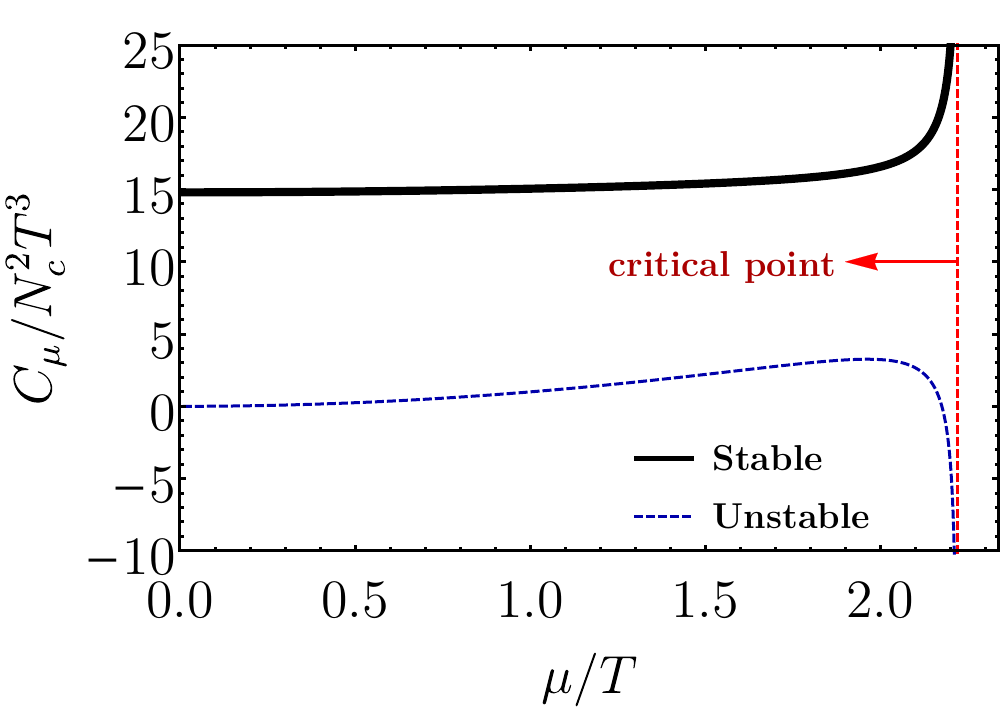}
		\caption{Heat capacity}
	\end{subfigure}
	\caption{(Color online) Thermodynamic quantities for the 1RCBH model. These figures are taken from Ref.\ \cite{Finazzo:2016psx}.} \label{fig:thermo}
\end{figure}

In equilibrium this model has the following analytical solution (written in a slightly different chart from Eq.\ \eqref{lineElement}, which we call the ``modified EF'' coordinates and denote with a tilde),
\begin{align}
d s^2 & = dv\left[2 e^{a(\tilde{r})+b(\tilde{r})}d\tilde{r} - e^{2a(\tilde{r})}h(\tilde{r})dv\right] + e^{2a(\tilde{r})}d\vec{x}^2 \label{eq:modEF}\\
a(\tilde{r}) & = \ln  \tilde{r}  + \frac{1}{6} \ln \left( 1 + \frac{Q^2}{\tilde{r}^2} \right), \label{eq:atil} \\
b(\tilde{r}) & = - \ln \tilde{r}  -  \frac{1}{3} \ln \left( 1 + \frac{Q^2}{\tilde{r}^2} \right), \label{btil} \\
h(\tilde{r}) & = 1 - \frac{M^2}{\tilde{r}^2(\tilde{r}^2+Q^2)},\label{eq:htil} \\
\phi (\tilde{r}) & = -\sqrt{\frac{2}{3}} \ln \left( 1 + \frac{Q^2}{\tilde{r}^2} \right),\label{eq:scalar field} \\
\Phi(\tilde{r})& = \left(-\frac{M Q}{\tilde{r}^2+Q^2} +\frac{M Q}{\tilde{r}_H^2+Q^2}\right),\label{eq:electromagnetic four-potential}
\end{align}
where $\tilde{r}_h$ is the radius of the black hole's event horizon given by
\begin{equation}
\tilde{r}_h = \sqrt{\frac{\sqrt{Q^4 + 4 M^2} - Q^2}{2}}.
\end{equation}
As discussed in Refs.\ \cite{DeWolfe:2011ts,Finazzo:2016psx} this model is characterized by two non-negative parameters $(Q,M)$ or, alternatively, $(Q,\tilde{r}_h)$. They are related to the Hawking temperature of the black hole  
\begin{align}
\label{eq:temperature}
T = \frac{Q^2 + 2 \tilde{r}_h^2}{2 \pi \sqrt{Q^2+\tilde{r}_h^2}},
\end{align}
and the $U(1)$ R-charge chemical potential
\begin{align}
\label{eq:chemical potential}
\mu = \lim\limits_{\tilde{r}\to\infty}\Phi(\tilde{r}) = \frac{Q \tilde{r}_h}{\sqrt{Q^2 + \tilde{r}_h^2}}.
\end{align}
In fact, standard algebraic manipulations show that
\begin{align}\label{eq:critical_point}
\dfrac{Q}{\tilde{r}_h}= \sqrt{2}\left(\dfrac{1\pm\sqrt{1-\left(\frac{\mu/T}{\pi/\sqrt{2}}\right)^2}}{\frac{\mu/T}{\pi/\sqrt{2}}}\right).
\end{align}
Since $Q/\tilde{r}_h$ is non-negative, Eq.\ \eqref{eq:critical_point} implies that $\mu/T\in\big[0,\pi/\sqrt{2}\big]$. It also follows from \eqref{eq:critical_point} that for every value of $\mu/T\in\big[0,\pi/\sqrt{2}\big)$ there are two different values of $Q/\tilde{r}_h$, which in turn parameterize two different branches of solutions. By analyzing the thermodynamics in Fig.\ \ref{fig:thermo}, one can see the two branches merge at $\mu/T=\pi/\sqrt{2}$, which defines a critical point marking a 2nd order phase transition.

Since $\frac{1}{\kappa_5^2} = \frac{N_c^2}{4\pi^2}$ for a SYM plasma \cite{Gubser:1996de}, the entropy density can be determined from Bekenstein-Hawking's relation \cite{Bekenstein:1973ur,Hawking:1974sw} as follows
\begin{align}\label{eq:entropy density}
\dfrac{s}{N_c^2T^3}=\dfrac{\pi^2}{16}\left(3\pm\sqrt{1-\left(\dfrac{\mu/T}{\pi/\sqrt{2}}\right)^2}\right)^2 \left(1\mp\sqrt{1-\left(\dfrac{\mu/T}{\pi/\sqrt{2}}\right)^2}\right).
\end{align}
On the other hand, the $U(1)$ R-charge density, $\rho_c=\lim_{\tilde{r}\to\infty}\delta S/\delta\Phi'$, may be expressed as
\begin{align}\label{eq:charge density}
\dfrac{\rho_c}{N_c^2T^3}=\dfrac{\mu/T}{16}\left(3\pm\sqrt{1-\left(\dfrac{\mu/T}{\pi/\sqrt{2}}\right)^2}\right)^2,
\end{align}
where the stable/unstable branches correspond to lower/upper signs, as in Eq.\ \eqref{eq:critical_point}. Furthermore, standard thermodynamic relations give us 
the pressure
\begin{align}\label{eq:pressure}
\dfrac{p}{N_c^2 T^4}= \dfrac{\pi^2}{128}\left(3\pm\sqrt{1-\left(\dfrac{\mu/T}{\pi/\sqrt{2}}\right)^2}\right)^3 \left(1\mp\sqrt{1-\left(\dfrac{\mu/T}{\pi/\sqrt{2}}\right)^2}\right).
\end{align}
which can be shown to be $p=\varepsilon/3$ with the internal energy density given by $\varepsilon=Ts-p+\mu\rho$. This is the expected result for a conformal field theory (CFT) in 4 spacetime dimensions.

\subsection{Mapping between FG \eqref{eq:FGline} and the modified EF \eqref{eq:modEF} coordinates}
\label{sec:EqSol:FGtoModEF}

Once the equilibrium solution is known, one may compute the one-point functions of the dual QFT by using Eqs.\ \eqref{eq:DilVEV}, \eqref{eq:Jmu}, and \eqref{eq:Tij} derived in Appendix \ref{sec:HoloRen}. However, we note that these equations are written in the so-called Fefferman-Graham (FG) coordinates and, thus, we need to find a relation between these coordinates and the modified EF coordinates in order to calculate $\langle T_{\mu\nu} \rangle$, $\langle J^{\mu} \rangle$, and $\langle \mathcal{O}_{\phi} \rangle$ in equilibrium.

First we consider the diagonal form of the line element \eqref{eq:modEF},
\begin{equation}\label{eq:ModEFtoFG}
ds^2  = e^{2a(\tilde{r})}(-h(\tilde{r})dt^2 + d\vec{x}^2) + \frac{e^{2b(\tilde{r})}}{h(\tilde{r})}d\tilde{r}^2.
\end{equation}
The task now is to find a relation between $\tilde{r}$ and $\rho$ (the radial coordinate of the FG chart discussed in Appendix \ref{sec:HoloRen}). This can be achieved by evaluating the following integral
\begin{equation}\label{eq:IntrTilde}
\int \frac{e^{b(\tilde{r})}}{\sqrt{h(\tilde{r})}}d\tilde{r} = -\frac{1}{2}\ln\rho.
\end{equation}
This integral can be solved perturbatively in $\tilde{r}$ close to the boundary in order to determine $\tilde{r}(\rho)$ as a series expansion. Such expansion will enable us to expand the bulk fields near the boundary in the form of Eqs.\ \eqref{eq:FGgamma}---\eqref{eq:FGphi}, which is what we need to fix the ultraviolet coefficients $\gamma_{(4)\mu\nu}$, $A_{(0)\mu}$, and $\phi_{(0)}$ by comparing these expansions with the equilibrium solution. With these ultraviolet coefficients fixed, we may proceed to calculate $\langle T^{\mu\nu} \rangle$, $\langle J^{\mu} \rangle$ and $\langle \mathcal{O}_{\phi} \rangle$ using Eqs.\ \eqref{eq:Tij}, \eqref{eq:Jmu}, and \eqref{eq:DilVEV}, respectively. Thus, by perturbatively evaluating the integral \eqref{eq:IntrTilde} close to the boundary using the equilibrium solution, we find that
\begin{equation}\label{eq:TilrToRho}
\tilde{r} = \frac{1}{\sqrt{\rho }}-\frac{Q^2 \sqrt{\rho }}{6}+\frac{1}{72} \left(9 M^2+Q^4\right) \rho ^{3/2} +\mathcal{O}(\rho)^{5/2}.
\end{equation}

The next step is to expand the bulk fields near the boundary and substitute $\tilde{r}$ by $\rho$ using Eq.\ \eqref{eq:TilrToRho}. This gives the following asymptotic results,
\begin{align}
\gamma_{tt}(\rho) & = -\frac{1}{\rho} - \frac{\rho}{2}\left(-\frac{3}{2}M^2-\frac{Q^4}{9} \right)+\mathcal{O}(\rho^2),\\
\gamma_{xx}(\rho) & = \frac{1}{\rho} + \frac{\rho}{2}\left(\frac{1}{2}M^2-\frac{Q^4}{9} \right)+\mathcal{O}(\rho^2),\\
\Phi(\rho) & = \frac{M Q}{Q^2+\tilde{r}_h^2}-MQ \rho +\mathcal{O}(\rho^2),\\
\phi(\rho) & = -\sqrt{\frac{2}{3}}Q^2\rho+\mathcal{O}(\rho^2).
\end{align}

By comparing the above expressions with Eqs.\ \eqref{eq:FGgamma}---\eqref{eq:FGphi} (taking into account that all the logarithmic terms vanish for the 1RCBH model, as stated before), and then using Eqs.\ \eqref{eq:Tij}, \eqref{eq:Jmu}, and \eqref{eq:DilVEV}, we find 
\begin{align}
\langle T_{tt} \rangle &= \frac{1}{\kappa_{5}^{2}}\frac{3M^2}{2},\\
\langle T_{xx} \rangle &= \frac{1}{\kappa_{5}^{2}}\frac{M^2}{2},\\
\langle J^{t} \rangle &= \frac{1}{\kappa_{5}^{2}} MQ, \\
\langle \mathcal{O}_{\phi} \rangle &= \frac{1}{\kappa_{5}^{2}}\sqrt{\frac{2}{3}}Q^2.
\end{align}

At this stage, it is useful to rewrite the energy density ($\varepsilon$), the equilibrium pressure ($p$), and the $U(1)$ R-charge density ($\rho_c$) as
\begin{align}\label{eq:RescQuant}
\varepsilon & \equiv \langle T_{tt} \rangle = \frac{N_{c}^{2}}{4\pi^2}\frac{3M^2}{2}, \ \ \ p  \equiv \langle T_{xx} \rangle =\frac{N_{c}^{2}}{4\pi^2} \frac{M^2}{2}, \ \ \ \rho_c \equiv \langle J^{t} \rangle  = \frac{N_{c}^{2}}{4\pi^2} MQ.
\end{align}
The above results hold for the 1RCBH plasma in equilibrium.

\subsection{Mapping between modified EF \eqref{eq:modEF} and the original EF \eqref{lineElement} coordinates}
\label{sec:EqSol:EFtoModEF}

The purpose of this subsection is to find relations between the coefficients of the non-equilibrium solution in the near-boundary expansions \eqref{eq:nearBdry1}---\eqref{eq:nearBdry5}, and the parameters of the equilibrium solution for the 1RCBH plasma \eqref{eq:modEF}. Evidently, we can only match the equilibrium solution with the asymptotically late time behavior of the non-equilibrium solution when thermodynamic equilibrium is achieved by the system. For this reason, before we derive an expression for $r(\tilde{r})$, we define the equilibrium coefficients of the non-equilibrium solution as follows,
\begin{equation}\label{eq:Xeq}
X^{(eq)}\equiv \lim_{v\rightarrow\infty}X(v), \ \ \ \text{with} \ \ \partial_{v} X^{(eq)}=0,
\end{equation}
where $X(v) \in \lbrace B_{4}(v),\phi_{2}(v),\Phi_{0}(v) \rbrace$. Note that the coefficients $H$ and $\Phi_2$ are not included in this definition because they are constant, as stated before.

In order to find the relation between $\tilde{r}$ and $r$, we need to solve the following equation
\begin{equation}
e^{a(\tilde{r})+b(\tilde{r})}d\tilde{r} = dr.
\end{equation}
The above expression renders an analytical relation between $r$ and $\tilde{r}$,
\begin{align}\label{eq:rTorTild}
r = \frac{3 \tilde{r}^{4/3} \, _2F_1\left(\frac{1}{6},\frac{2}{3};\frac{5}{3};-\frac{\tilde{r}^2}{Q^2}\right)}{4\sqrt[3]{Q}} + \xi,
\end{align}
where $_2F_1$ is the hypergeometric function and $\xi$ is an integration constant which we choose to be
\begin{equation}
\xi = \frac{3\sqrt{\pi}\,\Gamma\left[\frac{5}{3}\right]}{2\,\Gamma\left[\frac{1}{6}\right]} Q,
\end{equation}
in order to obtain $\lim_{\tilde{r}\rightarrow\infty} r(\tilde{r})-\tilde{r} =0$. Just as we did in Sec.\ \ref{sec:EqSol:FGtoModEF} to relate $\tilde{r}$ with $\rho$, we expand the above relation in powers of $\tilde{r}$ close to the boundary
\begin{equation}
r(\tilde{r}) = \tilde{r}+\frac{Q^2}{6 \tilde{r}}-\frac{7 Q^4}{216 \tilde{r}^3}+\frac{91 Q^6}{6480 \tilde{r}^5} +\mathcal{O}(\tilde{r}^{-7}).
\end{equation}

Next, we substitute the above expression into Eqs.\ \eqref{eq:nearBdry1}---\eqref{eq:nearBdry5} and take the asymptotic equilibrium limit defined by Eq.\ \eqref{eq:Xeq}. The result reads
\begin{subequations}
\begin{align}
g_{tt}(\tilde{r}) & = -2A(\tilde{r}) = -\tilde{r}^2-\frac{Q^2}{3}+\frac{Q^4-54 H+3 (\phi^{(eq)}_2)^2}{27 \tilde{r}^2} + \mathcal{O}(\tilde{r}^{-4}), \label{eq:AsymEq1} \\
g_{xx}(\tilde{r}) & = \Sigma(\tilde{r})^2 = \tilde{r}^2+\frac{Q^2}{3}+\frac{-Q^4-3 (\phi^{(eq)}_2)^2}{27 \tilde{r}^2} + \mathcal{O}(\tilde{r}^{-4}), \\
\phi(\tilde{r}) & = \frac{\phi^{(eq)}_2}{\tilde{r}^2}+\frac{-4 Q^2 \phi^{(eq)}_2+\sqrt{6}(\phi^{(eq)}_2)^2}{12 \tilde{r}^4} + \mathcal{O}(\tilde{r}^{-6}), \\ 
\Phi(\tilde{r}) & =  \Phi_{0}^{(eq)}+\frac{\Phi _2}{r^2}-\frac{\left(Q^2-\sqrt{6}\phi_{2}^{(eq)}\right) \Phi _2}{3 r^4}+ \mathcal{O}(\tilde{r}^{-6}) \label{eq:AsymEq4}.
\end{align}
\end{subequations}

On the other hand, the near-boundary behavior of the analytical static solution of the 1RCBH model obtained from Eqs.\ \eqref{eq:atil}---\eqref{eq:electromagnetic four-potential} is
\begin{subequations}
\begin{align}
g_{tt}(\tilde{r}) & = -\tilde{r}^2-\frac{Q^2}{3}+\frac{M^2+\frac{Q^4}{9}}{\tilde{r}^2}+\frac{-\frac{1}{3} \left(2 M^2 Q^2\right)-\frac{5Q^6}{81}}{\tilde{r}^4}+\frac{\frac{5 M^2 Q^4}{9}+\frac{10 Q^8}{243}}{\tilde{r}^6}+ \mathcal{O}(\tilde{r}^{-8}), \label{eq:AsymStat1}\\
g_{xx}(\tilde{r}) & = \tilde{r}^2+\frac{Q^2}{3}-\frac{Q^4}{9 \tilde{r}^2}+\frac{5 Q^6}{81 \tilde{r}^4}-\frac{10 Q^8}{243 \tilde{r}^6} + \mathcal{O}(\tilde{r}^{-8}),\\
\phi(\tilde{r}) & = -\frac{\sqrt{\frac{2}{3}} Q^2}{\tilde{r}^2}+\frac{Q^4}{\sqrt{6} \tilde{r}^4}-\frac{\sqrt{\frac{2}{3}} Q^6}{3 \tilde{r}^6} + \mathcal{O}(\tilde{r}^{-8}),\\ 
\Phi(\tilde{r}) & = \frac{M Q}{Q^2+\tilde{r}_{h}^2}-\frac{M Q}{\tilde{r}^2}+\frac{M Q^3}{\tilde{r}^4}-\frac{M Q^5}{\tilde{r}^6} + \mathcal{O}(\tilde{r}^{-8}). \label{eq:AsymStat4}
\end{align}
\end{subequations}
Comparing Eqs.\ \eqref{eq:AsymEq1}---\eqref{eq:AsymEq4} and Eqs.\ \eqref{eq:AsymStat1}---\eqref{eq:AsymStat4}, we find that
\begin{align}
\phi_{2}^{(eq)} & = -\sqrt{\frac{2}{3}} Q^2, \\
H & =-\frac{M^2}{2}, \\
\Phi_{0}^{(eq)} & = \frac{M Q}{Q^2+\tilde{r}_h^2}, \\
\Phi_{2} & = -M Q.
\end{align}

\subsection{Mapping between EF \eqref{lineElement} and FG \eqref{eq:FGline} coordinates}
\label{sec:EqSol:EFtoFG}

Now we want to relate the non-equilibrium metric in EF coordinates \eqref{lineElement} with the expression for the metric in FG coordinates \eqref{eq:FGline} in order to obtain expressions for the one-point functions $\langle T_{\mu\nu} \rangle$, $\langle J^{\mu}\rangle$, and $\langle \mathcal{O}_{\phi} \rangle$ far-from-equilibrium. To do so, we need to solve the integral below
\begin{equation}
\int \frac{dr}{\sqrt{2A(v,r)}} = -\frac{1}{2}\ln\rho,
\end{equation}
which, again, may be solved perturbatively near the boundary with the help of Eq.\ \eqref{eq:nearBdry1}. The result is, 
\begin{equation}
r(\rho) = \frac{1}{\sqrt{\rho }}+\rho ^{3/2} \left(-\frac{H}{4}+\frac{1}{72} \phi _2(v){}^2\right)+\frac{1}{90}\rho ^2 \phi _2(v) \dot{\phi}_2(v) +\mathcal{O}(\rho^{5/2}).
\end{equation}

Plugging $r(\rho)$ into Eqs.\ \eqref{eq:nearBdry1}---\eqref{eq:nearBdry5}, one obtains
\begin{subequations}
\begin{align}
\gamma_{tt}(\rho) & = -\frac{1}{\rho }+\frac{1}{12} \rho  \left(-18 H+\phi_{2}(v)^{2}\right) + \mathcal{O}(\rho^{2}), \\
\gamma_{xx}(\rho) & = \frac{1}{\rho }+\rho  \left(-\frac{H}{2}+B_4(v)-\frac{1}{12} \phi _2(v){}^2\right) + \mathcal{O}(\rho^{2}),\\
\gamma_{zz}(\rho) & = \frac{1}{\rho }+\rho  \left(-\frac{H}{2}-2 B_4(v)-\frac{1}{12} \phi _2(v){}^2\right) + \mathcal{O}(\rho^{2}),\\
\phi(\rho) & = \rho \phi _2(v) + \mathcal{O}(\rho^{2}),\\ 
\Phi(\rho) & = \Phi _0(v)+\rho  \Phi _2(v) + \mathcal{O}(\rho^{2}). 
\end{align}
\end{subequations}

With the above set of equations at hand, we obtain $\langle T^{\mu\nu} \rangle$, $\langle J^{\mu}\rangle$, and $\langle \mathcal{O}_{\phi} \rangle$ from Eqs.\ \eqref{eq:Tij}, \eqref{eq:Jmu}, and \eqref{eq:DilVEV}, respectively,
\begin{align}
\langle T_{tt} \rangle & = \frac{1}{\kappa_{5}^{2}}(-3 H),\\
\langle T_{xx} \rangle & = \frac{1}{\kappa_{5}^{2}}(-H+2B_{4}(v)),\\
\langle T_{zz} \rangle & = \frac{1}{\kappa_{5}^{2}}(-H-4B_{4}(v)),\\
\langle J^{t} \rangle & = - \frac{1}{\kappa_{5}^{2}} \Phi_{2}, \\
\langle \mathcal{O}_{\phi} \rangle &= -\frac{1}{\kappa_{5}^{2}}\phi_{2}(v).
\end{align}

Analogously to what was done in Eqs.\ \eqref{eq:RescQuant}, one may recast the physical observables of the boundary QFT as follows
\begin{subequations}
\begin{align}
\varepsilon & = \langle T_{tt} \rangle = \frac{N_{c}^{2}}{4\pi^2}(-3 H), \\
\Delta p(v) & \equiv \langle T_{xx} \rangle - \langle T_{zz}\rangle =\frac{N_{c}^{2}}{4\pi^2} 6B_{4}(v) , \label{eq:DeltaP}\\
\rho_{c} & = \langle J^t \rangle = -\frac{N_{c}^{2}}{4\pi^2}\Phi_{2}, \\
\langle \mathcal{O}_{\phi} \rangle(v) &= -\frac{N_{c}^{2}}{4\pi^2}\phi_{2}(v), \label{eq:DilCondensate}
\end{align}
\end{subequations}
where $\Delta p(v)$ is the pressure anisotropy. The above expressions hold for the far-from-equilibrium 1RCBH plasma undergoing a spatially homogeneous equilibration process. Moreover, in terms of the charge density $\rho_c$, the bulk electric field defined in Eq.\ \eqref{eq:Efield2} is given by
\begin{equation}
\mathcal{E}(v,r) =-\frac{8\pi^2\rho_c}{N_c^2}\Sigma^{-3}(v,r)e^{2\sqrt{\frac{2}{3}}\phi(v,r)}.
\end{equation}

Thus, we finish here the discussion of the quantities that we need to follow when solving the full nonlinear PDEs of the 1RCBH model far-from-equilibrium. Next, we discuss how to proceed with the numerics to obtain the time evolution of the 1RCBH plasma.

%%%%%%%%%%%%%%%%%%%%%%%%%
\section{Far-from-equilibrium solutions}
\label{sec:FarSol}

\subsection{Field redefinitions}
\label{sec:FarSol:FieldRed}

First, we express the system of PDEs in a form where the numerical analysis becomes as simple as possible, mapping the infinite radial domain to a finite one. This may be accomplished by redefining the radial coordinate as follows
\begin{equation}\label{eq:uDef}
u = \frac{1}{r},
\end{equation}
which means that the boundary now lies at $u=0$. How far we go into the bulk in a numerical simulation is an issue which shall be discussed in Sec.\ \ref{sec:FarSol:HolePosit}. Moreover, since all the nontrivial dynamics of the system depends on the subleading terms of the near-boundary expansion of the bulk fields, as indicated in Eqs.\ \eqref{eq:expA}---\eqref{eq:expPhi}, we define the following subtracted fields\footnote{Our redefinitions are similar to the ones given in Ref.\ \cite{vanderSchee:2014qwa}. We also remark that $d_{+}(\Sigma_s)\neq (d_{+}\Sigma)_{s}$.}
\begin{align}\label{eq:Redefinitions}
u^2 A_{s} = A -\frac{1}{2u^2}, \ \ \ u^4 B_{s} = B, \ \ \ u^2 \Sigma_{s} = \Sigma -\frac{1}{u}, \ \ \ u^2\phi_{s} = \phi, \ \ \ \mathcal{E}_s = \mathcal{E}, \notag \\
u^2(d_{+}\Sigma)_{s} =d_{+}\Sigma  -\frac{1}{2u^2},\ \ \  u^3(d_{+}B)_{s} = d_{+}B, \ \ \  u(d_{+}\phi)_{s} = d_{+}\phi.
\end{align}

The equations of motion \eqref{PDE4}---\eqref{PDE1} rewritten in terms of these subtracted fields read
\begin{subequations}
\begin{align}
& u^2\Sigma_s'' +6u\Sigma_s' + \frac{\Sigma_s}{6} \left(36+u^4 \left(4 \phi_s^2+u \left(3 u^3 \left(4 B_s+u B_s'\right)^2+u (\phi_s')^2+4\phi_s'\phi_s \right)\right)\right)\notag \\
&+ \frac{1}{6} u \left(4 \phi_s^2+u \left(3 u^3 (4 B_s+u B_s')^2+u (\phi_s')^2+4\phi_s'\phi_s  \right)\right)= 0, \label{eq:nPDE1} \\
& \mathcal{E} = \frac{2 e^{2 \sqrt{\frac{2}{3}} u^2 \phi_s } u^3 \Phi_2}{\left(1+ u^3\Sigma_s\right)^3}, \\
& -(1+u^3 \Sigma_s)(d_+\Sigma)_{s}'-2 u^2 (3 \Sigma_s+u \Sigma_s')(d_+\Sigma)_{s} \notag \\ 
&+ \frac{12 \left(1+\Sigma_s u^3\right)-12 \left(-1+2 \Sigma_s u^3+u^4 \Sigma_s'\right)+\left(1+\Sigma_s u^3\right)^2 \left(\mathcal{E}^2 f\left(u^2 \phi_s \right)+2 V\left(u^2 \phi_s \right)\right)}{12u^5}=0, \\
&-u \left(1+\Sigma_s u^3\right)(d_{+}B)_s' - \frac{3}{2}\left(1+4 \Sigma_s  u^3+u^4 \Sigma_s '\right)(d_{+}B)_s \notag \\
&-\frac{3}{4} \left(1+2 u^4(d_+\Sigma)_{s}\right) (4 B_s+u B_s')= 0,\\
& -4 \left(u+u^4\Sigma_s\right)(d_{+}\phi)_s' +(2-2 u^3 (8 \Sigma_s+3 u \Sigma_s'))(d_{+}\phi)_s  \notag \\
& + \frac{-3 u^3 \left(1+2 u^4(d_+\Sigma)_{s}\right) \phi_s' -6 \left(u^2+2u^6(d_+\Sigma)_{s}\right) \phi_s}{u^2} \notag \\
&+\frac{\left(1+\Sigma_s u^3\right) \left(\mathcal{E}^2 \partial_{\phi}f\left(u^2 \phi_s \right)-2 \partial_{\phi}V\left(u^2 \phi_s \right)\right)}{u^2} =0, \\
& u^2\left(1+u^3\Sigma_s\right)^2 A_s'' + 6 u \left(1+u^3\Sigma_s\right)^2A_s' +6 \left(1+u^3\Sigma_s\right)^2 A_s \notag \\
& - \frac{-\left(1+u^3\Sigma_s\right)^2-3 \left(1+2 (d_+ \Sigma)_s u^4\right) \left(-1+2 u^3\Sigma_s+u^4 \Sigma_s'\right)}{ u^4} \notag \\
& - \frac{6 (d_+ B)_s u^8 \left(1+u^3\Sigma_s\right)^2 (4 B_s+u B_s')+2(d_+ \phi)_s u^4 \left(1+u^3\Sigma_s\right)^2 (u\phi_s' +2 \phi_s )}{4 u^4} \notag \\
& +\frac{7 \left(1+u^3\Sigma_s\right)^2 \mathcal{E}^2 f\left(u^2 \phi_s \right)+2 \left(1+u^3\Sigma_s\right)^2 V\left(u^2 \phi_s \right)}{12 u^4} = 0, \label{eq:nPDE6}
\end{align}
\end{subequations}
with the prime now representing $\partial_u$. The price we paid to remove the redundant information from the bulk fields is that the equations of motion for them are now longer. Moreover, to solve these equations we need to specify the boundary conditions which, in light of Eqs.\ \eqref{eq:nearBdry1}---\eqref{eq:nearBdry5} and Eqs.\ \eqref{eq:Redefinitions}, are given by
\begin{align}
A_s(v,u) &= H-\frac{\phi _2(v){}^2}{18}-\frac{\phi_2(v)\dot{\phi}_2(v)}{18}u+\mathcal{O}(u^2),\\
\Sigma_s(v,u) &= -\frac{\phi _2(v){}^2}{18}u-\frac{\phi _2(v) \dot{\phi}_2(v)}{10}u^2 + \mathcal{O}(u^3),\\
(d_{+}\Sigma)_s(v,u) &= H+\frac{\phi _2(v){}^2}{36}+\mathcal{O}(u),\\
B_s(v,u) &= B_4(v)+ \mathcal{O}(u),\\
(d_{+}B)_s(v,u) &= -2B_4(v)+ \mathcal{O}(u),\\
\phi_s(v,u) &= \phi _2(v)+\dot{\phi}_2(v)u+  \mathcal{O}(u^2),\\
(d_{+}\phi)_s(v,u) &= -\phi _2(v)+ \mathcal{O}(u).
\end{align}

The time evolution equations for $B_s$ and $\phi_{s}$, which are obtained from the definitions of $d_+ B$ and $d_+\phi $ using the redefined fields \eqref{eq:Redefinitions}, are given by
\begin{align}
\partial_{v}B_{s} & = \frac{2 B_s}{u}+\frac{(\text{d}_+B)_{s}}{u}+4 A_s B_s u^3+\frac{B_s'}{2}+A_s u^4 B_s', \label{eq:timevolB}\\
\partial_{v}\phi_{s} & = \frac{(\text{d}_+\phi)_{s}}{u}+\frac{\phi_{s}}{u}+2 u^3 A_{s}\phi_{s}+\frac{1}{2}\phi_{s}' +u^4 A_{s} \phi_{s}'.\label{eq:timevolphi}
\end{align}
We discuss in Sec.\ \ref{sec:FarSol:NumTec} the numerical scheme we use to evolve in time the EMD fields provided $\partial_{v}B_{s}$ and $\partial_{v}\phi_{s}$ are known.

\subsection{Radial position of the black hole event horizon}
\label{sec:FarSol:HolePosit}

A delicate aspect of the numerical solution of the far-from-equilibrium equations of motion is how deep into the bulk one should go when integrating the radial dependence of the PDEs. Since we are working with a finite temperature setup, there is a black hole inside the bulk and one needs to cover the entire portion of the bulk between the black hole's event horizon and the boundary of the asymptotically AdS$_5$ spacetime, since this is the region of the bulk geometry causally connected to the boundary QFT. In the case of time-dependent backgrounds, one main difficulty is that the radial position of the event horizon is unknown \textit{a priori} since it depends on the time evolution of the system. If one cuts off the radial integration before reaching the horizon, then the obtained numerical solution will be inevitably inaccurate; on the other hand, if the radial integration proceeds too deep inside the horizon, the numerical simulation will probably break down due to possible singularities (caustics) associated with strong curvatures in this deep infrared region.

One possible way to deal with this issue is to use the radial shift function $\lambda(v)$ associated with the residual diffeomorphism invariance of the metric and regard it as an auxiliary field to fix the (apparent) horizon position at a specific value of the radial coordinate \cite{Chesler:2013lia}. However, we do not adopt this strategy here. Instead, as mentioned above Eq. \eqref{eq:nearBdry1} we set $\lambda(v) = 0$ and let the horizon position fluctuate in each simulation. We then follow the reasoning of Ref.\ \cite{Heller:2013oxa} in which an infrared radial cutoff corresponding to $u_{IR}\approx 1.01$ is used in the numerical simulations. Moreover, since we are interested here in analyzing the equilibration dynamics of the system in terms of the dimensionless ``time combination'' $vT$ for different values of $\mu/T$, we choose to set the equilibrium temperature to $1/\pi$ for any value of $\mu/T$. This gives the following relation between $\tilde{r}_{h}$ and $Q$ coming from Eq.\ \eqref{eq:temperature},
\begin{equation}\label{eq:TempFix}
\frac{1}{\pi} = \frac{Q^2 + 2 \tilde{r}_h^2}{2 \pi \sqrt{Q^2+\tilde{r}_h^2}}.
\end{equation}

\begin{figure}[t]
\centering
\begin{subfigure}{0.49\textwidth}
\includegraphics[width=\textwidth]{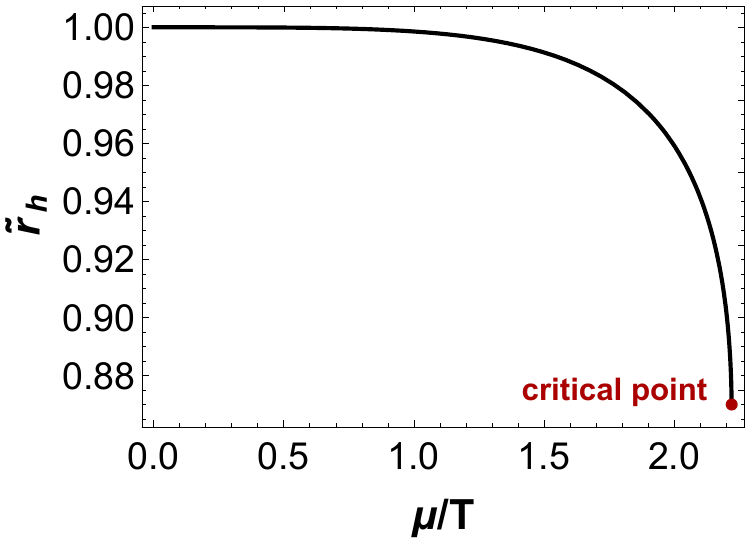}
\caption{}
\end{subfigure}
\begin{subfigure}{0.49\textwidth}
\includegraphics[width=\textwidth]{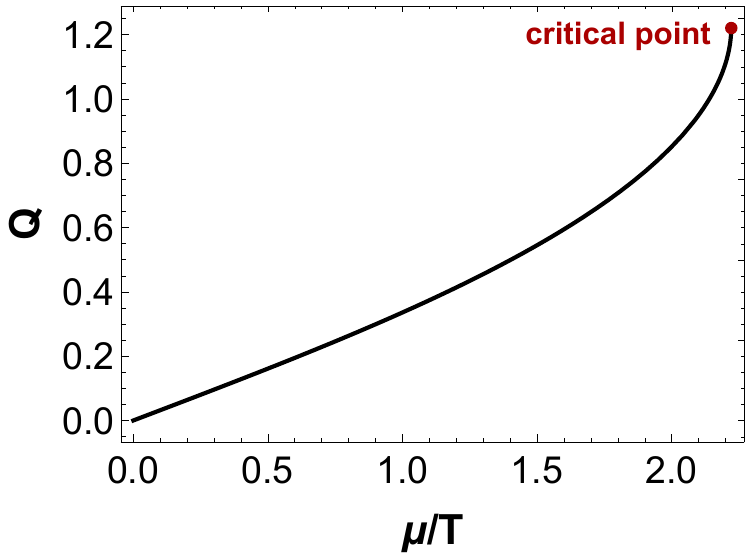}
\caption{}
\end{subfigure}
\begin{subfigure}{0.49\textwidth}
\includegraphics[width=\textwidth]{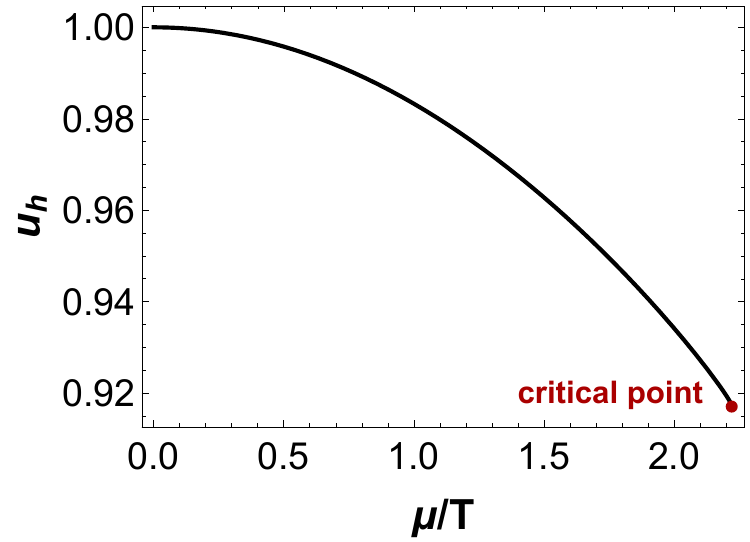}
\caption{}
\end{subfigure}
\begin{subfigure}{0.49\textwidth}
\includegraphics[width=\textwidth]{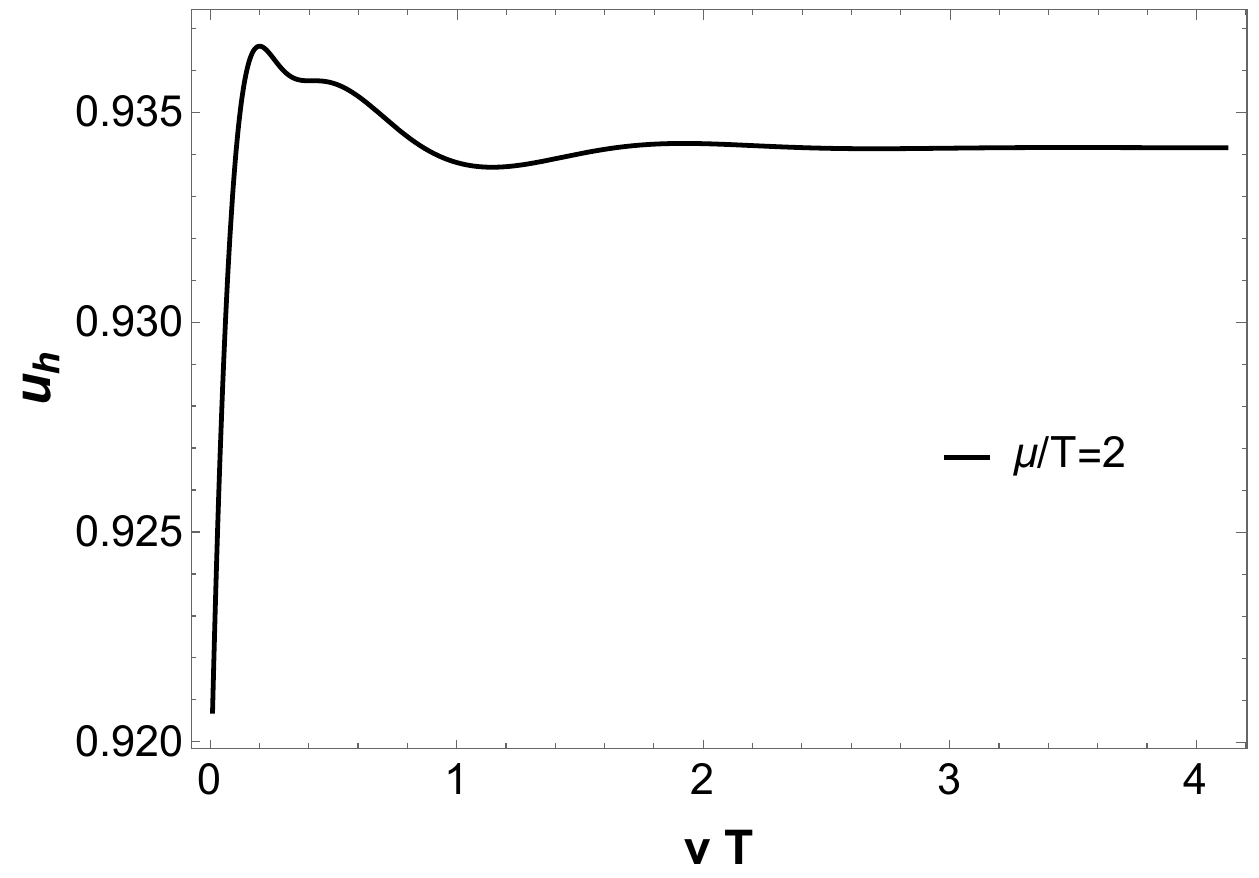}
\caption{}
\end{subfigure}
\caption{(Color online) (a) Radius of the equilibrium black hole expressed in the modified EF coordinates \eqref{eq:modEF}, (b) the parameter $Q$ of the equilibrium solution, and (c) the radius of the equilibrium black hole expressed in the numerical radial coordinate \eqref{eq:uDef}, all of them plotted as functions of the chemical potential $\mu/T$. (d) Time evolution of the radius of the far-from-equilibrium black hole solution obtained using constant profiles for the initial metric anisotropy and dilaton field at $\mu/T=2$.}
\label{fig:rmuTQmuT}
\end{figure}

Therefore, given some value of $\mu/T$ we use Eqs. \eqref{eq:critical_point} and \eqref{eq:TempFix} to determine the corresponding values of $\tilde{r}_{h}$ and $Q$. In Fig. \ref{fig:rmuTQmuT} we show the plots for $\tilde{r}_{h}$, $Q$, and $u_{h}$ as functions of $\mu/T$.\footnote{In order to go from $\tilde{r}_h$ to $r_h$ and then to $u_h$ one must use Eqs. \eqref{eq:rTorTild} and \eqref{eq:uDef}.} In particular, Fig. \ref{fig:rmuTQmuT} (c) indicates that it is fine to adopt $u_{IR}\approx 1.01$ as the inner radial cutoff for numerical integration since this prescription already works well at zero density and the horizon position in equilibrium gets closer to the boundary as the chemical potential is increased. Moreover, in Fig. \ref{fig:rmuTQmuT} (d) we illustrate a typical case with the time evolution of the event horizon\footnote{To obtain the event horizon, one just needs to solve the equation for the radial null geodesic (by setting $ds^2=0$ in Eq. \eqref{lineElement} and discarding the piece which does not depend on $dv$),
\begin{equation}
0 = 2dvdr - 2A(v,r) dv^2 \Rightarrow \frac{dr}{dv} = A(v,r(v)),
\end{equation}
subjected to the boundary condition $r(v\rightarrow\infty)=r_h$.}, showing that also for transient states our inner radial cutoff is always beyond the horizon.

\subsection{Initial states}
\label{sec:FarSol:InitialState}

The last step required before we numerically simulate the far-from-equilibrium evolution of the 1RCBH plasma is to choose the initial data. In the context of the homogeneous equilibration of the 1RCBH model, we need to specify three inputs to start the numerical integration of the PDEs \eqref{eq:nPDE1}---\eqref{eq:nPDE6}:
\begin{itemize}
\item The initial metric anisotropy function, $B_{s}(v_0,u)$;

\item The initial profile for the dilaton, $\phi_{s}(v_0,u)$;

\item The value of the equilibrium chemical potential, $\mu/T$. 
\end{itemize}

The initial metric anisotropy will be assumed to have either the form of a ``pulse'' with a Gaussian parameterization \cite{Fuini:2015hba}, or simply a constant profile
\begin{align}
(a) \ \ B_{s}(v_0,u) & = \mathcal{A}\,e^{\frac{1}{2}(u-u_0)^2/\sigma^2}, \label{eq:Bini1} \\
(b)\ \ B_{s}(v_0,u) & = \mathcal{A}. \label{eq:Bini2}
\end{align}
The parameter $\mathcal{A}$ controls the initial amplitude of the metric anisotropy whereas the parameters $\sigma$ and $u_{0}$ are related to the width and how deep into the bulk the pulse is centered.

For the initial dilaton profile, we shall consider again the constant and Gaussian forms and, additionally, we will also consider the case where the initial dilaton profile is already at its equilibrium solution, i.e.\footnote{Note that these are initial profiles for the subtracted dilaton field defined in Eqs.\ \eqref{eq:Redefinitions}. Therefore, the full initial dilaton profile $\phi(v_0,u)=u^2\phi_s(v_0,u)$ vanishes at the boundary $u=0$ for all these initial data, as it should be.}
\begin{align}
(a) \ \ \phi_{s}(v_0,u) & = -\kappa_{5}^{2}\langle \mathcal{O}_{\phi} \rangle = -\sqrt{\frac{2}{3}}Q^2, \label{eq:phiini1}\\
(b) \ \ \phi_{s}(v_0,u) & =\mathcal{A}_\phi e^{\frac{1}{2}(u-u_0)^2/\sigma^2}, \label{eq:phiini2}\\
(c) \ \ \phi_{s}(v_0,u) & = -\frac{1}{u^2}\sqrt{\frac{2}{3}} \ln \left( 1 + \tilde{u}(u)Q^2 \right),\label{eq:phiini3}
\end{align}
where $\tilde{u} = 1/\tilde{r}$, and Eq.\ \eqref{eq:phiini3} has this form to match the equilibrium solution \eqref{eq:scalar field}. Note that for $B(v_0,u)=0$ (corresponding to a pure thermalization process with no initial anisotropy), the use of the last initial condition for the dilaton field automatically gives the equilibrium solution for the 1RCBH plasma.

Finally, one needs to choose a value for the equilibrium chemical potential $\mu/T\in [0,\pi/\sqrt{2}]$. The value $\mu/T=0$ corresponds to the standard $\mathcal{N}=4$ SYM plasma whereas $\mu/T=\pi/\sqrt{2}$ corresponds to the critical point. Moreover, by fixing $\mu/T$ we also fix the values of $Q$ and $u_h$ as explained above in Eq.\ \eqref{eq:TempFix}.

Regarding the initial energy density (and the pressure, since $\varepsilon=3p$), one may find it using the expression obtained in Eq.\ \eqref{eq:pressure} from the thermodynamics of the 1RCBH plasma, i.e.
\begin{equation}
\varepsilon = -3H = \dfrac{3\pi^2 N_c^2 T^4}{128}\left(3 - \sqrt{1-\left(\dfrac{\mu/T}{\pi/\sqrt{2}}\right)^2}\right)^3 \left(1 + \sqrt{1-\left(\dfrac{\mu/T}{\pi/\sqrt{2}}\right)^2}\right),
\label{eq:trololo}
\end{equation}
where the temperature $T$ is given by Eq.\ \eqref{eq:temperature}. Hence, far-from-equilibrium solutions with different values of $\mu/T$ will have different energy density and equilibrium pressure.

Furthermore, since the 1RCBH model is a conformal theory, one cannot use only $v$ to measure the time evolution of the fields, i.e. one needs to compose this EF time with another quantity to produce a dimensionless time measure. Thus, since the temperature $T$ will be the same for all the numerical simulations, as explained in Sec.\ \ref{sec:FarSol:HolePosit}, the dimensionless quantity that we use to measure the time flow is $vT$. A different approach, which we do not explore in this manuscript, would be to fix the energy density for all the values of $\mu/T$ and let the temperature $T$ vary among the solutions; in this case, one would use the dimensionless quantity $v\varepsilon^{1/4}$ to compare the time evolutions between different values of $\mu/T$. This is done, for instance, in Ref.\ \cite{Fuini:2015hba}.

\subsection{Numerical techniques}
\label{sec:FarSol:NumTec}

In this work, in order to deal with the radial part of the system of PDEs \eqref{eq:nPDE1}---\eqref{eq:nPDE6} we developed a new numerical code that employs the so-called pseudospectral (or collocant) method \cite{boyd01,Fornberg,Lloyd}, which is a widespread technique used to solve the equations of motion in the characteristic formulation of numerical relativity due to its accuracy and rapid convergence.

The main idea behind the pseudospectral method is to expand the numerical solution in terms of functions that form a complete basis, converge, and are easy to compute. One family of functions that accomplish these demands are the Chebyshev polynomials of the first kind, $T_{n}(x)$. Indeed, for well-behaved functions, the convergence of the numerical solution is exponential with respect to the number of added polynomials \cite{boyd01,Fornberg}. A simple example where the solution is not so well-behaved occurs when there are logarithmic terms in the near-boundary expansion of the bulk fields; this situation led the authors of Ref.\ \cite{Ishii:2015gia} to favor the Runge-Kutta method to solve the radial part of the PDEs.

Assuming that the domain of interest is $u \in [0,u_{IR}]$, where $u_{IR}$ is the infrared radial cutoff, we write the numerical approximation $X_{N}(u)$ of a function $X(u)$ as,
\begin{equation}
X(u)\approx X_{N}(u) = \sum_{k=0}^{N-1} a_{k} T_{k}\left(\frac{2}{u_{IR}} u-1\right),
\end{equation}
where $\{a_{k}\}$ is the set of spectral coefficients. In the problem that we are interested to solve here, $X(u)$ represents the elements of the set $\lbrace \Sigma_{s},(d_{+}\Sigma)_{s}, \mathcal{E},B_s, (d_{+}B)_{s}, \phi_s,(d_{+}\phi)_{s}, A_{s}  \rbrace$.

In the numerical calculation one needs to specify the set of grid points $\{u_{i}\}$ where the discretized equations of motion take place. For this work, we adopt the Chebyshev-Gauss-Lobatto grid, which is given by\footnote{As discussed in Section \ref{sec:FarSol:HolePosit}, we used in the numerical simulations an infrared cutoff corresponding to $u_{IR}\approx 1.01$.}
\begin{equation}
u_{k} = \frac{u_{IR}}{2}\left(1+\cos\left(\frac{k\pi}{N-1}\right)\right), \ \ \ k=0,\dots , N-1,
\label{eq:ugrid}
\end{equation}
where $N$ is the number of grid points, also known as the collocant points.

Another important representation of $X_{N}$ is given in terms of the values it assumes at the grid points, i.e.
\begin{equation}\label{eq:CardinalBasis}
X_{N}(u_{i}) = \sum_{j=0}^{N-1}C_{j}\left(\frac{2}{u_{IR}} u_{i}-1\right)X_{j},
\end{equation}
where $C_{i}\left(\frac{2}{u_{IR}} u-1\right)$ is the cardinal function defined as
\begin{equation}
C_{i}\left(\frac{2}{u_{IR}} u-1\right) = \frac{2}{(N-1)p_i}\sum_{j=0}^{N-1}\frac{1}{p_j}T_j\left(\frac{2}{u_{IR}} u_i-1\right)T_j\left(\frac{2}{u_{IR}} u-1\right),
\end{equation}
with $p_{0}=p_{N-1}=2$ and $p_{i}=1$ otherwise, satisfying the condition,
\begin{equation}
C_{j}\left(\frac{2}{u_{IR}} u_{i}-1\right) = \delta_{ij}.
\end{equation}
It is then evident that the set of coefficients $\lbrace X_{i} \rbrace$ are precisely the values of $X_{N}(u_i)$. Moreover, the description of the problem in terms of $\lbrace X_{i} \rbrace$ is completely equivalent to the description in terms of the spectral coefficients $\lbrace a_{i} \rbrace$, the choice between them being just a matter of convenience. In this work, in particular, we will solve the PDEs \eqref{eq:nPDE1}-\eqref{eq:nPDE6} using the cardinal basis representation \eqref{eq:CardinalBasis} to get $\lbrace X_{i} \rbrace$.

The last element that we need to define before numerically solving the equations of motion is the pseudospectral differentiation matrix, $D_{ij}$, which provides the discrete version of $X'(u)$, i.e. $X_{i}'=D_{ij}X_{j}$. To obtain this matrix, one needs to go through the derivation of the cardinal functions. Here we just give the final form of the differentiation matrix \cite{boyd01}:\footnote{One can obtain the differentiation matrix in Mathematica using the command:
\texttt{NDSolve`FiniteDifferenceDerivative[1,ugrid,DifferenceOrder->{N-1}][``DifferentiationMatrix'']}
where \texttt{ugrid} denotes the radial grid and \texttt{N} is the number of collocant points.}
\begin{equation}
D_{ij} = \left.\frac{d C_j\left(\frac{2}{u_{IR}} u-1\right)}{du}\right\vert_{u=u_i} = \begin{cases} 
      (1+2(N-1)^2)/(3u_{IR}) & i=j=0, \\
      -(1+2(N-1)^2)/(3u_{IR}) & i=j=N-1,  \\
      -u_j/(u_{IR}(1-u_{j}^{2})) & i=j;0<j<N-1, \\
      2(-1)^{i+j}p_{i}/(u_{IR}\,p_j(u_i-u_j)) & i \neq j, 
   \end{cases}
\end{equation}
A nice property of $D_{ij}$ is that, to obtain higher order derivatives, one just needs to exponentiate this matrix
\begin{equation}
D_{ij}^{(n)} = (D^{n})_{ij},
\end{equation}
where $n$ is the order of the desired derivative. The discretized version of the differential equations acquire then the following generic form
\begin{equation}\label{eq:DiscreteEq}
Q_{X} \vec{X} = \vec{f}_{X},
\end{equation}
where $Q_{X}$ is a $N\times N$ matrix and $\vec{f}_{X}$ is a vector with $N$ components. Hence, we have transformed a continuum differential problem into a linear algebra problem, i.e. an inversion matrix problem since the solution of Eq.\ \eqref{eq:DiscreteEq} is given by
\begin{equation}\label{eq:DiscreteSol}
\vec{X} = Q_{X}^{-1} \vec{f}_{X}.
\end{equation}
We note that when going from Eq.\ \eqref{eq:DiscreteEq} to Eq.\ \eqref{eq:DiscreteSol} we assumed that $Q_{X}$ is invertible. This is not always true. Indeed, in many cases one needs to add an initial/boundary condition to the original Eq.\ \eqref{eq:DiscreteEq} to obtain an invertible matrix.

At this stage it is instructive to give a simple example to see what should be the form of $Q_{X}$ and $\vec{f}_{X}$. For instance, the following second order differential equation
\begin{equation}
X''(u) = 2,
\end{equation}
assumes a discretized version given by,
\begin{equation}
(D^{2})_{ij}X_{j} = (2,\dots,2)_i,
\end{equation}
which means that $Q_{X}=D^2$ and $\vec{f}_{X}=(2,\dots,2)^{T}$. Now the discussion regarding the solution of the radial part of the equations of motion is almost done, we only have to discuss the filtering process.

In our calculations, we have found the common ``aliasing'' (a.k.a. ``spectral blocking'') problem where high frequency modes have spurious growth and contaminate the numerical computation until it eventually breaks down. Such issue is typical in problems involving nonlinear equations \cite{boyd01,Fornberg}. To circumvent this problem, we need to access the spectral coefficients ${a_{i}}$ and perform a ``damping'' on the higher modes. To go from $\lbrace X_{i}\rbrace$ to $\lbrace a_{i}\rbrace$ we use the Matrix Multiplication Transform (MMT) as defined in Ref.\ \cite{boyd01},
\begin{equation}
a_{i} = M_{ij}X_{j},
\end{equation} 
where
\begin{equation}
M_{ij} = \frac{w(j)\,T_{i}\left(\frac{2}{u_{IR}} u_{j+1}-1\right)}{(T_{i},T_{i})},
\end{equation}
with
\begin{align}
w(i) = \frac{\pi}{N-1} \left[1 - \frac{\delta_{i, 0} + \delta_{i, N-1}}{2}\right],
\end{align}
denoting the Gaussian quadrature weight function and $(T_i,T_i)$ denoting the scalar product,
\begin{equation}
(T_i,T_i)\equiv \begin{cases} 
      \pi & i \in \{0,N-1\}, \\
      \pi/2 & i \notin \{0,N-1\}. 
   \end{cases}
\end{equation}

Our damping process, in particular, is very efficient. For instance, if we take $N=40$ as the number of radial grid points, setting the last three spectral coefficients to zero at each time step suffices to produce a well behaved numerical evolution. We illustrate this process schematically below\footnote{Another common way to do the filtering process is to use the Fast Fourier Transform (FFT). This is explained in the review \cite{Chesler:2013lia} and it is also used, e.g., in Ref.\ \cite{Attems:2017zam}.}
\begin{equation}
\lbrace X_i \rbrace \xrightarrow{M}\lbrace a_{i}^{\textrm{aliased}} \rbrace \xrightarrow{\textrm{damp}} \lbrace a_{i}^{\textrm{anti-aliased}} \rbrace \xrightarrow{M^{-1}} \lbrace X_{i}^{\textrm{filtered}} \rbrace.
\end{equation}

For the time evolution of Eqs. \eqref{eq:timevolB} and \eqref{eq:timevolphi} we have used the Adams-Bashforth (AB) method, which is a well-known explicit multi-step method employed to solve ordinary differential equations. In particular, the third order AB formula is given by,
\begin{equation}\label{eq:AB3}
\vec{X}_{n+1} = \vec{X}_{n}+\frac{\Delta v}{12}\left(23\, \partial_v\vec{X}_{n}-16\, \partial_v\vec{X}_{n-1}+5\, \partial_v\vec{X}_{n-2} \right),
\end{equation}
where the subscript $n$ denotes the position of the variable at the nth time step and $\Delta v$ is the time step value. Furthermore, since the AB method \eqref{eq:AB3} requires the time derivative of a few previous steps it needs to be initialized with another method. In this work we used the Euler method for the first three steps.\footnote{Note that to evolve Eqs. \eqref{eq:timevolB} and \eqref{eq:timevolphi} to the next time slice using the Euler method we just need to know the RHS of these equations on the current time slice.}

\section{Equilibration dynamics: results for different initial data}
\label{sec:results}

After the detailed discussion carried out in the previous sections, we are finally in position to perform the full numerical evolution of the nonlinear PDEs of the 1RCBH model in their numerical form \eqref{eq:nPDE1}---\eqref{eq:nPDE6}. Next, we present the results for the homogeneous equilibration of the 1RCBH plasma for different initial data given in Eqs.\ \eqref{eq:Bini1}, \eqref{eq:Bini2} and Eqs.\ \eqref{eq:phiini1}---\eqref{eq:phiini3}.

\subsection{Constant metric anisotropy and dilaton profiles}
\label{sec:results:const}

We begin by discussing the results for the initial data specified in Eqs.\ \eqref{eq:Bini2} and \eqref{eq:phiini1}, i.e. when the initial profiles for the metric anisotropy $B_s(v_0,u)$ and the dilaton field $\phi_s(v_0,u)$ are both constant with respect to the radial coordinate $u$. We are going to consider here the values
\begin{align}\label{eq:BinConCon1}
B_s(v_0,u) =2 \ \ \text{and} \ \ \phi_s(v_0,u) = -\sqrt{\frac{2}{3}}Q^2,
\end{align}
which are similar to the first condition probed in Ref.\ \cite{Heller:2013oxa}.

\begin{figure}[h]
\centering
\begin{subfigure}{0.5\textwidth}
\includegraphics[width=\textwidth]{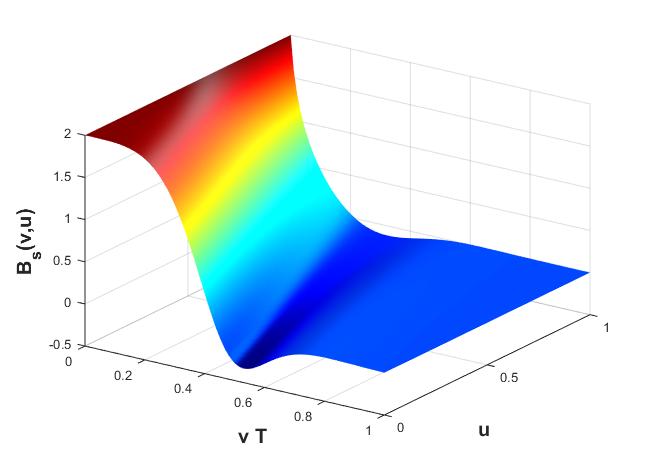}
\caption{}
\end{subfigure}
\begin{subfigure}{0.49\textwidth}
\includegraphics[width=\textwidth]{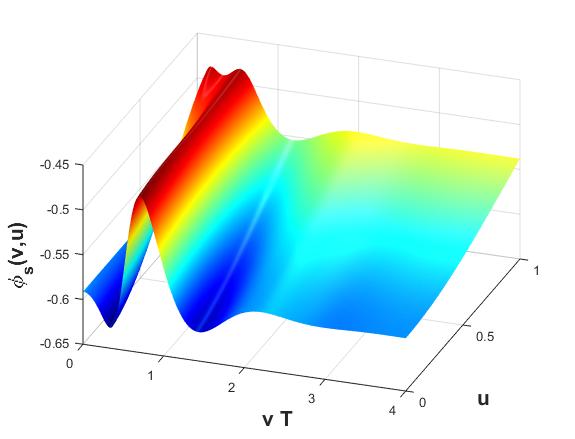}
\caption{}
\end{subfigure}
\caption{(Color online) Results for the time evolution of some fields involved in the 1RCBH setup for the initial condition \eqref{eq:BinConCon1} with $\mu/T=2$: (a) the subtracted metric anisotropy function $B_{s}(v,u)$, and (b) the subtracted dilaton field $\phi_{s}(v,u)$.}
\label{fig:3DConCon}
\end{figure}

\begin{figure}[h]
\centering
\begin{subfigure}{0.49\textwidth}
\includegraphics[width=\textwidth]{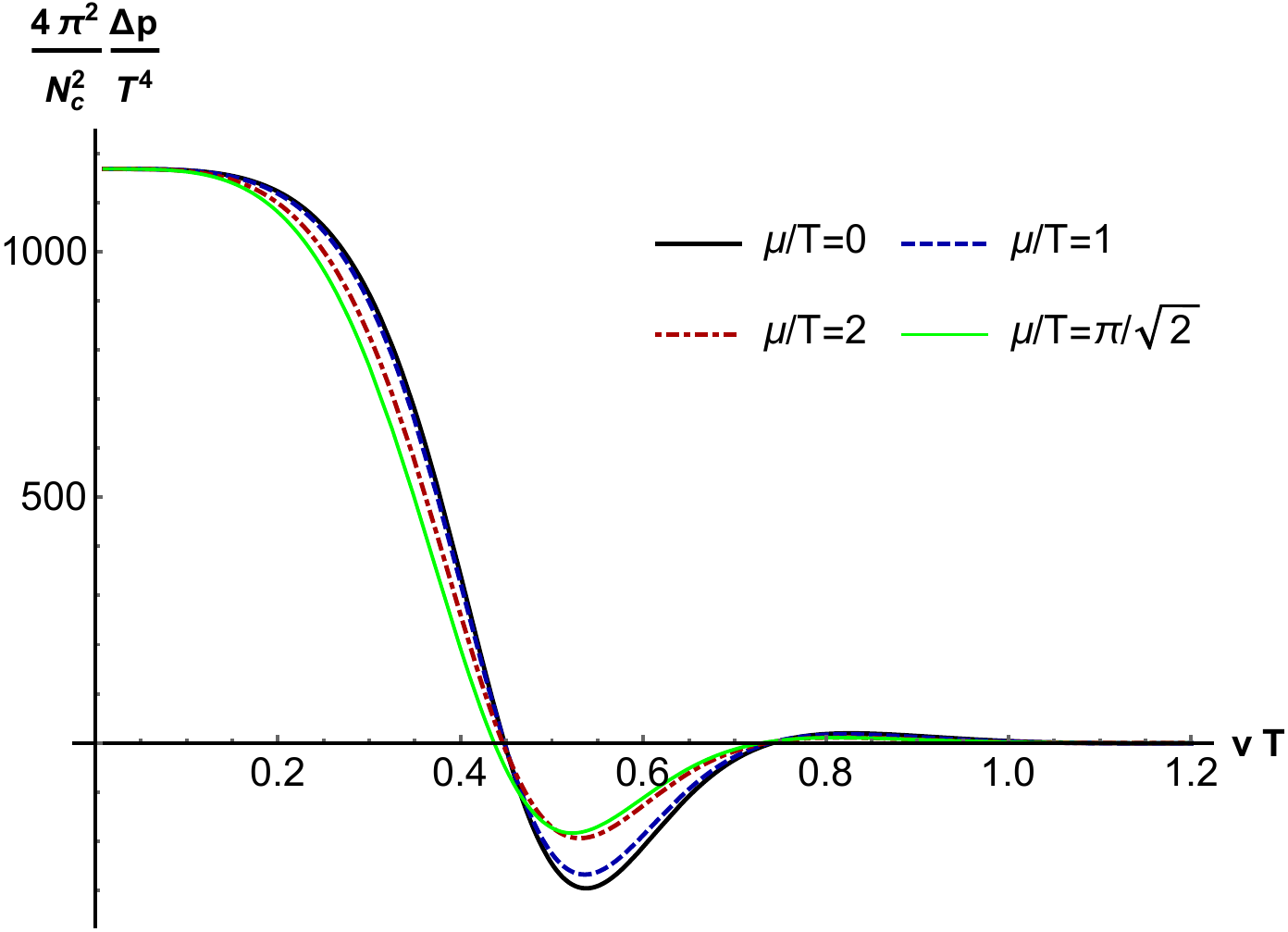}
\caption{}
\end{subfigure}
\begin{subfigure}{0.49\textwidth}
\includegraphics[width=\textwidth]{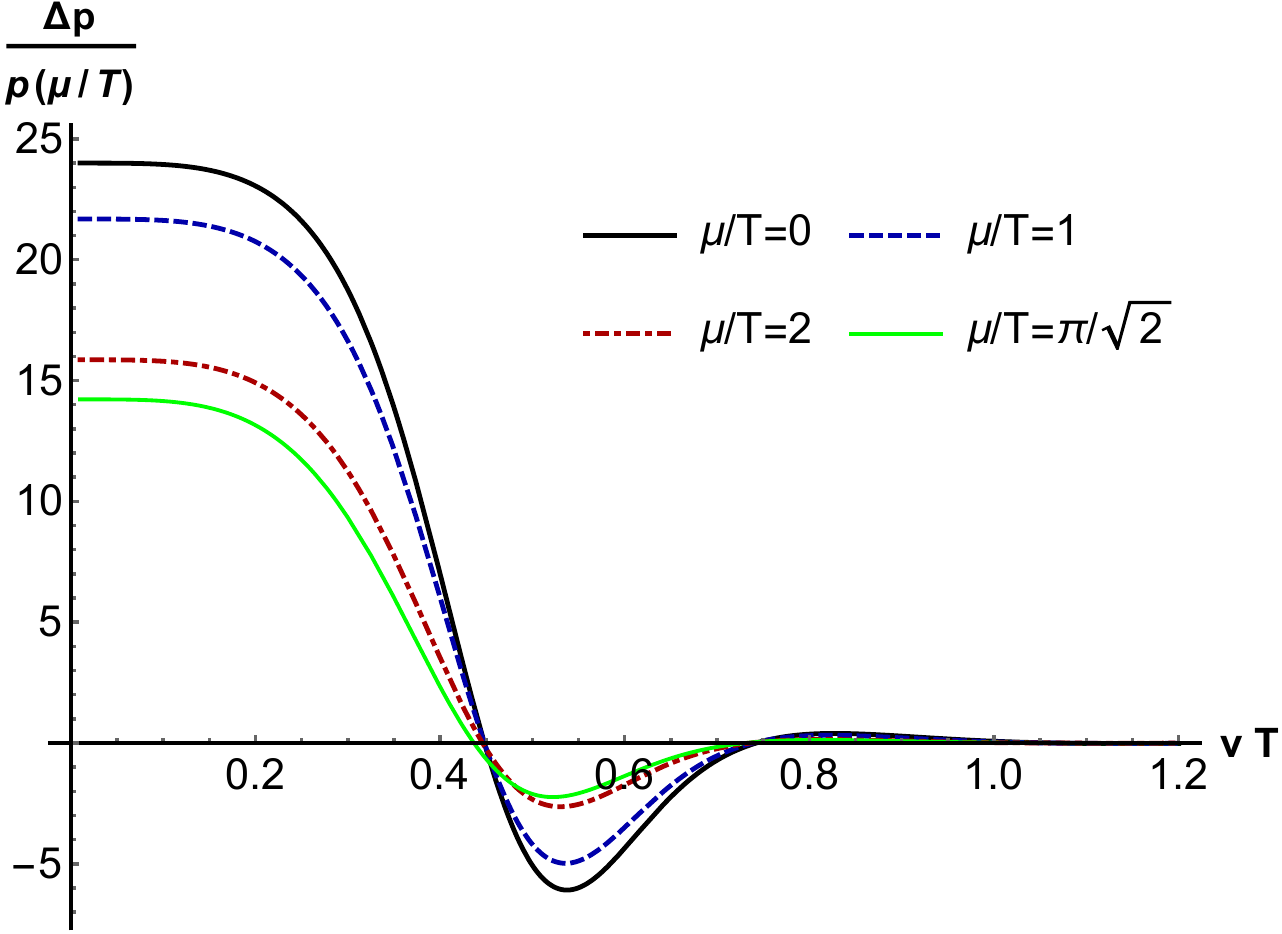}
\caption{}
\end{subfigure}
\caption{(Color online) Time evolution of the pressure anisotropy for several values of the chemical potential using the initial data \eqref{eq:BinConCon1}: (a) $\Delta p$ normalized by the (equilibrium) temperature to the fourth, and (b) $\Delta p$ normalized by the equilibrium pressure. }
\label{fig:PressureAni1}
\end{figure}

\begin{figure}[h]
\centering
\begin{subfigure}{0.48\textwidth}
\includegraphics[width=\textwidth]{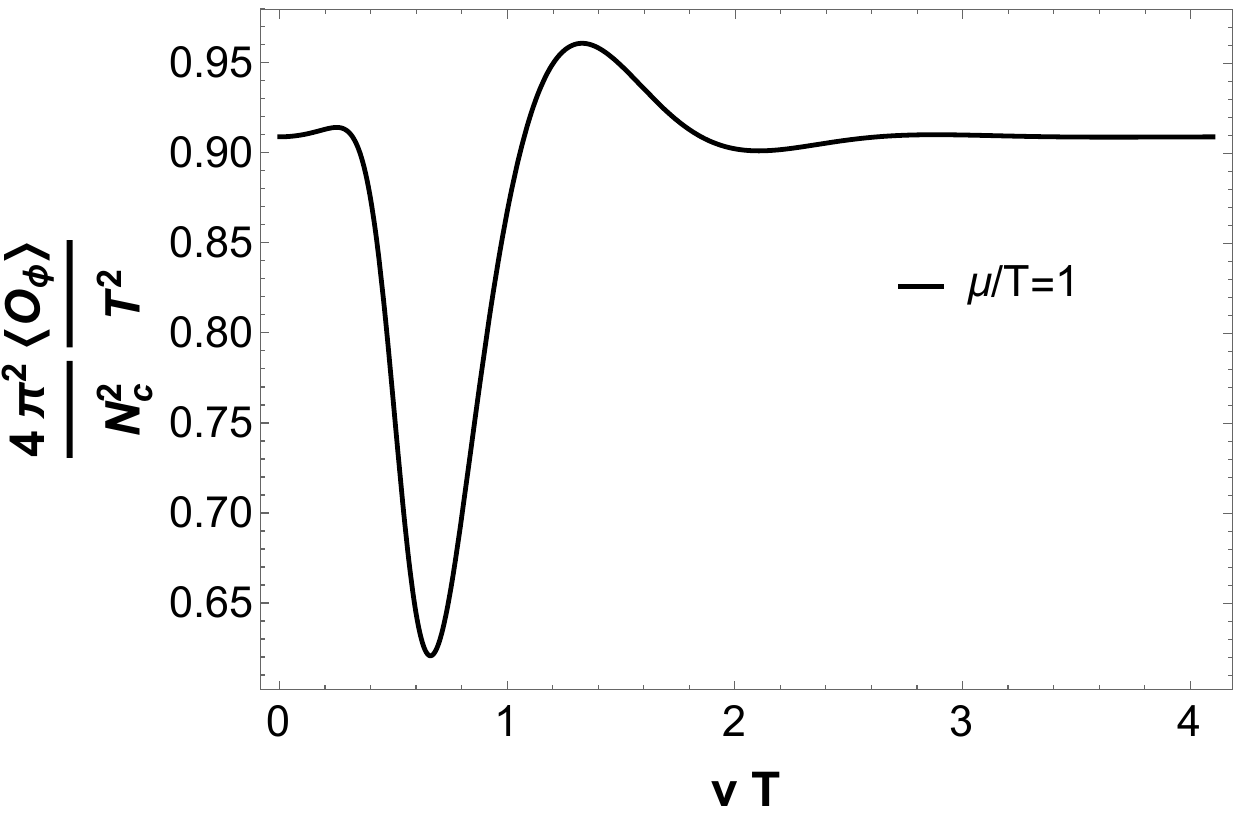}
\caption{}
\end{subfigure}
\begin{subfigure}{0.48\textwidth}
\includegraphics[width=\textwidth]{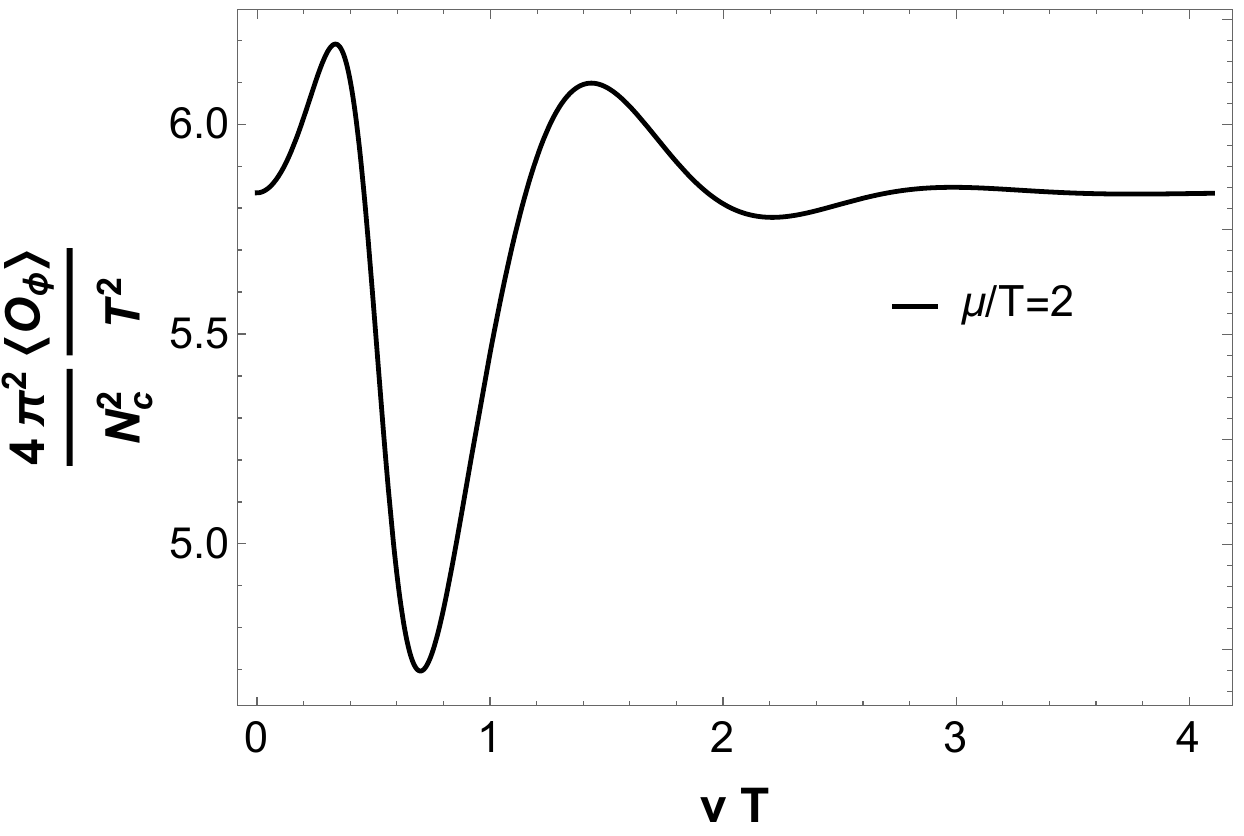}
\caption{}
\end{subfigure}
\begin{subfigure}{0.48\textwidth}
\includegraphics[width=\textwidth]{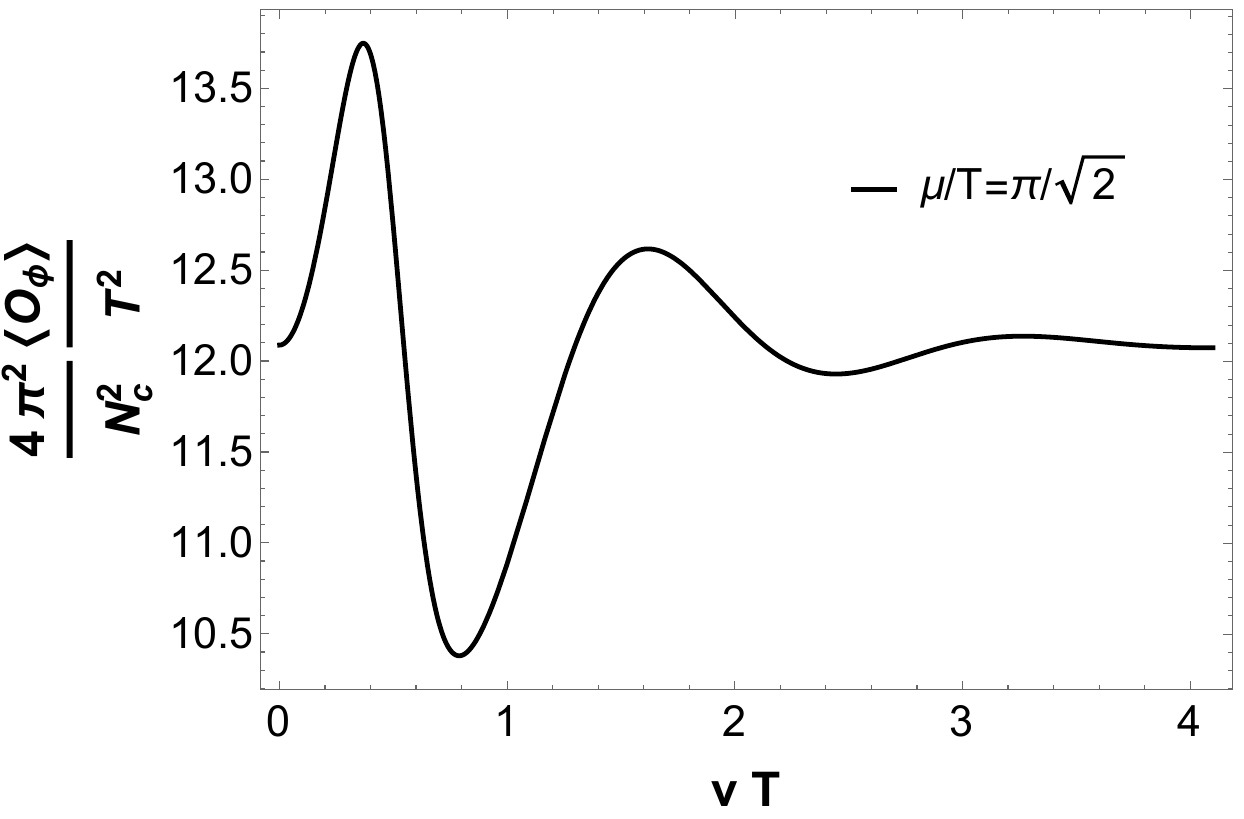}
\caption{}
\end{subfigure}
\caption{Time evolution of the scalar condensate $\langle\mathcal{O}_{\phi}\rangle$ for different $\mu/T$ using the initial data \eqref{eq:BinConCon1}.}
\label{fig:OphiConCon}
\end{figure}

In Fig.\ \ref{fig:3DConCon} we show a 3D plot with the time evolution of $B_s$ and $\phi_s$ for $\mu/T=2$. As discussed before, we choose the dimensionless quantity $vT$ to represent the time flow in the CFT. With $B_s$ and $\phi_s$ now fully determined, one may compute the time evolution of the pressure anisotropy $\Delta p$ \eqref{eq:DeltaP} and the scalar condensate $\langle\mathcal{O}_{\phi}\rangle$ \eqref{eq:DilCondensate}. 

The results for the isotropization process describing how the pressure an\-isot\-ropy $\Delta p$ goes to zero as the time evolves, taking into account the initial condition \eqref{eq:BinConCon1}, are shown in Fig.\ \ref{fig:PressureAni1}. In Fig.\ \ref{fig:PressureAni1} (a), where $\Delta p$ is normalized by the (equilibrium) temperature to the fourth, one notes that at the earliest times the pressure anisotropy in insensitive to the value of the dimensionless chemical potential $\mu/T$ which, however, becomes relevant as the time evolves. In fact, as we shall discuss in Sec.\ \ref{sec:MatQNM}, the isotropization time (which cannot be adequately resolved by eyeball in the scale of the plot shown in Fig.\ \ref{fig:PressureAni1} (a)) has a non-monotonic dependence on the value of $\mu/T$, which is a direct consequence of the presence of a critical point in the phase diagram of the 1RCBH model. Namely, as it will become clear in the analysis of Sec.\ \ref{sec:MatQNM}, the isotropization time may decrease or increase for increasing values of $\mu/T$ depending on whether $\mu/T$ is far or close enough to the critical point, respectively. This is very different from what happens e.g. in the case of the AdS-Reissner-Nordstrom background investigated in Ref.\ \cite{Fuini:2015hba}, which does not feature a critical point in its phase diagram. It is also interesting to note that, as anticipated around Eq.\ \eqref{eq:trololo}, the equilibrium pressure (and energy density) depends on the chosen value of $\mu/T$, which is clear from the curves shown in Fig.\ \ref{fig:PressureAni1} (b), where $\Delta p$ is normalized by the equilibrium pressure. Moreover, we have verified the effect on the isotropization time obtained by varying the value of the initial metric anisotropy and the effect was negligible. For instance, if one doubles the initial anisotropy, the only effect is to have a steeper downfall of $\Delta p$ before it reaches the first minimum. The explanation for this is the surprising low nonlinearity of the system \cite{Heller:2012km, Heller:2013oxa}, which is also quantitatively explored at length in Ref.\ \cite{Fuini:2015hba}.

The time evolution of the scalar condensate $\langle\mathcal{O}_\phi \rangle$ for different values of $\mu/T$ is presented in Fig.\ \ref{fig:OphiConCon}. The first remarkable feature of the time evolution of the scalar condensate is that it takes a much longer time to relax to equilibrium than the pressure anisotropy. This is the reason why we adopted the equilibration time associated with the relaxation of the scalar condensate toward equilibrium as the true thermalization time of the 1RCBH plasma. We also observe a qualitatively different behavior for the dependence of the thermalization time with $\mu/T$ when compared to the corresponding dependence of the isotropization time (note also that in the case of the AdS-Reisser-Nordstrom background studied in Ref.\ \cite{Fuini:2015hba}, spatially homogeneous isotropization corresponds already to the true thermalization of the system since there is no scalar condensate in that case). The thermalization time associated with the relaxation of the scalar condensate always increase with increasing $\mu/T$. It is also interesting to note the qualitative similarity between the time evolution of $\langle\mathcal{O}_\phi \rangle$ and the thermalization process of confining theories studied in Ref.\ \cite{Ishii:2015gia} using holographic quenches, even though the 1RCBH model is non-confining.

\subsection{Constant metric anisotropy profile and Gaussian dilaton profile}
\label{sec:results:ConsGauss}

In this subsection we study the time evolution of the system given the initial data:
\begin{align}\label{eq:BinConGauss}
B_s(v_0,u) =2 \ \ \text{and} \ \ \phi_s(v_0,u) = -0.5\, e^{-100(u-0.3)^2},
\end{align}
where we kept the same constant initial metric anisotropy as before but changed the initial condition for the dilaton field and considered now a Gaussian profile. This is a way to look for possible new features depending mostly on the choice for the initial dilaton profile.

\begin{figure}[h]
\centering
\begin{subfigure}{0.5\textwidth}
\includegraphics[width=\textwidth]{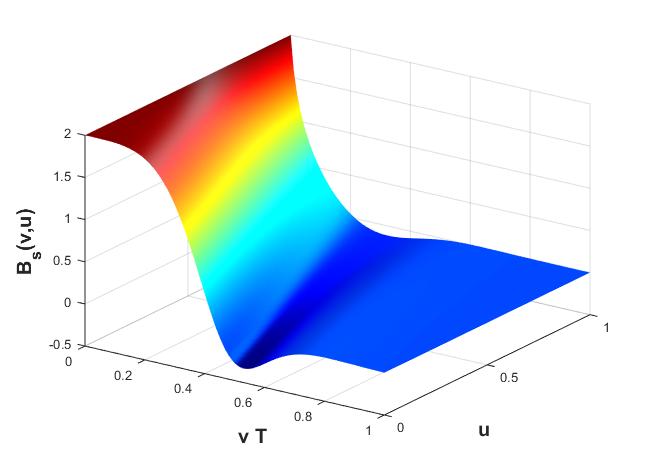}
\caption{}
\end{subfigure}
\begin{subfigure}{0.49\textwidth}
\includegraphics[width=\textwidth]{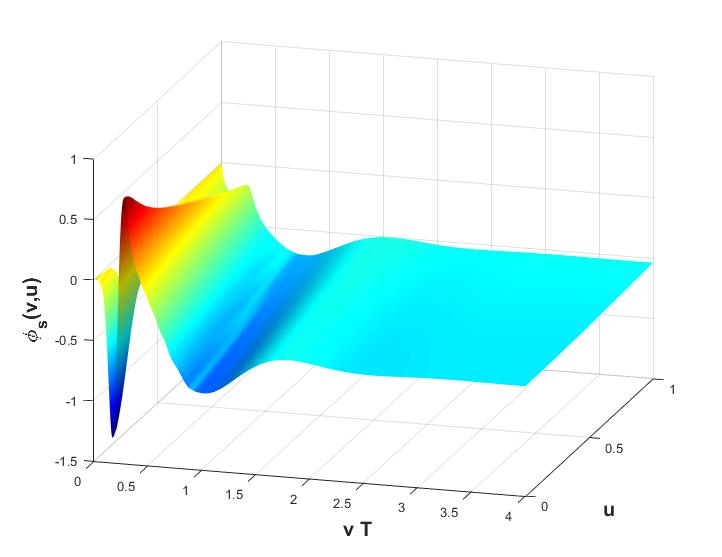}
\caption{}
\end{subfigure}
\caption{(Color online) Results for the time evolution of some fields involved in the 1RCBH setup for the initial condition \eqref{eq:BinConGauss} with $\mu/T=2$: (a) the subtracted metric anisotropy function $B_{s}(v,u)$, and (b) the subtracted dilaton field $\phi_{s}(v,u)$.}
\label{fig:3DConGauss}
\end{figure}

\begin{figure}[h]
\centering
\begin{subfigure}{0.49\textwidth}
\includegraphics[width=\textwidth]{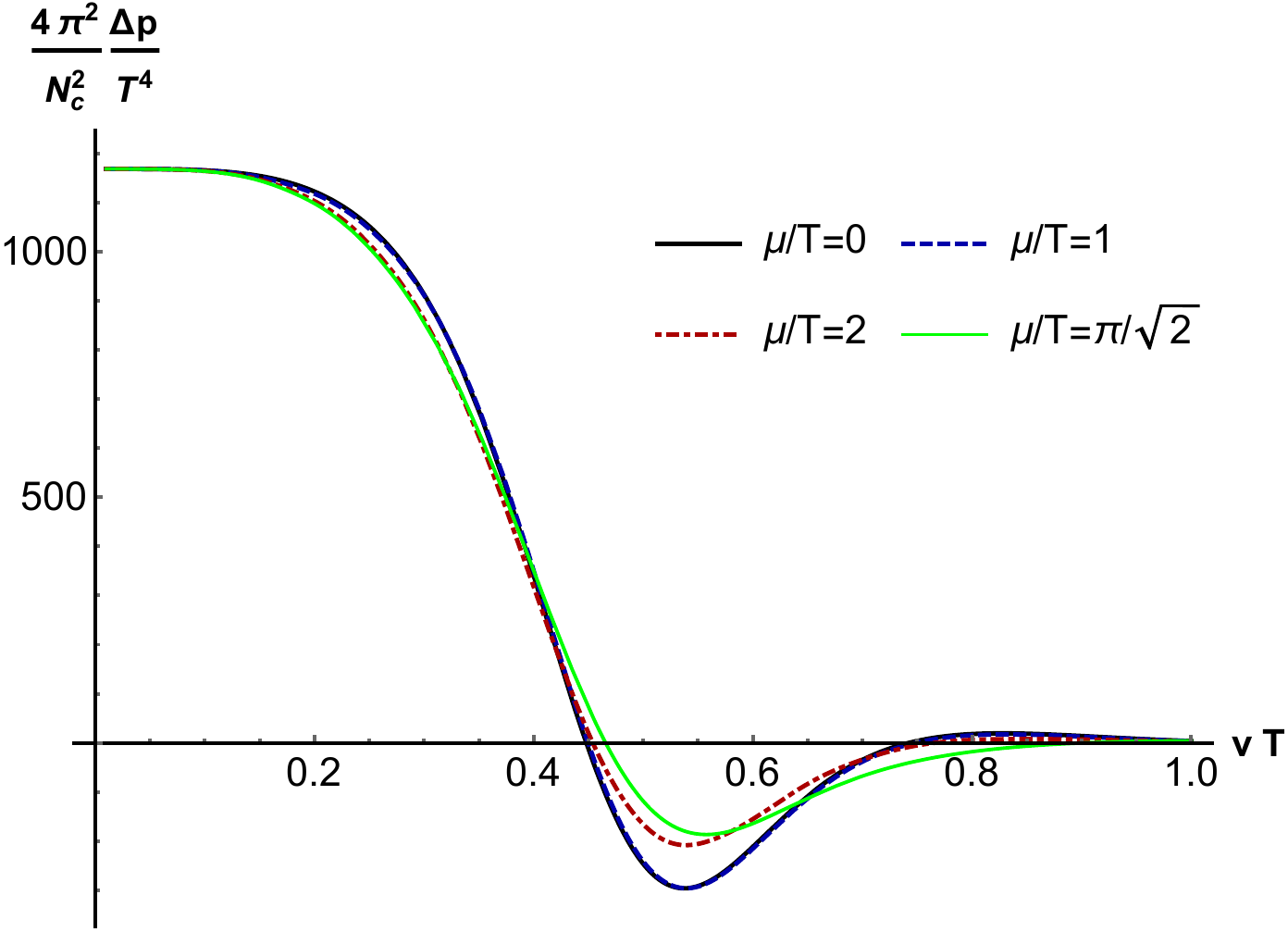}
\caption{}
\end{subfigure}
\begin{subfigure}{0.49\textwidth}
\includegraphics[width=\textwidth]{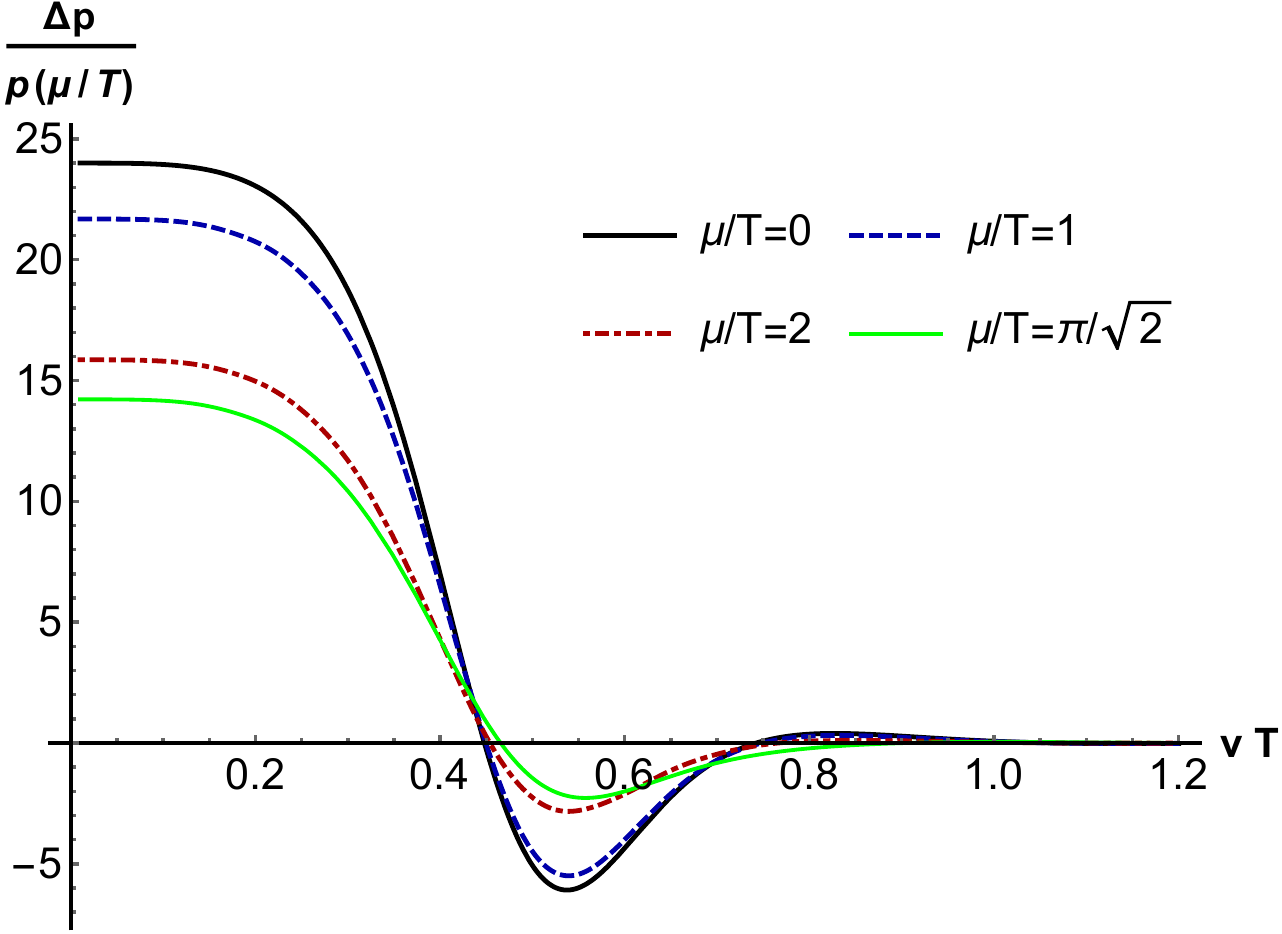}
\caption{}
\end{subfigure}
\caption{(Color online) Time evolution of the pressure anisotropy for several values of the chemical potential using the initial data \eqref{eq:BinConGauss}: (a) $\Delta p$ normalized by the (equilibrium) temperature to the fourth, and (b) $\Delta p$ normalized by the equilibrium pressure. }
\label{fig:PressureAniConGauss}
\end{figure}

\begin{figure}[h]
\centering
\begin{subfigure}{0.48\textwidth}
\includegraphics[width=\textwidth]{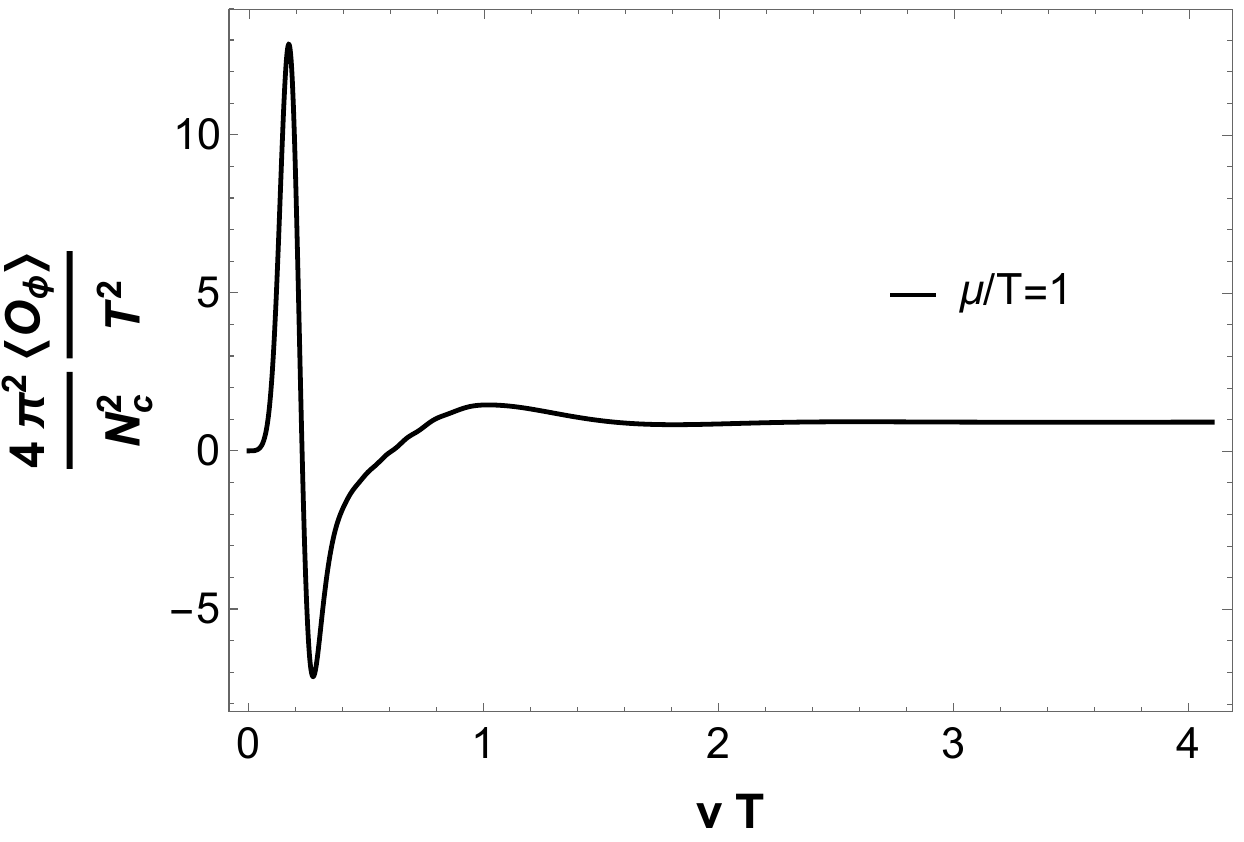}
\caption{}
\end{subfigure}
\begin{subfigure}{0.48\textwidth}
\includegraphics[width=\textwidth]{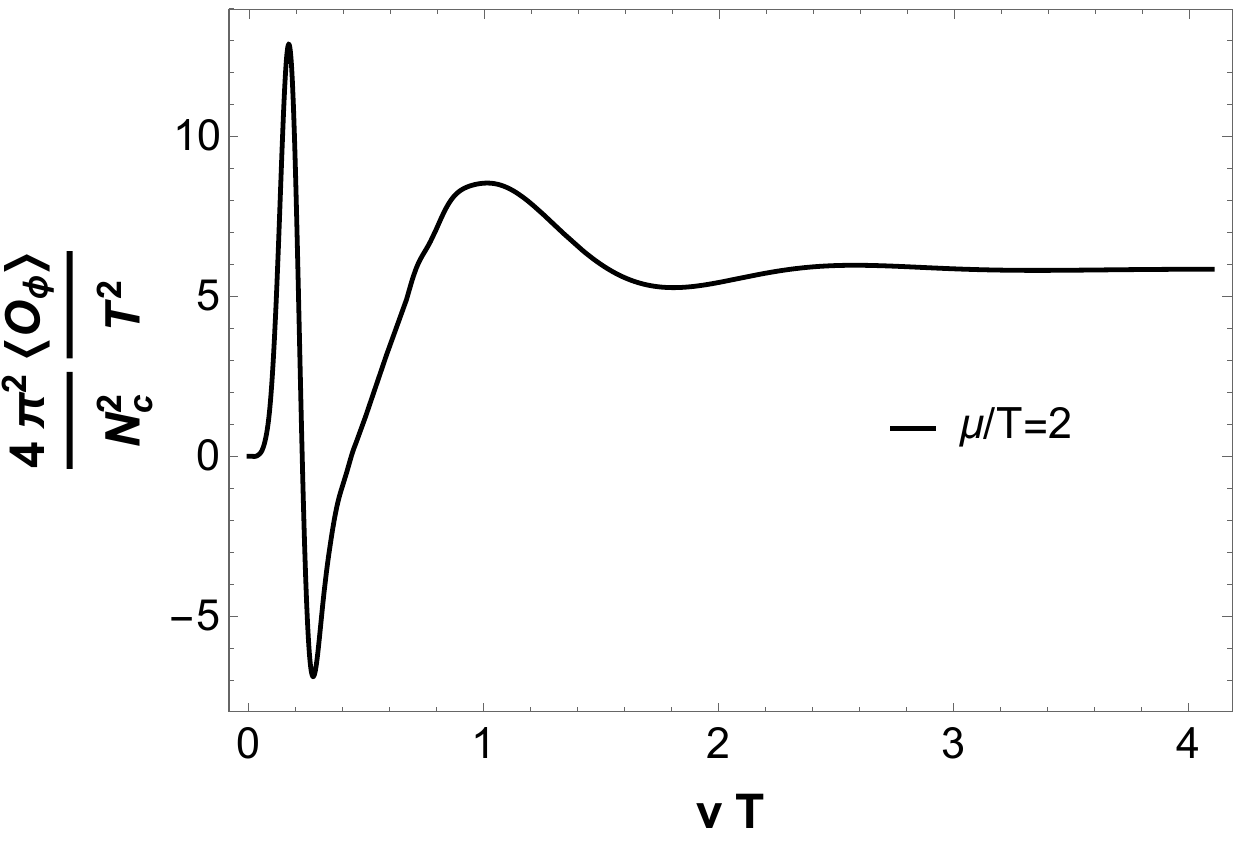}
\caption{}
\end{subfigure}
\begin{subfigure}{0.48\textwidth}
\includegraphics[width=\textwidth]{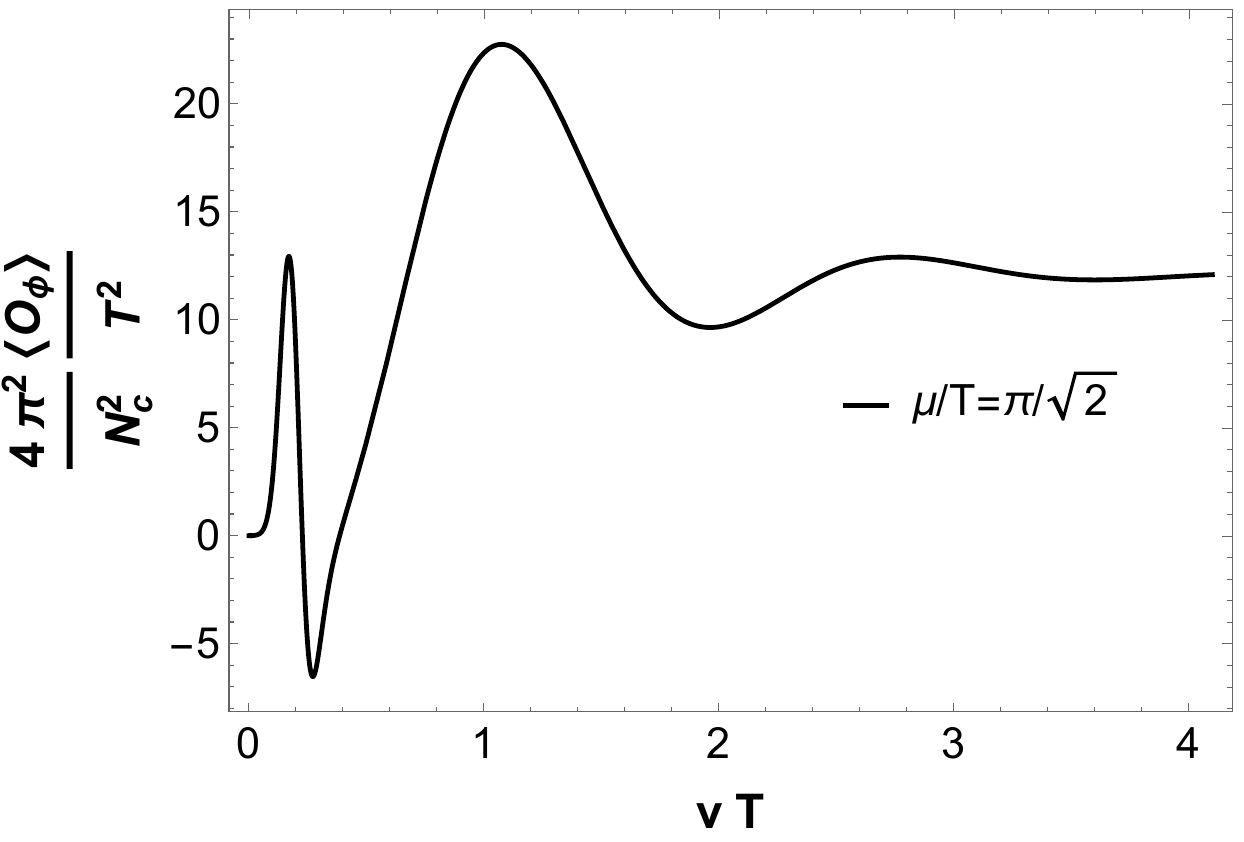}
\caption{}
\end{subfigure}
\caption{Time evolution of the scalar condensate $\langle\mathcal{O}_{\phi}\rangle$ for different $\mu/T$ using the initial data \eqref{eq:BinConGauss}.}
\label{fig:OphiConGauss}
\end{figure}

In Fig.\ \ref{fig:3DConGauss} we display a 3D plot with the time evolution of $B_s(v,u)$ and $\phi_s(v,u)$ for $\mu/T=2$ and the initial conditions set by Eq.\ \eqref{eq:BinConGauss}. By comparing with Fig.\ \ref{fig:3DConCon}, one notes that the evolution of the metric anisotropy is not very sensitive to the variation of the initial profile for the dilaton field. On the other hand, as expected, the evolution of the dilaton field is significantly affected by the choice of its initial value. This seems to suggest that the backreaction produced on the metric anisotropy by varying the initial profile for the dilaton is small, which will be confirmed in the course of the next subsections.

In Fig.\ \ref{fig:PressureAniConGauss} we show our results for the time evolution of the pressure anisotropy $\Delta p$ for different values of $\mu/T$ and the initial conditions given in Eq.\ \eqref{eq:BinConGauss}. By comparing with the results obtained in the previous subsection, displayed in Fig.\ \ref{fig:PressureAni1}, one notes that the pressure anisotropy does not change much by varying the initial condition for the dilaton, if we keep fixed the constant initial metric anisotropy. This is a direct consequence of the previously mentioned robustness of the metric anisotropy against variations of the initial dilaton profile. On the other hand, the early time dynamics of the scalar condensate $\langle\mathcal{O}_\phi\rangle$ displayed in Fig.\ \ref{fig:OphiConGauss} is very different from the result obtained in the previous subsection and shown in Fig.\ \ref{fig:OphiConCon}, while its late time dynamics is remarkably similar for the two different set of initial conditions given in Eqs.\ \eqref{eq:BinConGauss} and \eqref{eq:BinConCon1}. Moreover, the same observations done in the previous subsection, that the thermalization time associated with the equilibration of the scalar condensate only happens significantly after the system has already relaxed to an (approximately) isotropic state, and that this thermalization time always increase with increasing $\mu/T$, also hold for the initial conditions given in Eq.\ \eqref{eq:BinConGauss}. As we are going to see in the next subsections, these qualitative trends hold for all the different initial conditions considered in the present work, suggesting that they are general features of the equilibration process of the 1RCBH plasma.

\subsection{Constant metric anisotropy profile and equilibrium dilaton profile}
\label{sec:results:ConsEqui}

In this subsection we study the time evolution of the system given the initial data:
\begin{align}\label{eq:BinConEqui}
B_s(v_0,u) =2 \ \ \text{and} \ \ \phi_s(v_0,u) =-\frac{1}{u^2}\sqrt{\frac{2}{3}} \ln \left( 1 + \tilde{u}(u)Q^2 \right),
\end{align}
where we kept the same constant initial metric anisotropy as in the two previous subsections but considered now the initial dilaton profile discussed in Eq.\ \eqref{eq:phiini3}, corresponding to the equilibrium value of the dilaton field.

\begin{figure}[h]
\centering
\begin{subfigure}{0.5\textwidth}
\includegraphics[width=\textwidth]{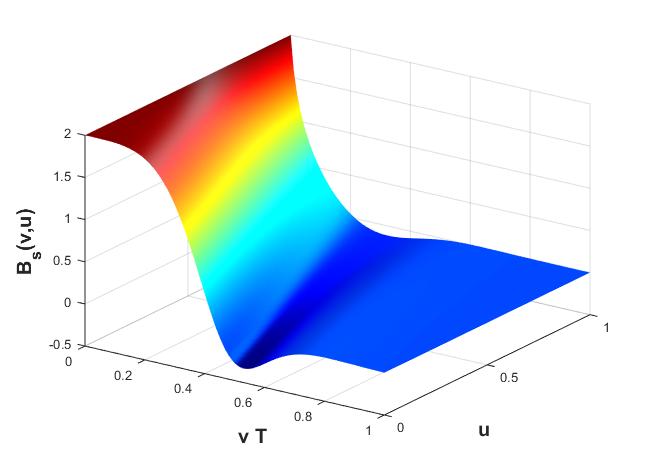}
\caption{}
\end{subfigure}
\begin{subfigure}{0.49\textwidth}
\includegraphics[width=\textwidth]{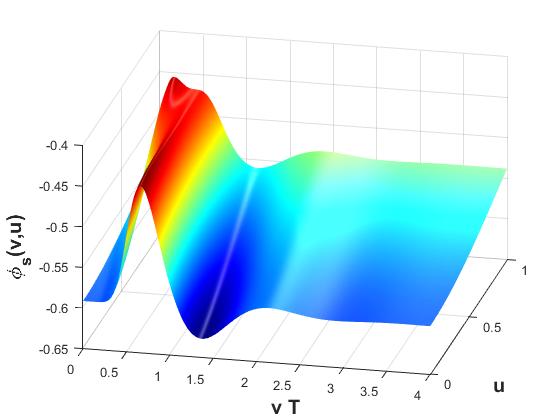}
\caption{}
\end{subfigure}
\caption{(Color online) Results for the time evolution of some fields involved in the 1RCBH setup for the initial condition \eqref{eq:BinConEqui} with $\mu/T=2$: (a) the subtracted metric anisotropy function $B_{s}(v,u)$, and (b) the subtracted dilaton field $\phi_{s}(v,u)$.}
\label{fig:3DConEqui}
\end{figure}

\begin{figure}[h]
\centering
\begin{subfigure}{0.49\textwidth}
\includegraphics[width=\textwidth]{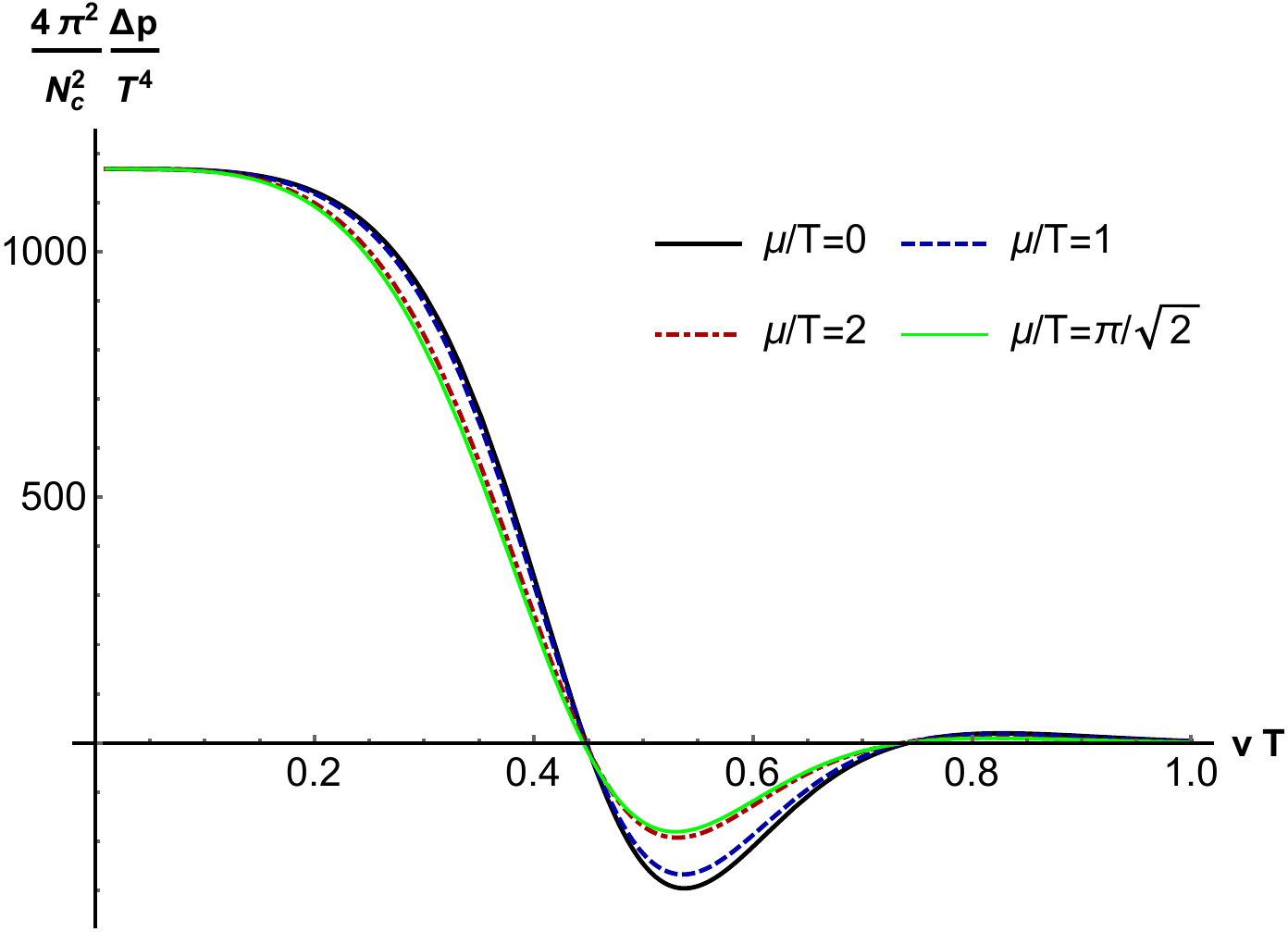}
\caption{}
\end{subfigure}
\begin{subfigure}{0.49\textwidth}
\includegraphics[width=\textwidth]{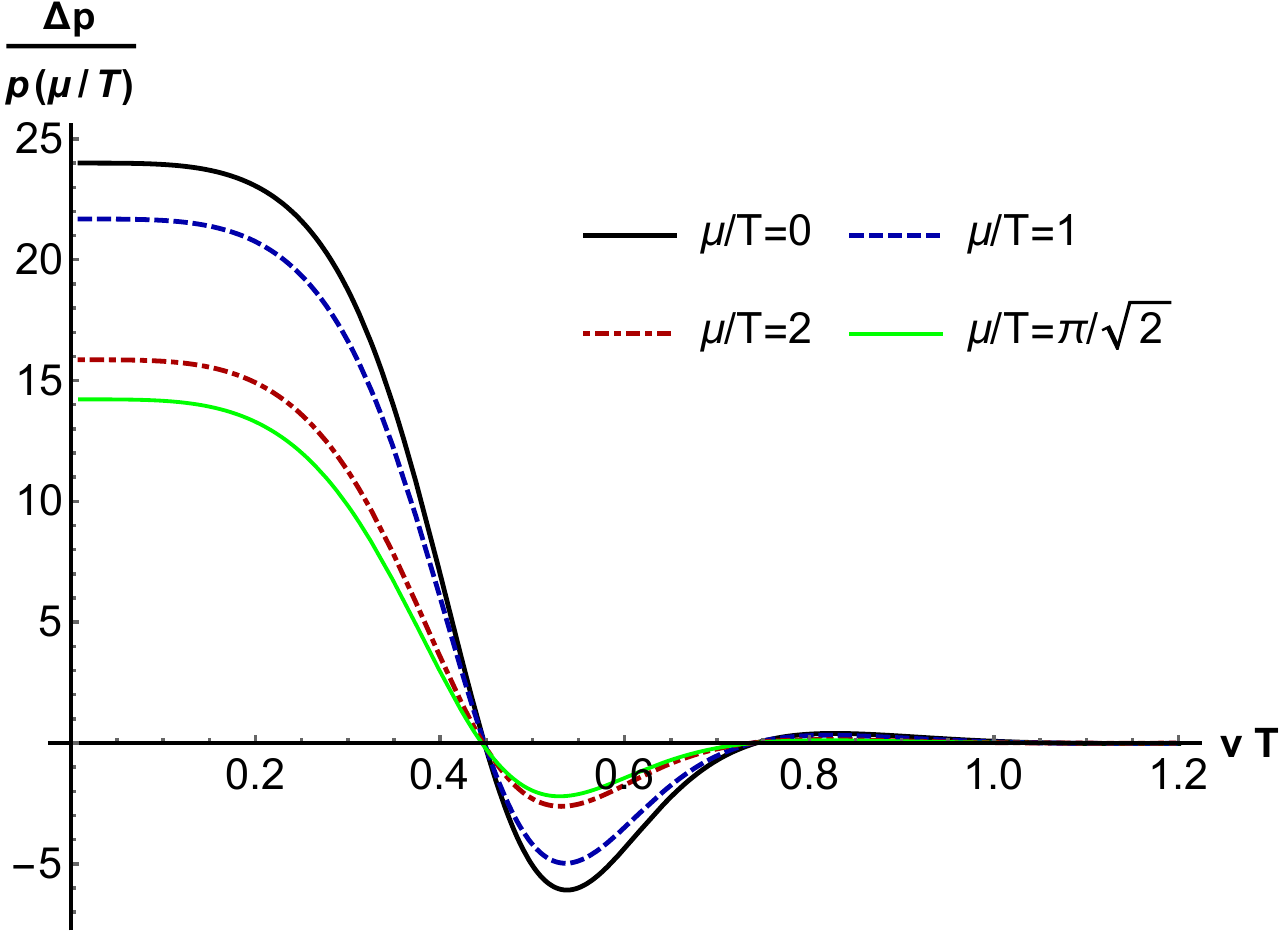}
\caption{}
\end{subfigure}
\caption{(Color online) Time evolution of the pressure anisotropy for several values of the chemical potential using the initial data \eqref{eq:BinConEqui}: (a) $\Delta p$ normalized by the (equilibrium) temperature to the fourth, and (b) $\Delta p$ normalized by the equilibrium pressure. }
\label{fig:PressureAniConEqui}
\end{figure}

\begin{figure}[h]
\centering
\begin{subfigure}{0.48\textwidth}
\includegraphics[width=\textwidth]{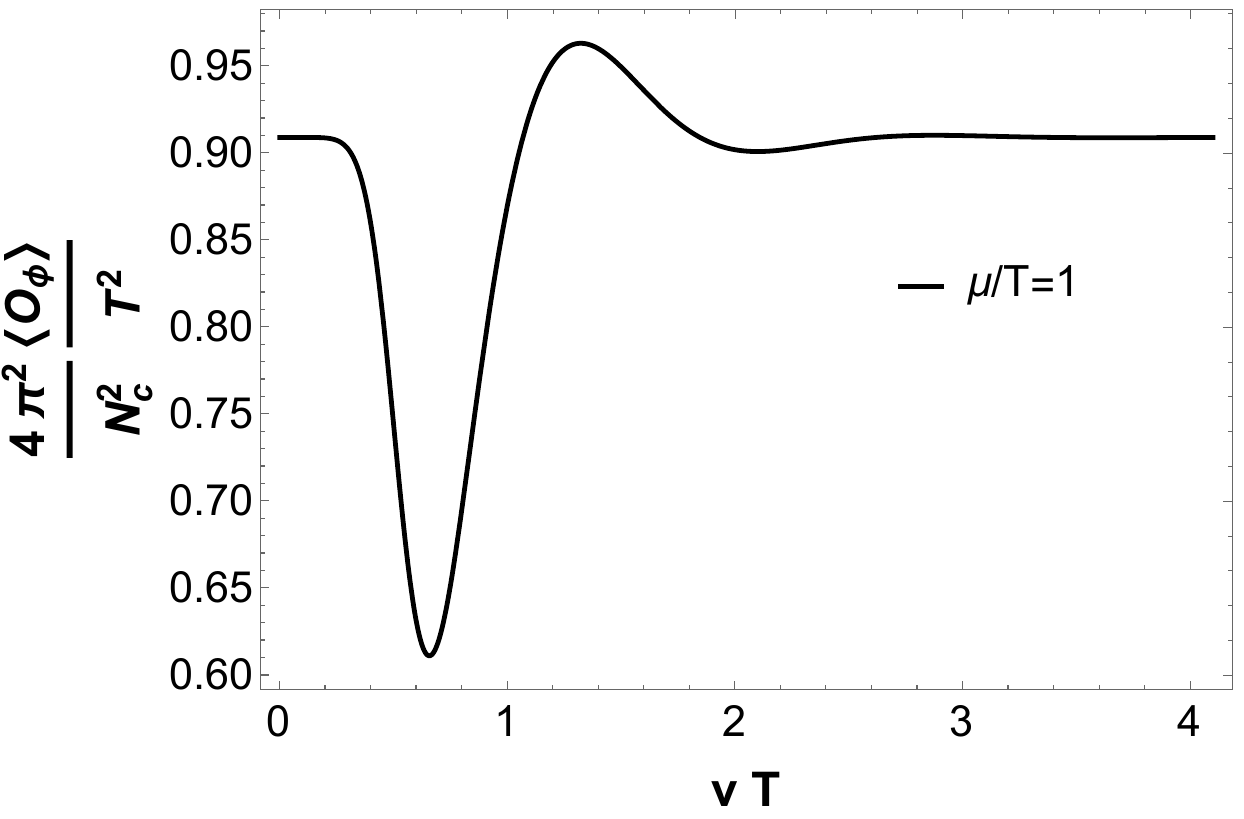}
\caption{}
\end{subfigure}
\begin{subfigure}{0.48\textwidth}
\includegraphics[width=\textwidth]{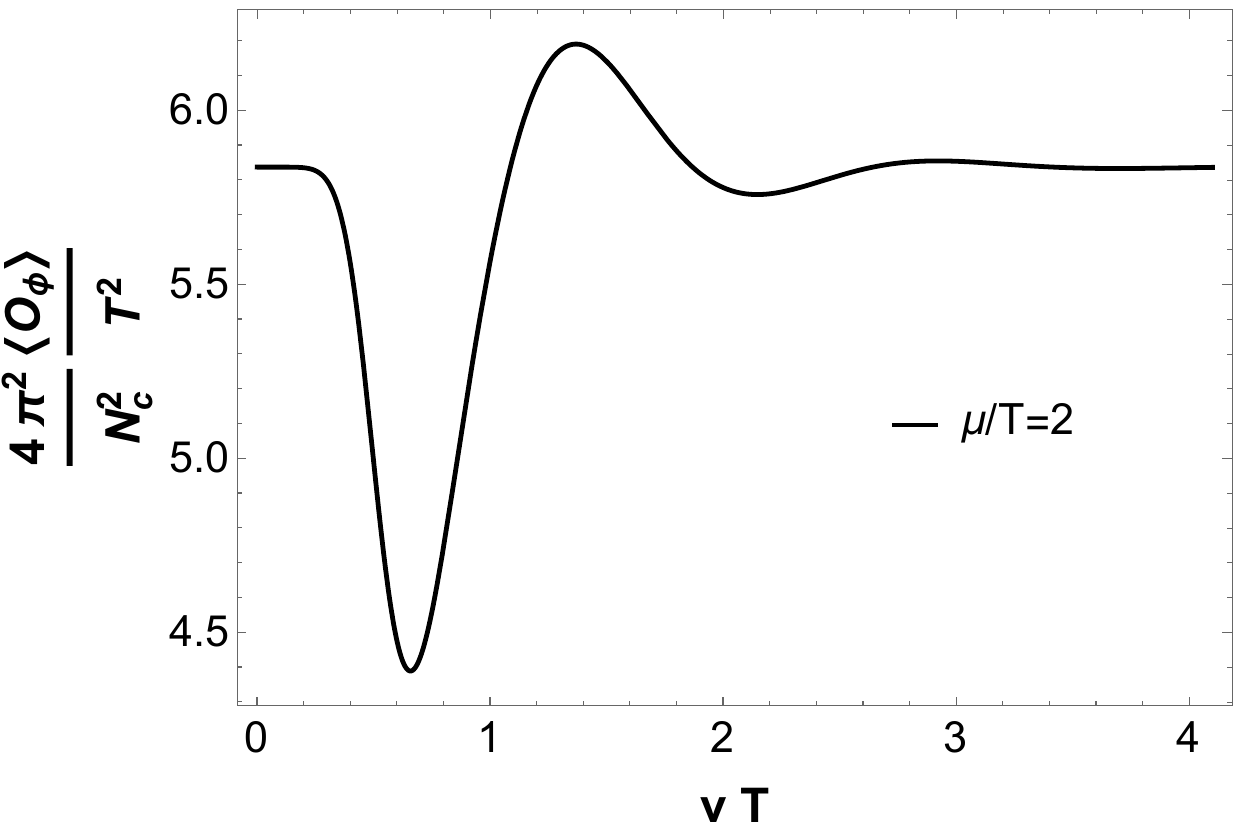}
\caption{}
\end{subfigure}
\begin{subfigure}{0.48\textwidth}
\includegraphics[width=\textwidth]{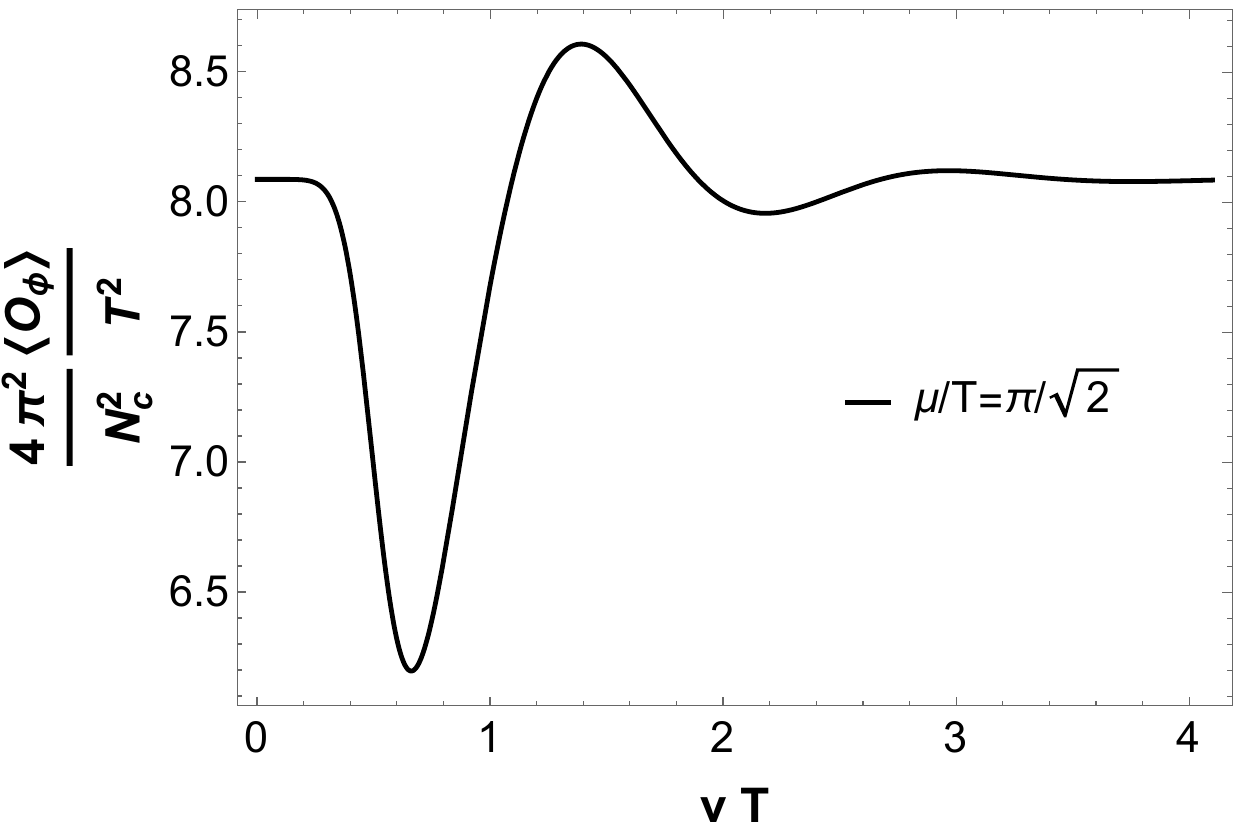}
\caption{}
\end{subfigure}
\caption{Time evolution of the scalar condensate $\langle\mathcal{O}_{\phi}\rangle$ for different $\mu/T$ using the initial data \eqref{eq:BinConEqui}.}
\label{fig:OphiConEqui}
\end{figure}

In Fig.\ \ref{fig:3DConEqui} we display a 3D plot with the time evolution of $B_s(v,u)$ and $\phi_s(v,u)$ for $\mu/T=2$ and the initial conditions set by Eq.\ \eqref{eq:BinConEqui}. By comparing with the results in the two previous subsections, one confirms once again the robustness of the time evolution of the metric anisotropy against different choices for the initial dilaton profile.

In Fig.\ \ref{fig:PressureAniConEqui} we display our results for the time evolution of the pressure anisotropy, which are similar to what we have found in the two previous subsections. Regarding the time evolution of the scalar condensate shown in Fig.\ \ref{fig:OphiConEqui}, we see that its early time dynamics is different from the previous cases considered here, although its late time dynamics is very similar to those found before. Furthermore, one also notes that in the present case where the initial condition for the dilaton is already its equilibrium value that the fluctuations in the value of the dilaton as time evolves are generally small. This is expected since in the present case the far-from-equilibrium dynamics of the system is being initially driven solely by the metric anisotropy.

\subsection{Gaussian metric anisotropy profile and constant dilaton profile}
\label{sec:results:GaussCon}

Now we consider a different set of initial conditions to scan out possible new features in the equilibration dynamics of the 1RCBH plasma as a function of the chosen initial data. We change the initial metric anisotropy profile $B_s(v_0,u)$ but use the same constant initial profile for the dilaton field $\phi_s(v_0,u)$ as in subsection \ref{sec:results:const}, which is a way to probe if there are some new features depending mostly on the chosen initial anisotropy. We present in this subsection the results for a Gaussian initial metric anisotropy \eqref{eq:Bini1}, which is perhaps the most common initial condition chosen for this field, combined with a constant initial profile for the dilaton field \eqref{eq:phiini1}
\begin{align}\label{eq:BinGaussCon}
B_s(v_0,u) =0.5\, e^{-100(u-0.4)^2}  \ \ \text{and} \ \ \phi_s(v_0,u) = -\sqrt{\frac{2}{3}}Q^2,
\end{align}
where the Gaussian initial anisotropy was chosen to be more or less half-way between the black hole horizon and the boundary.

\begin{figure}[h]
\centering
\begin{subfigure}{0.49\textwidth}
\includegraphics[width=\textwidth]{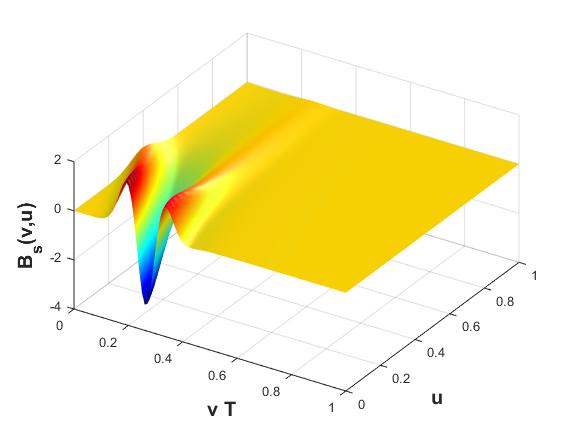}
\caption{}
\end{subfigure}
\begin{subfigure}{0.49\textwidth}
\includegraphics[width=\textwidth]{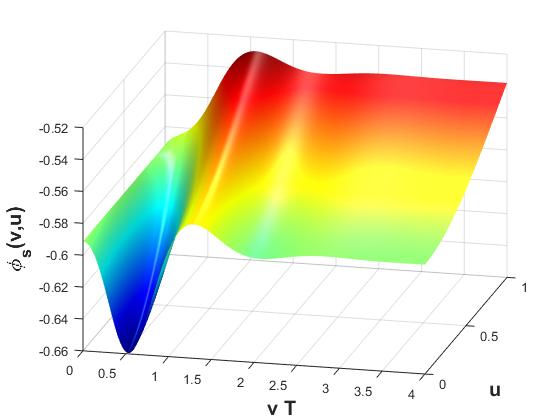}
\caption{}
\end{subfigure}
\caption{(Color online) Results for the time evolution of some fields involved in the 1RCBH setup for the initial condition \eqref{eq:BinGaussCon} with $\mu/T=2$: (a) the subtracted metric anisotropy function $B_{s}(v,u)$, and (b) the subtracted dilaton field $\phi_{s}(v,u)$.}
\label{fig:3DGaussCon}
\end{figure}

\begin{figure}[h]
\centering
\begin{subfigure}{0.49\textwidth}
\includegraphics[width=\textwidth]{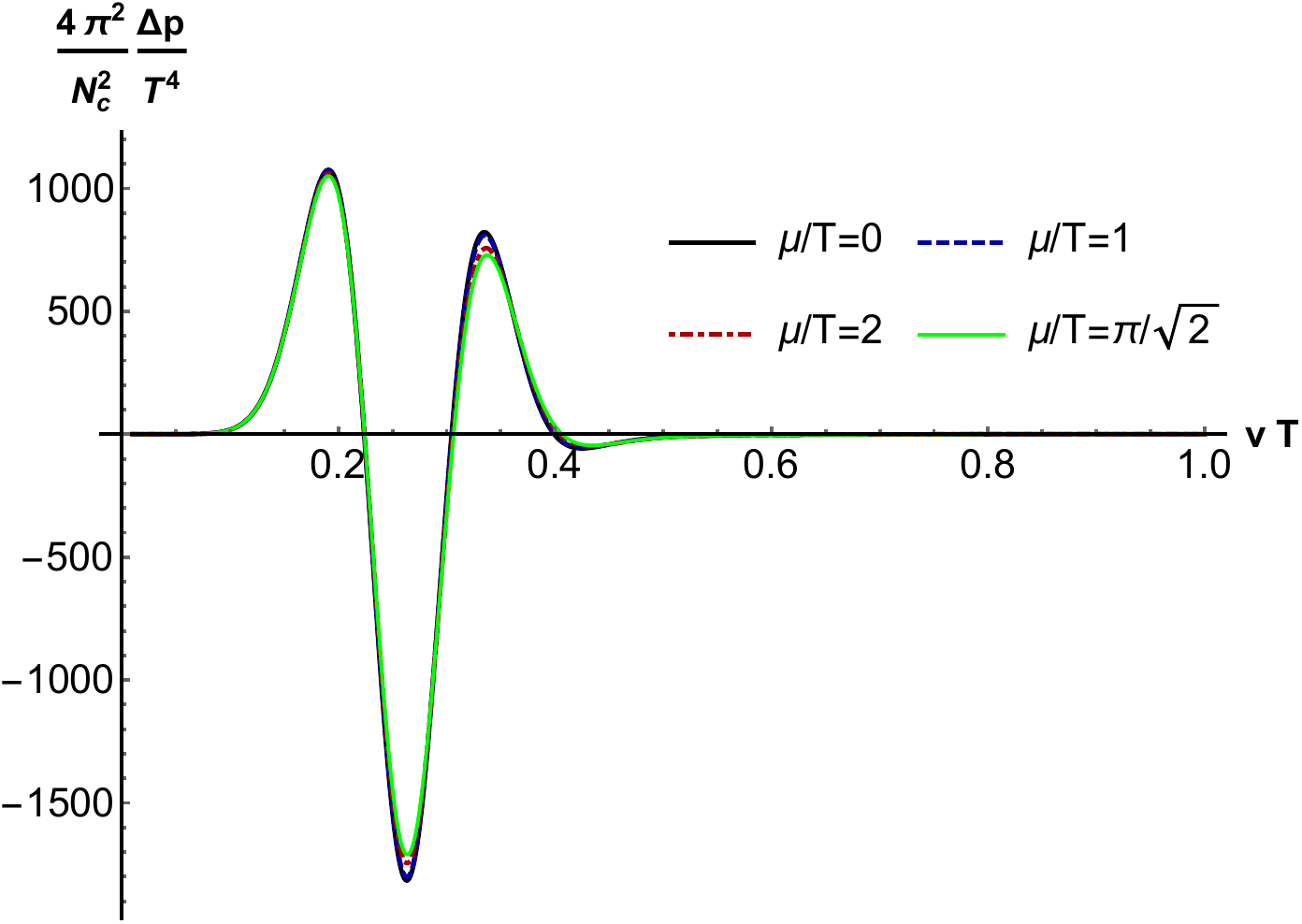}
\caption{}
\end{subfigure}
\begin{subfigure}{0.49\textwidth}
\includegraphics[width=\textwidth]{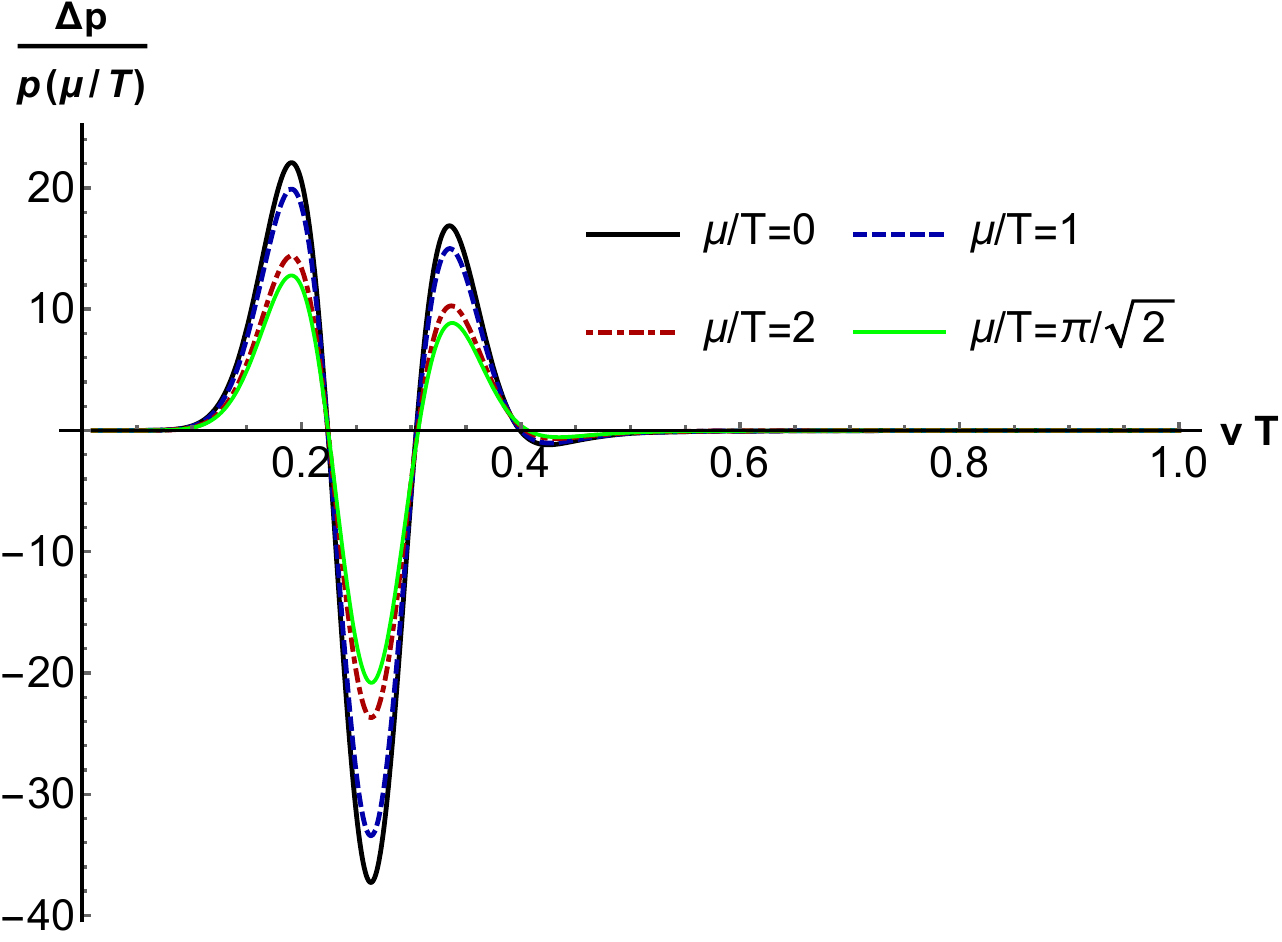}
\caption{}
\end{subfigure}
\caption{(Color online) Time evolution of the pressure anisotropy for several values of the chemical potential using the initial data \eqref{eq:BinGaussCon}: (a) $\Delta p$ normalized by the (equilibrium) temperature to the fourth, and (b) $\Delta p$ normalized by the equilibrium pressure.}
\label{fig:PressureAni2}
\end{figure}

\begin{figure}[h]
\centering
\begin{subfigure}{0.48\textwidth}
\includegraphics[width=\textwidth]{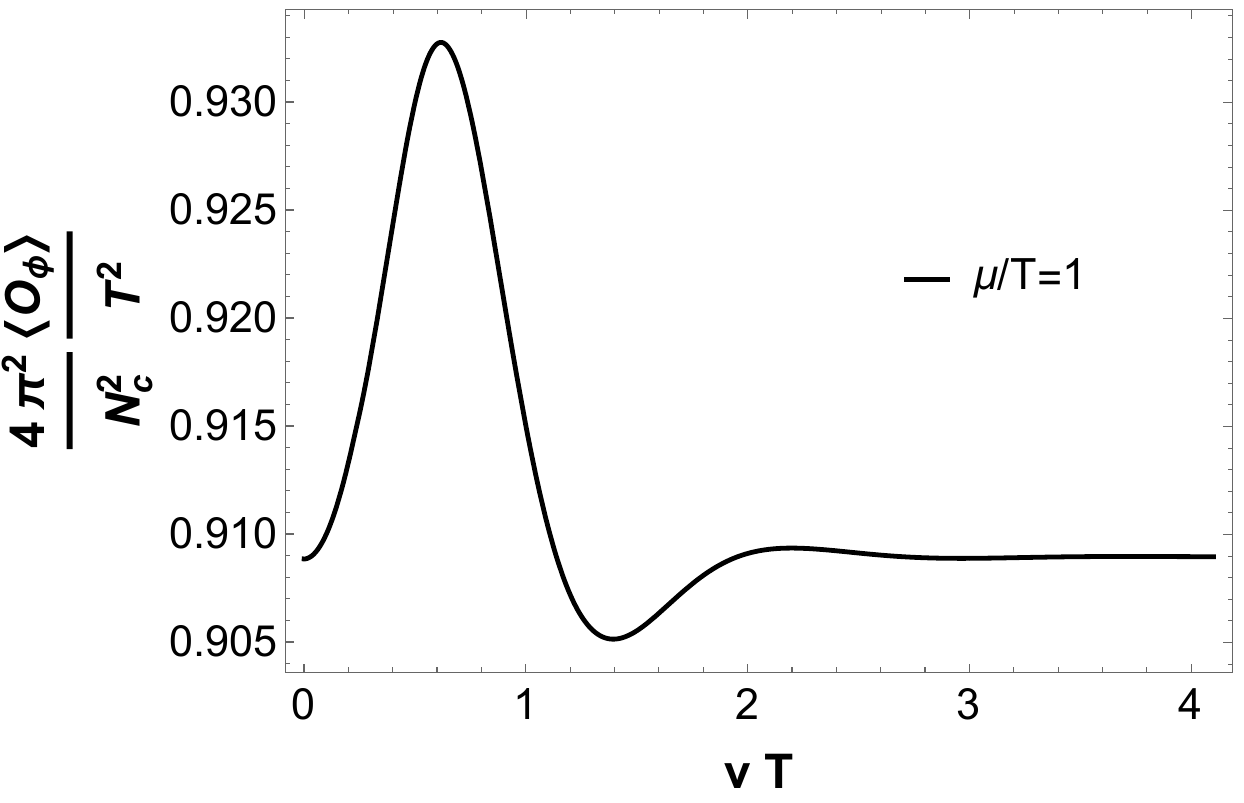}
\caption{}
\end{subfigure}
\begin{subfigure}{0.48\textwidth}
\includegraphics[width=\textwidth]{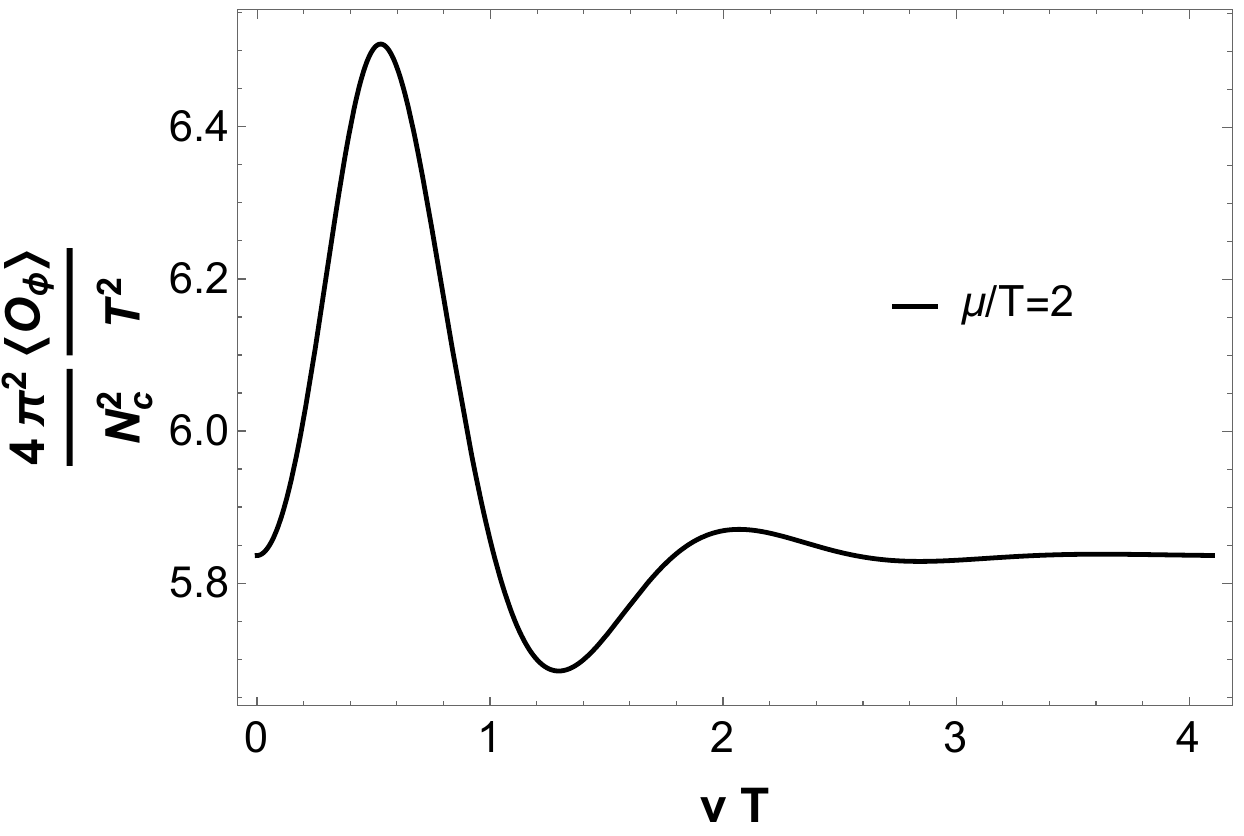}
\caption{}
\end{subfigure}
\begin{subfigure}{0.48\textwidth}
\includegraphics[width=\textwidth]{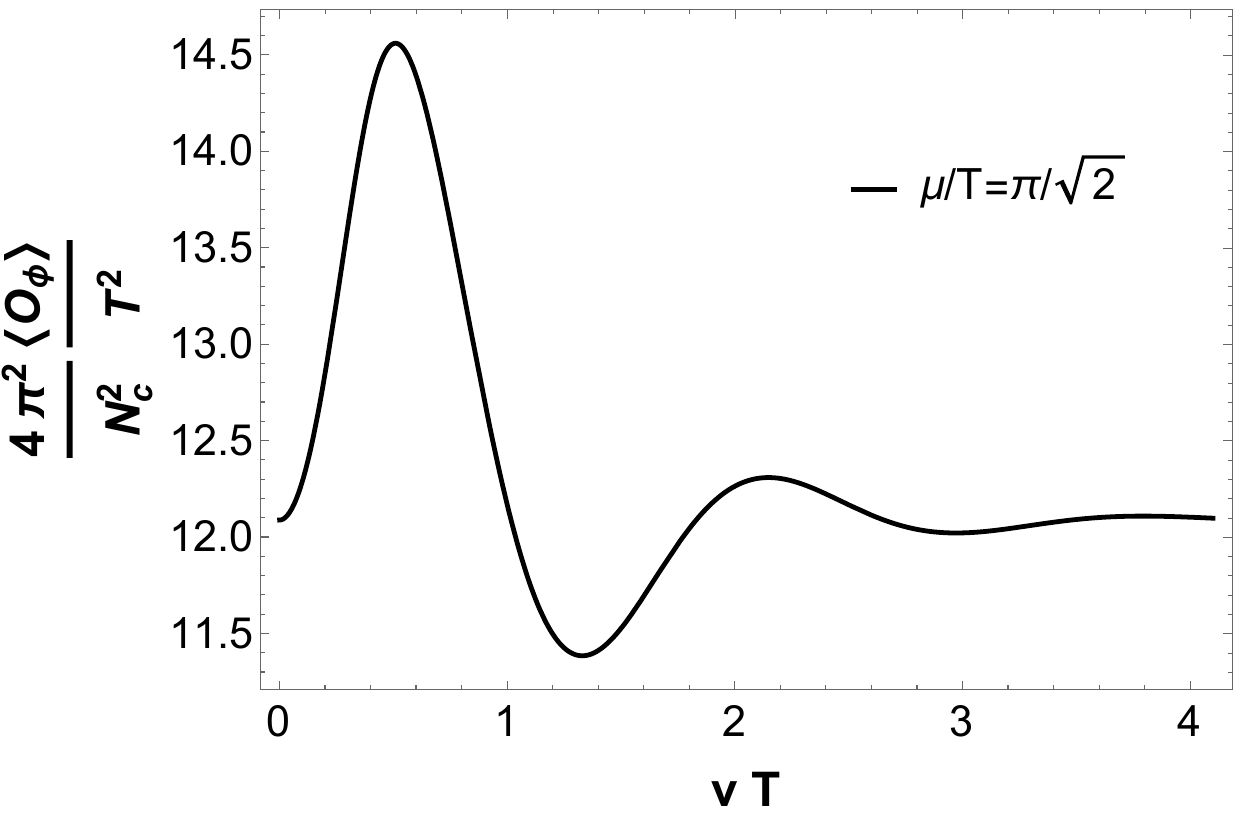}
\caption{}
\end{subfigure}
\caption{Time evolution of the scalar condensate $\langle\mathcal{O}_{\phi}\rangle$ for different $\mu/T$ using the initial data \eqref{eq:BinGaussCon}.}
\label{fig:OphiGaussCon}
\end{figure}

In Fig.\ \ref{fig:3DGaussCon} we show a 3D plot with the time evolution of $B_s(v,u)$ and $\phi_s(v,u)$ using the initial condition given in Eq.\ \eqref{eq:BinGaussCon} and $\mu/T=2$. The shape of both functions are very different than the case considered in subsection \ref{sec:results:const}, whose results are displayed in Fig.\ \ref{fig:3DConCon} and where the initial condition for the dilaton was the same as in Eq.\ \eqref{eq:BinGaussCon} but the initial metric anisotropy was constant. This signalizes a strong backreaction induced on the dilaton field by changing the initial metric anisotropy, in contrast to the general trend observed in the previous subsections that the backreaction induced on the metric anisotropy by changing the initial dilaton profile is quite small. These general features of the far-from-equilibrium 1RCBH plasma will be further confirmed in subsections \ref{sec:results:GaussGauss} and \ref{sec:results:GaussEq}.

The time evolution of the pressure anisotropy $\Delta p$ that we obtain from the boundary value of $B_s(v,u)$ in the present case is shown in Fig.\ \ref{fig:PressureAni2}. One can see that the early time dynamics of the pressure anisotropy in the case of the initial condition set in Eq.\ \eqref{eq:BinGaussCon} is completely different than the one obtained in Fig.\ \ref{fig:PressureAni1} by considering the initial condition given by Eq.\ \eqref{eq:BinConCon1}. In particular, the isotropization time has significantly decreased by considering a Gaussian initial metric anisotropy when compared to the isotropization times associated with a constant initial metric anisotropy analyzed in the previous subsections.

The time evolution of the scalar condensate $\langle\mathcal{O}_\phi \rangle$ for the initial condition \eqref{eq:BinGaussCon} is shown in Fig.\ \ref{fig:OphiGaussCon}. Compared with the result presented in Fig.\ \ref{fig:OphiConCon} using a constant initial anisotropy \eqref{eq:BinConCon1}, we note that the early time dynamics of $\langle\mathcal{O}_\phi \rangle$ is very different depending on the chosen initial data. However, the late time dynamics of the scalar condensate and the associated thermalization time are actually very similar for both sets of initial conditions. These observations are in agreement with the general trend observed in the previous subsections and will be further confirmed in the course of the next subsections.

\subsection{Gaussian metric anisotropy and dilaton profiles}
\label{sec:results:GaussGauss}

Now we change the initial dilaton profile $\phi_s(v_0,u)$ but maintain the prior parametrization for the initial metric anisotropy $B_s(v_0,u)$
\begin{align}\label{eq:BinGaussGauss}
B_s(v_0,u) =0.5\, e^{-100(u-0.4)^2}  \ \ \text{and} \ \ \phi_s(v_0,u) = -0.5\, e^{-100(u-0.3)^2},
\end{align}
where $B_s(v_0,u)$ is kept the same as in the analysis of the previous subsection. Moreover, we also considered different values for $B_s(v_0,u)$ and $\phi_s(v_0,u)$ with the same functional forms provided in Eq.\ \eqref{eq:BinGaussGauss} and the results follow the same general behavior as the ones obtained with this equation.

\begin{figure}[h]
\centering
\begin{subfigure}{0.49\textwidth}
\includegraphics[width=\textwidth]{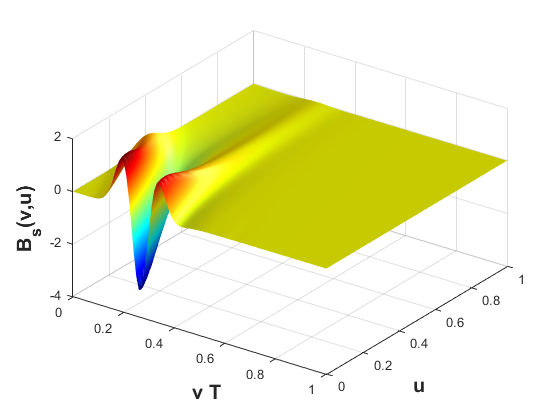}
\caption{}
\end{subfigure}
\begin{subfigure}{0.49\textwidth}
\includegraphics[width=\textwidth]{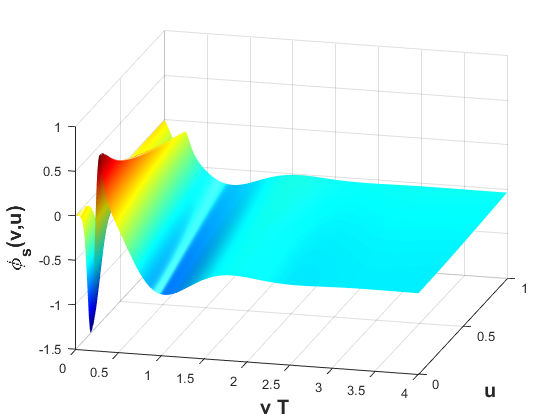}
\caption{}
\end{subfigure}
\caption{(Color online) Results for the time evolution of some fields involved in the 1RCBH setup for the initial condition \eqref{eq:BinGaussGauss} with $\mu/T=2$: (a) the subtracted metric anisotropy function $B_{s}(v,u)$, and (b) the subtracted dilaton field $\phi_{s}(v,u)$.}
\label{fig:3DGaussGauss}
\end{figure}

\begin{figure}[h]
\centering
\begin{subfigure}{0.49\textwidth}
\includegraphics[width=\textwidth]{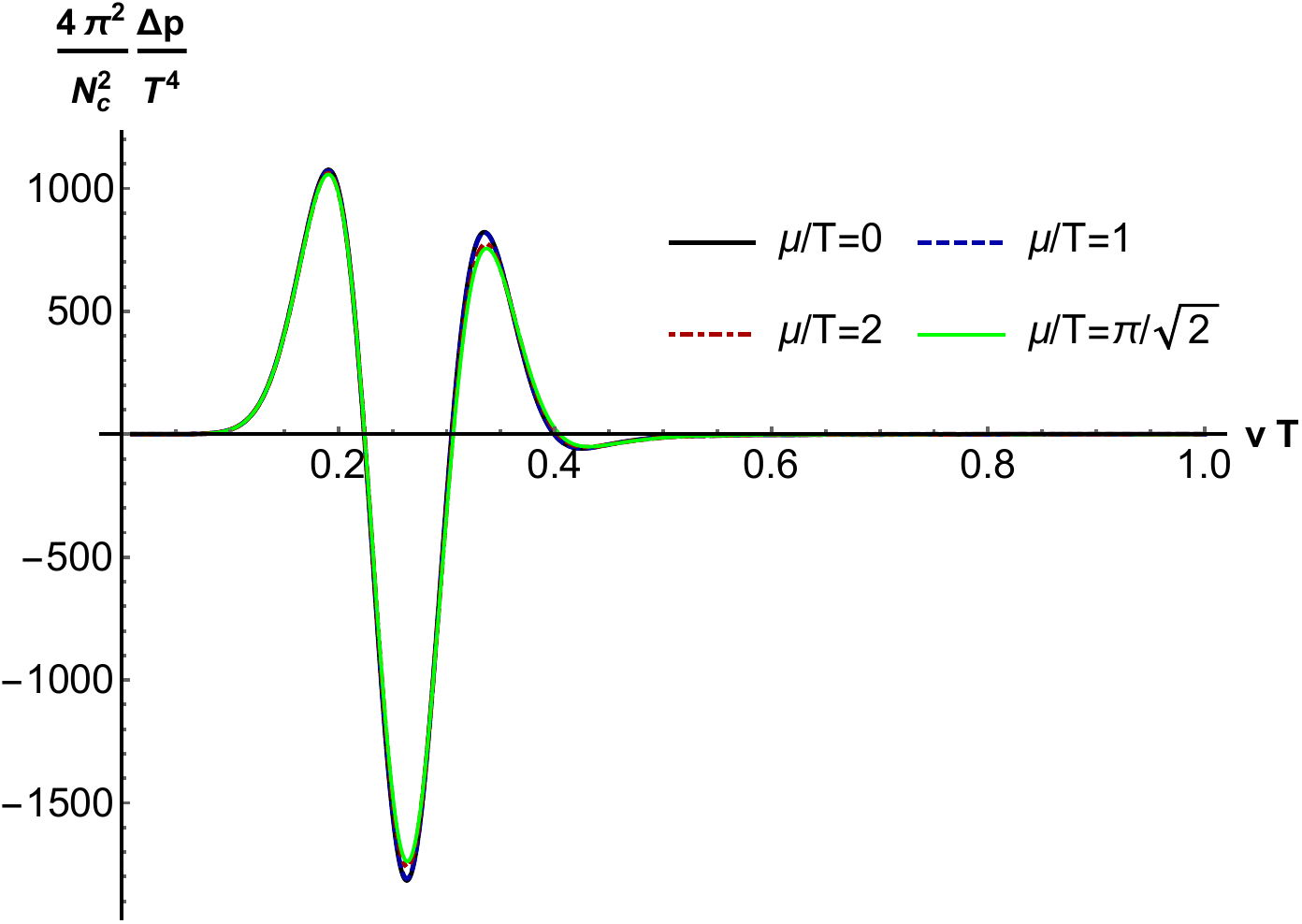}
\caption{}
\end{subfigure}
\begin{subfigure}{0.49\textwidth}
\includegraphics[width=\textwidth]{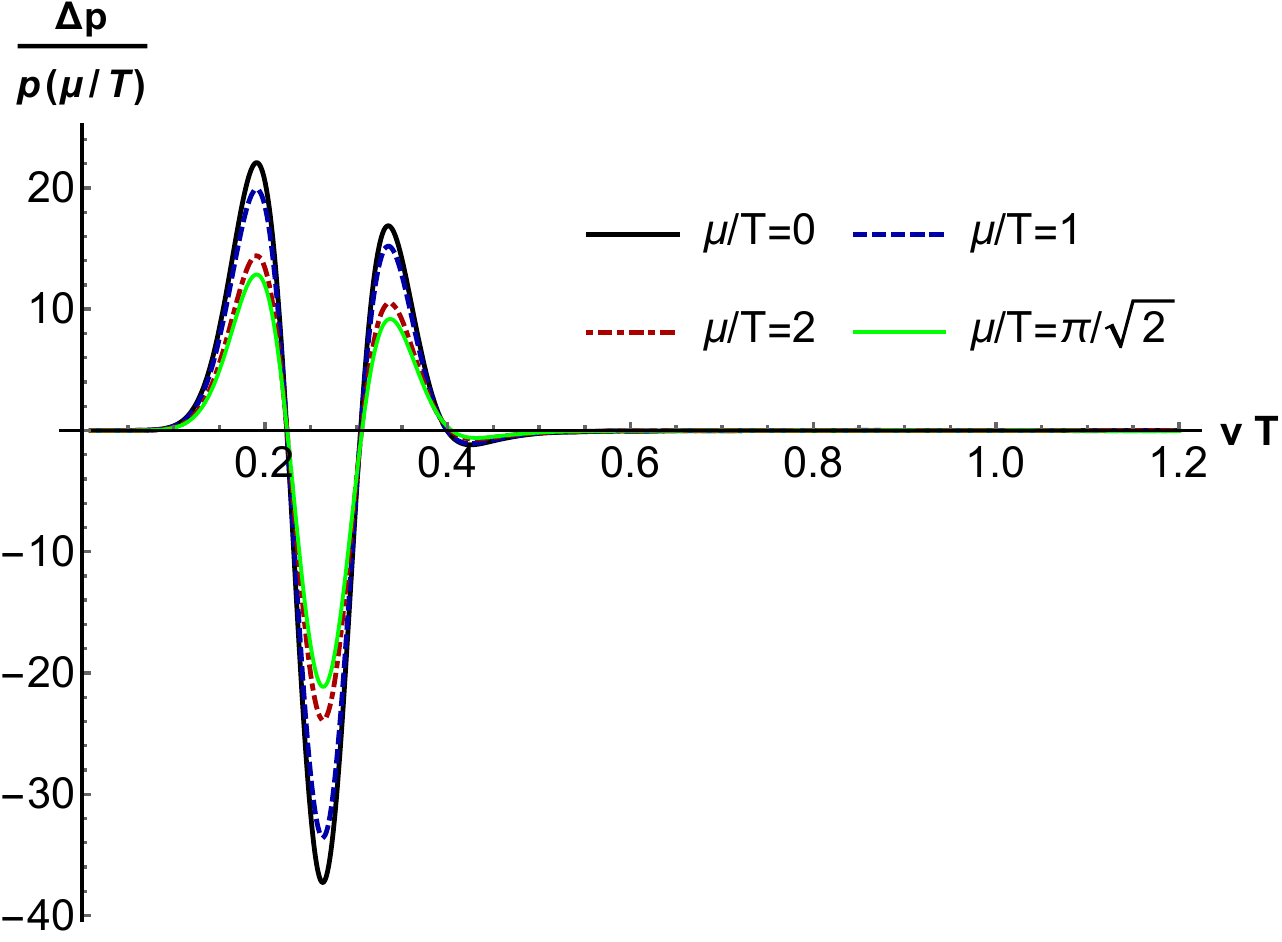}
\caption{}
\end{subfigure}
\caption{(Color online) Time evolution of the pressure anisotropy for several values of the chemical potential using the initial data \eqref{eq:BinGaussGauss}: (a) $\Delta p$ normalized by the (equilibrium) temperature to the fourth, and (b) $\Delta p$ normalized by the equilibrium pressure.}
\label{fig:PressureAni3}
\end{figure}

\begin{figure}[h]
\centering
\begin{subfigure}{0.48\textwidth}
\includegraphics[width=\textwidth]{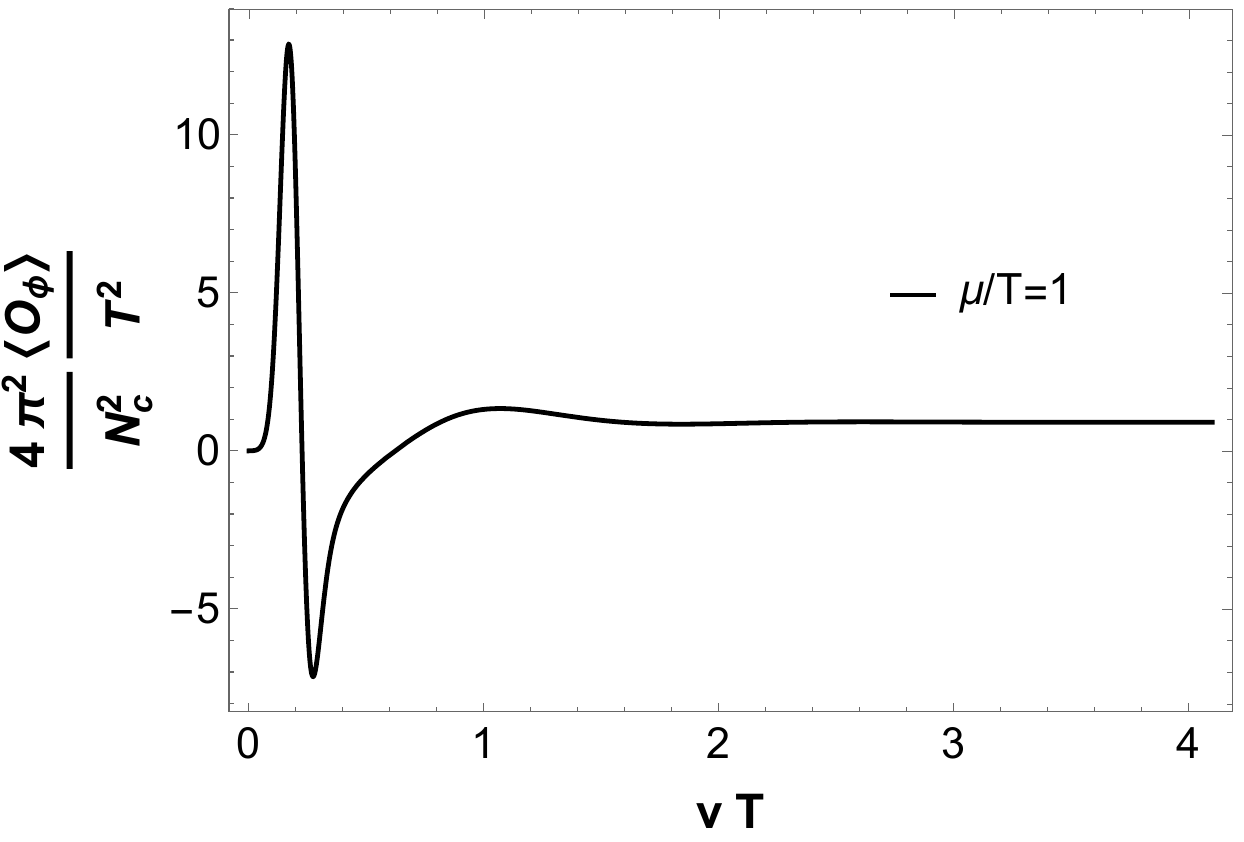}
\caption{}
\end{subfigure}
\begin{subfigure}{0.48\textwidth}
\includegraphics[width=\textwidth]{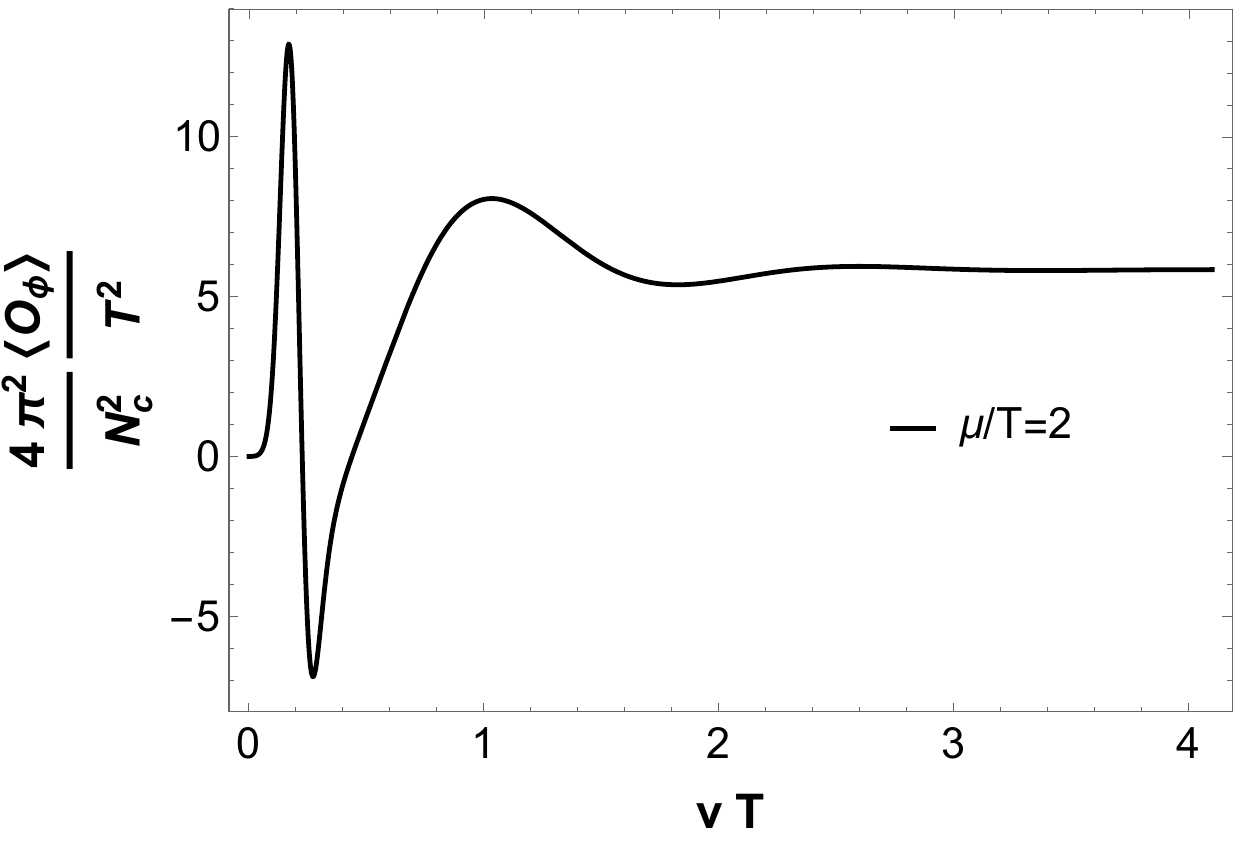}
\caption{}
\end{subfigure}
\begin{subfigure}{0.48\textwidth}
\includegraphics[width=\textwidth]{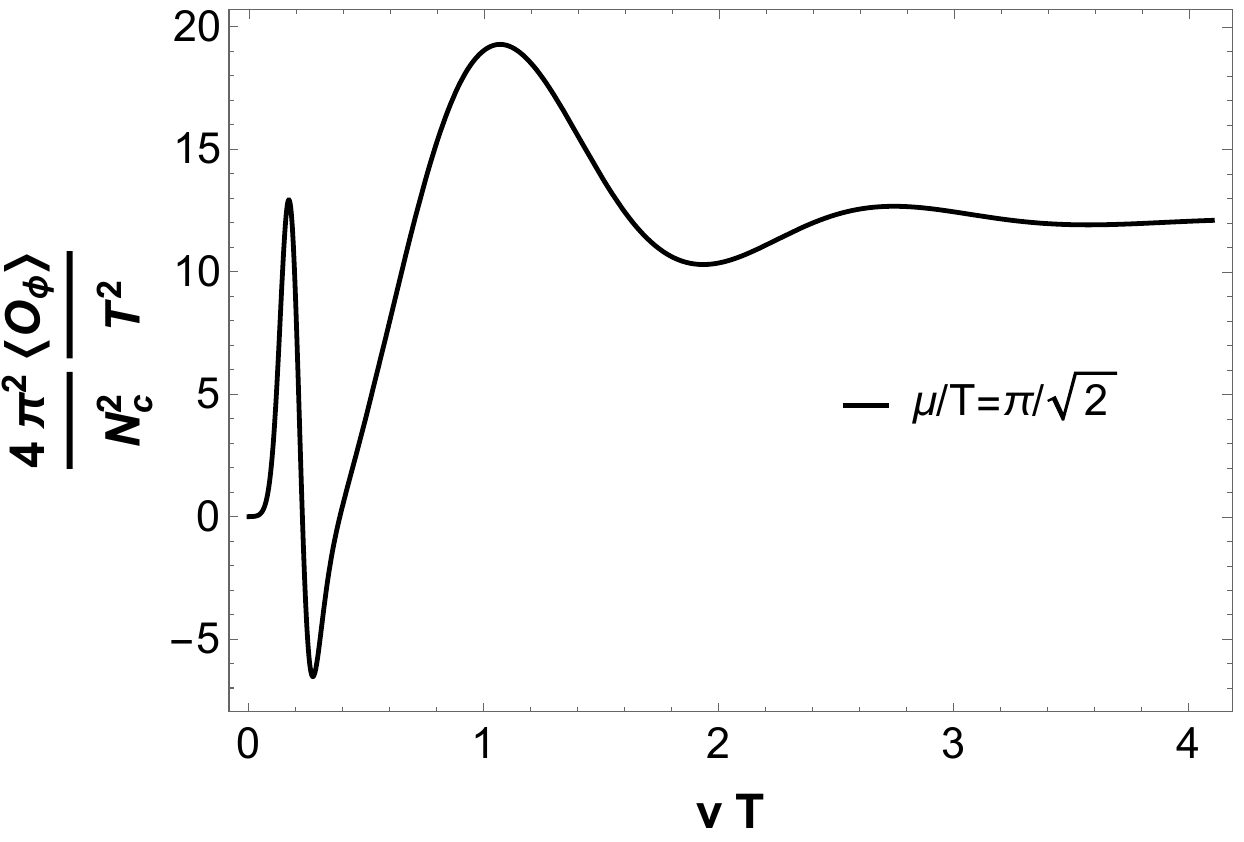}
\caption{}
\end{subfigure}
\caption{Time evolution of the scalar condensate $\langle\mathcal{O}_{\phi}\rangle$ for different $\mu/T$ using the initial data \eqref{eq:BinGaussGauss}.}
\label{fig:OphiGaussGauss}
\end{figure}

The 3D plot with the time evolution of $B_s(v,u)$ and $\phi_s(v,u)$ is shown in Fig.\ \ref{fig:3DGaussGauss} for the initial condition \eqref{eq:BinGaussGauss} and $\mu/T=2$. From this plot, it is evident the similarity of the shape of the metric anisotropy $B_s(v,u)$ with the previous result shown in \ref{fig:3DGaussCon}, i.e. a different parameterization for the initial dilaton profile $\phi_s(v_0,u)$ has essentially no effect on the time evolution of the metric anisotropy. On the other hand, one notes a very different time evolution for the dilaton $\phi_s(v,u)$, which is expected since we changed the initial parameterization of this field.

As a direct consequence of the features above, the time evolution of the pressure anisotropy $\Delta p$ depicted in Fig.\ \ref{fig:PressureAni3} is essentially the same as in the previous case shown in Fig.\ \ref{fig:PressureAni2}. The difference between them can be barely seen (what configures an even stronger test of robustness than observed before by considering initial conditions with a fixed constant initial metric anisotropy), indicating that the pressure anisotropy is remarkably robust against the addition of other fields in the gravitational action besides the metric. On the other hand, the early time dynamics of the scalar condensate $\langle\mathcal{O}_\phi \rangle$ presented in Fig.\ \ref{fig:OphiGaussGauss} is very different from the previous case shown in Fig.\ \ref{fig:OphiGaussCon}, although its late time dynamics and the associated thermalization time are very similar to the ones obtained in the previous subsections.

\subsection{Gaussian metric anisotropy profile and equilibrium dilaton profile}
\label{sec:results:GaussEq}

Finally, the last case which remains to be analyzed correspond to the initial condition with Gaussian initial metric anisotropy and an initial profile for the dilaton field equal to its equilibrium value
\begin{align}\label{eq:BinGaussEqui}
B_s(v_0,u) =0.5\, e^{-100(u-0.4)^2}  \ \ \text{and} \ \ \phi_s(v_0,u) =-\frac{1}{u^2}\sqrt{\frac{2}{3}} \ln \left( 1 + \tilde{u}(u)Q^2 \right).
\end{align}

\begin{figure}[h]
\centering
\begin{subfigure}{0.49\textwidth}
\includegraphics[width=\textwidth]{BGaussGauss5}
\caption{}
\end{subfigure}
\begin{subfigure}{0.49\textwidth}
\includegraphics[width=\textwidth]{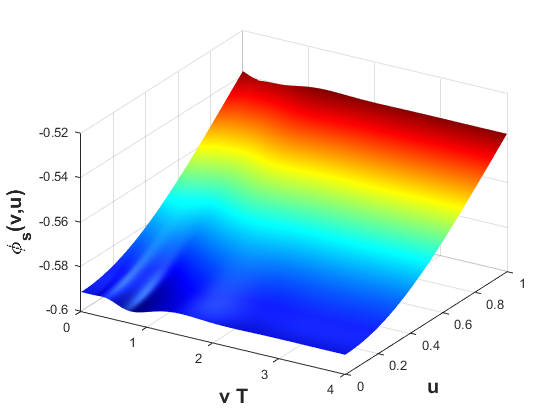}
\caption{}
\end{subfigure}
\caption{(Color online) Results for the time evolution of some fields involved in the 1RCBH setup for the initial condition \eqref{eq:BinGaussEqui} with $\mu/T=2$: (a) the subtracted metric anisotropy function $B_{s}(v,u)$, and (b) the subtracted dilaton field $\phi_{s}(v,u)$.}
\label{fig:3DGaussEqui}
\end{figure}

\begin{figure}[h]
\centering
\begin{subfigure}{0.49\textwidth}
\includegraphics[width=\textwidth]{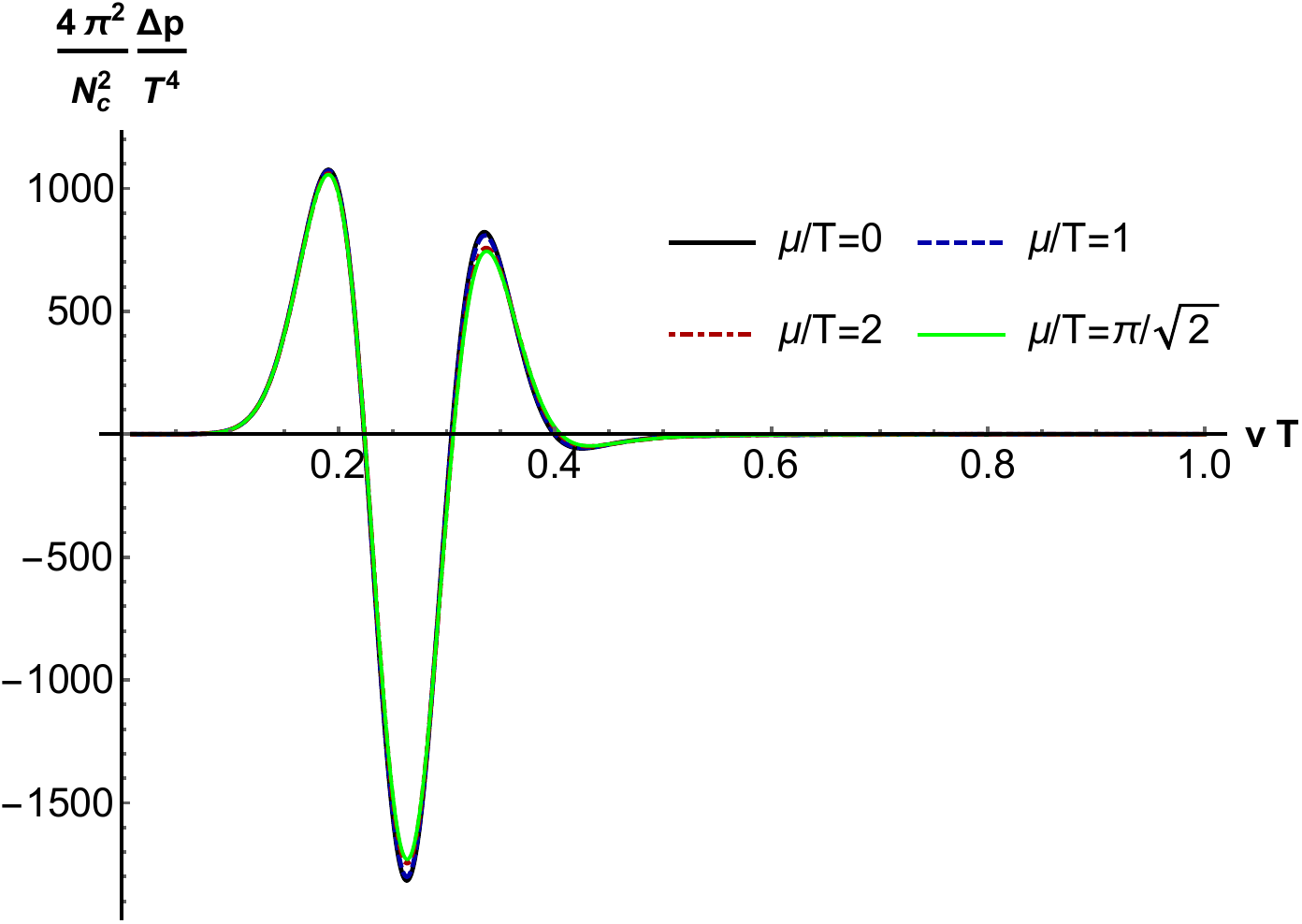}
\caption{}
\end{subfigure}
\begin{subfigure}{0.49\textwidth}
\includegraphics[width=\textwidth]{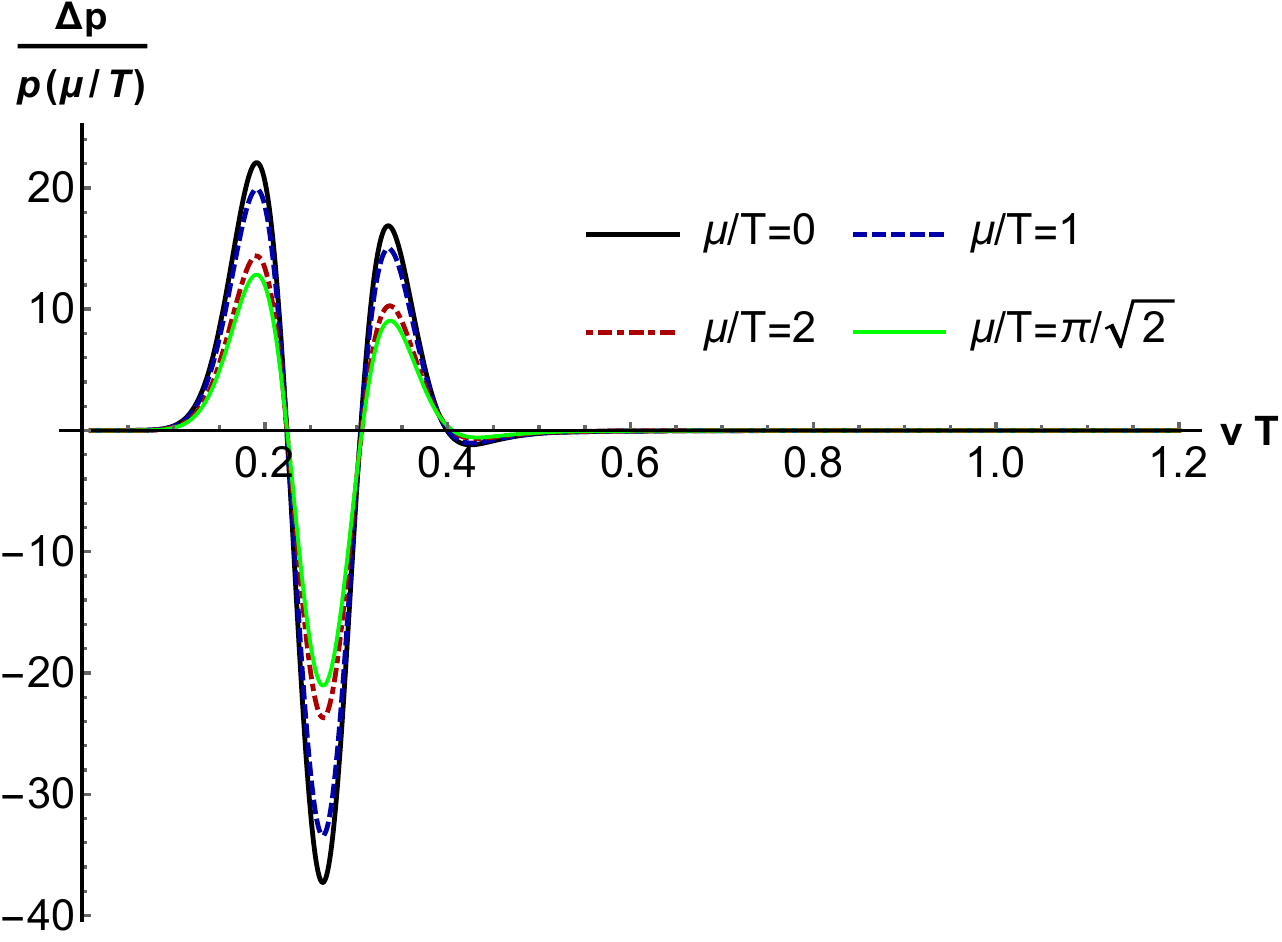}
\caption{}
\end{subfigure}
\caption{(Color online) Time evolution of the pressure anisotropy for several values of the chemical potential using the initial data \eqref{eq:BinGaussEqui}: (a) $\Delta p$ normalized by the (equilibrium) temperature to the fourth, and (b) $\Delta p$ normalized by the equilibrium pressure.}
\label{fig:PressureAni4}
\end{figure}

\begin{figure}[h]
\centering
\begin{subfigure}{0.48\textwidth}
\includegraphics[width=\textwidth]{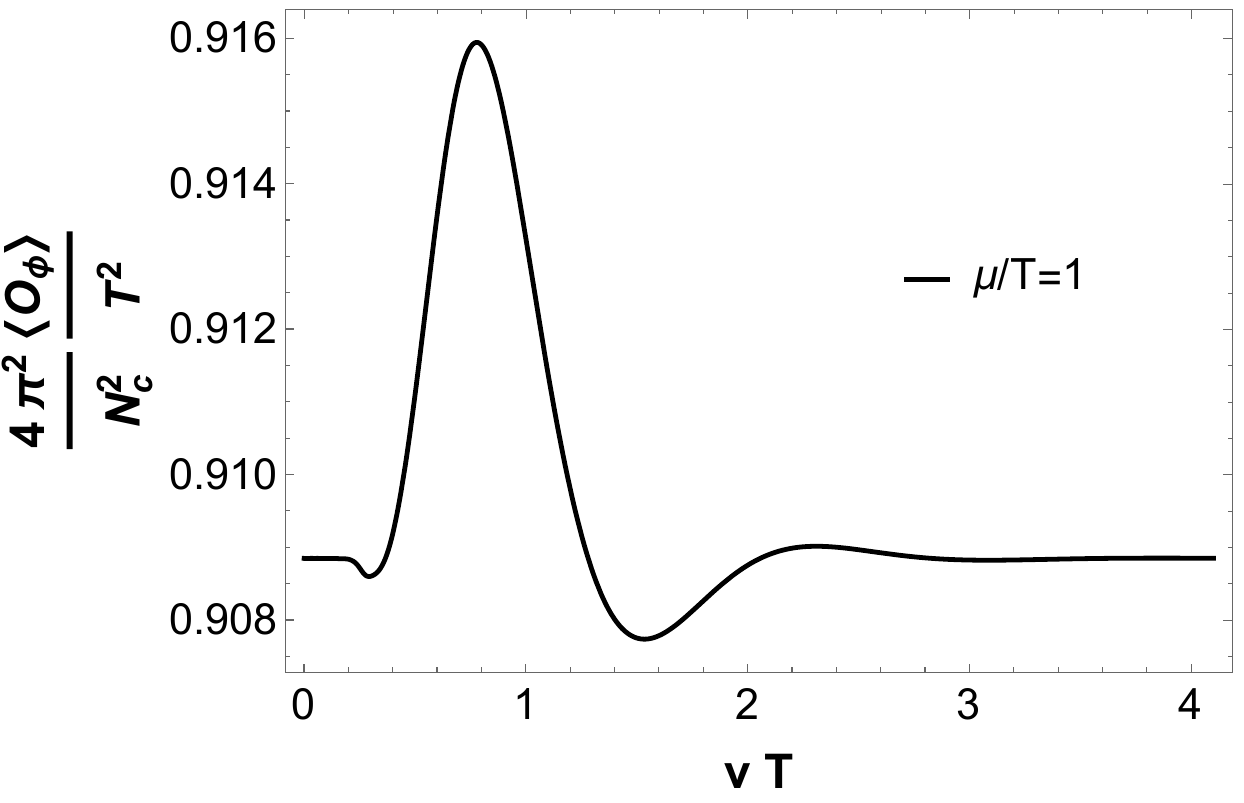}
\caption{}
\end{subfigure}
\begin{subfigure}{0.48\textwidth}
\includegraphics[width=\textwidth]{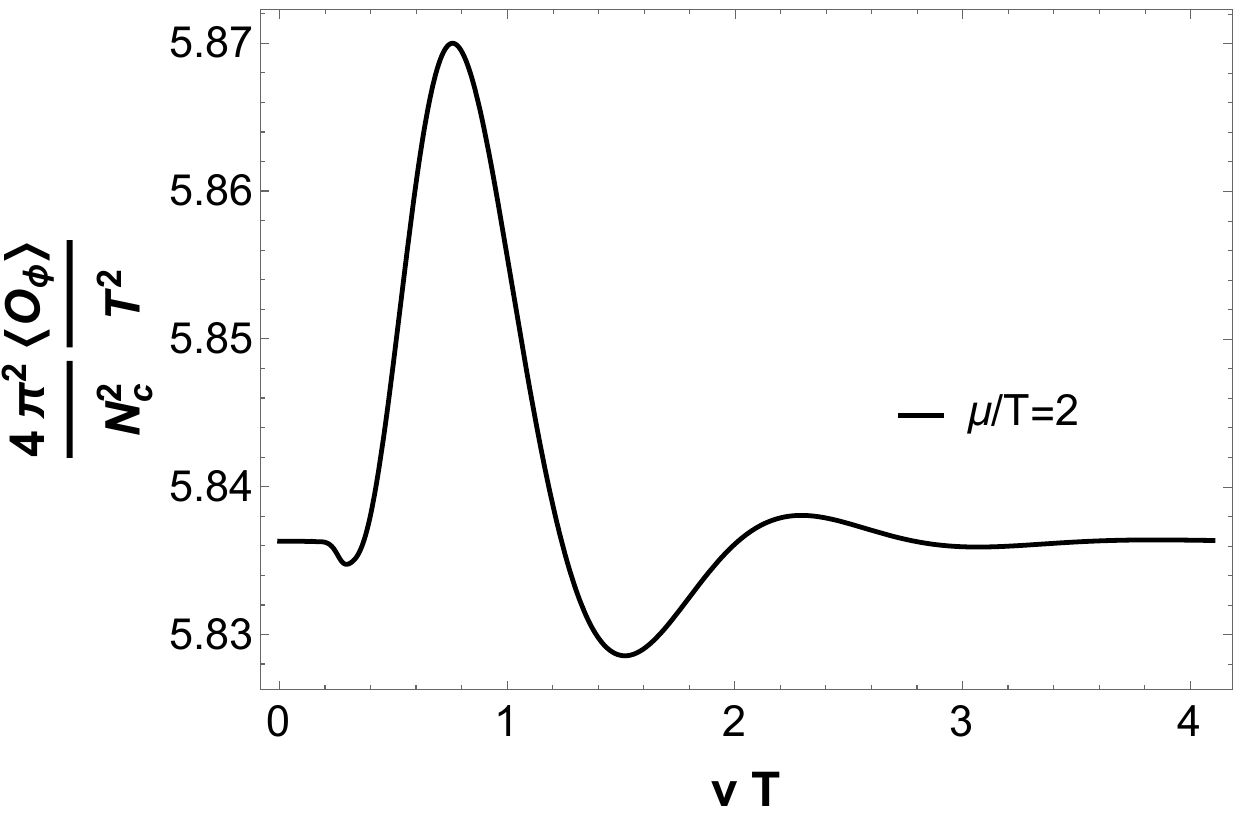}
\caption{}
\end{subfigure}
\begin{subfigure}{0.48\textwidth}
\includegraphics[width=\textwidth]{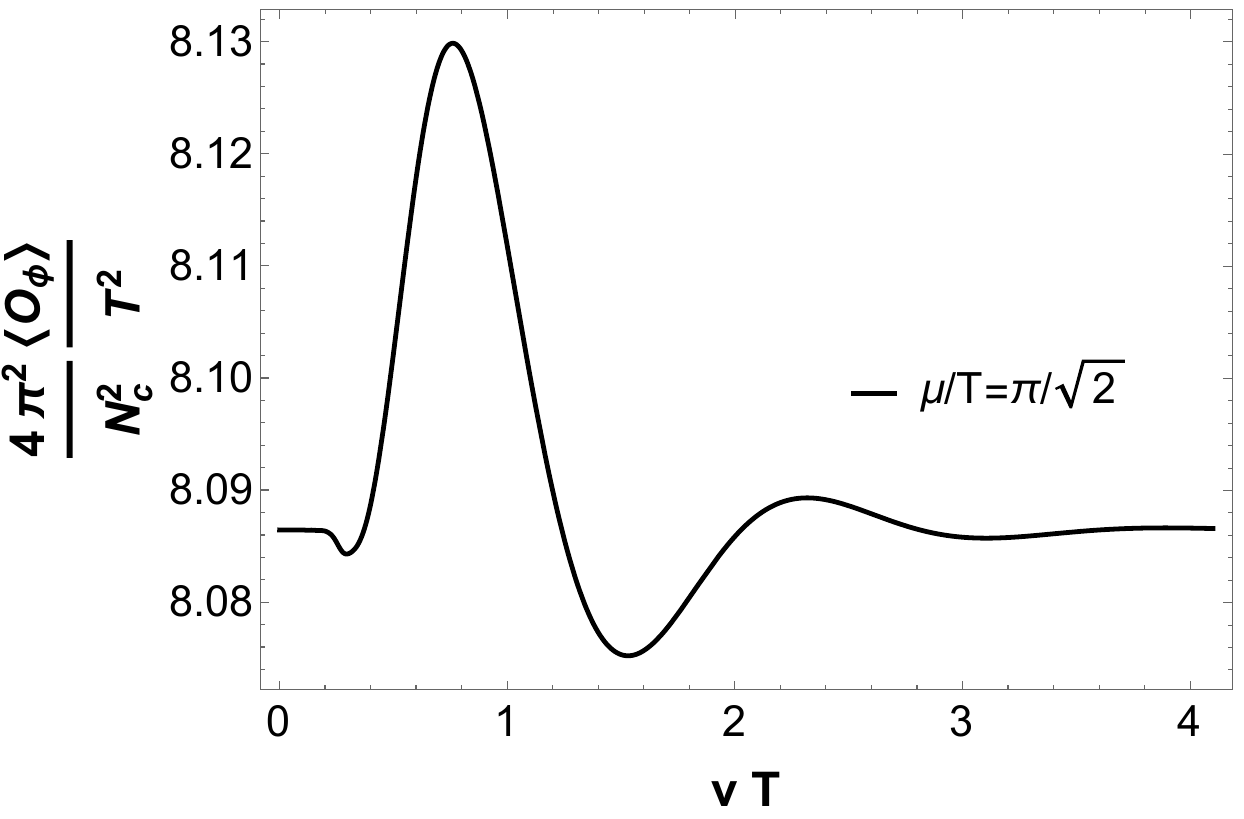}
\caption{}
\end{subfigure}
\caption{Time evolution of the scalar condensate $\langle\mathcal{O}_{\phi}\rangle$ for different $\mu/T$ using the initial data \eqref{eq:BinGaussEqui}.}
\label{fig:OphiGaussEqui}
\end{figure}

In Fig.\ \ref{fig:3DGaussEqui} we present the 3D plot with the time evolution of $B_s(v,u)$ and $\phi_s(v,u)$ using the initial condition \eqref{eq:BinGaussEqui} and $\mu/T=2$. As before, the metric anisotropy $B_s(v,u)$ is effectively unaffected by the change done in the initial condition for the dilaton field, while the dilaton profile $\phi_s(v,u)$ has only very small fluctuations in time, which is expected since we have already started the numerical simulations in the present case with an initial dilaton profile equal to its value in thermodynamic equilibrium. The pressure anisotropy evolves in time according to Fig.\ \ref{fig:PressureAni4}, which displays essentially the very same behavior found in the last two subsections. On the other hand, the early time dynamics of the scalar condensate depicted in Fig.\ \ref{fig:OphiGaussEqui} is different than what we have seen in the previous subsection, displaying only very small fluctuations in time. In any case, the thermalization time associated with the equilibration of the scalar condensate is found to be very similar to what has been observed in all the previous cases considered here.

Therefore, after we have exhausted the analysis of the selected sets of initial conditions, we may draw some general conclusions for the homogeneous equilibration dynamics of the 1RCBH plasma analyzed here:

\begin{enumerate}
\item The backreaction produced in the time evolution of the dilaton field by changing the initial metric anisotropy is extremely large, in contrast to what happens with the backreaction produced on the time evolution of the metric anisotropy by changing the initial dilaton profile, which is small. This implies that the pressure anisotropy is robust against the addition of other fields in the gravitational action besides the metric;

\item The isotropization (associated with the approximate vanishing of the pressure an\-isot\-ropy $\Delta p$) always happen (well) before the true thermalization of the system (associated with the equilibration of the scalar condensate $\langle\mathcal{O}_\phi \rangle$);

\item Regarding the late time dynamics of the scalar condensate and the associated thermalization time, these are robust features of the medium which remain almost unchanged regardless of the initial conditions considered. In particular, the thermalization time always increases with increasing chemical potential. On the other hand, the early time dynamics of the scalar condensate is strongly dependent on the set of initial data chosen to seed the time evolution of the far-from-equilibrium 1RCBH plasma;

\item The time evolution of the pressure anisotropy and the associated isotropization time of the system strongly depend on the chosen initial condition for the metric anisotropy.
\end{enumerate}

However, in the different cases considered above, due to the large scales plotted to cover the large amplitudes of oscillation observed in the time evolution of the pressure anisotropy, it was not possible yet to clearly reveal the crucial role played by the critical point in the isotropization of the system. We analyze this issue in detail in the next subsection, where we confront the late time dynamics of the pressure anisotropy with the lowest non-hydrodynamic QNM of the external scalar channel of the 1RCBH plasma obtained in Ref. \cite{Finazzo:2016psx}. We also compare the late time behavior of the scalar condensate with the lowest non-hydrodynamic QNM of the dilaton channel, to be derived in Appendix \ref{sec:QNMs}.

\subsection{Matching the quasinormal modes}
\label{sec:MatQNM}

In this section we compare the late time behavior of the full nonlinear evolution of the equations of motion with the quasinormal modes of the system, which describe exponentially damped collective excitations produced in response to disturbances of a black hole background (see e.g. \cite{Berti:2009kk,Konoplya:2011qq} for some recent reviews). The QNM spectra of the theory are responsible for the so-called ``quasinormal ringdown'' phenomenon describing the linear part of the decaying perturbations of a disturbed black hole at late times.

The purpose of this comparison is threefold: it can give a measure of the degree of nonlinearity of the full out-of-equilibrium solutions of the equations of motion\footnote{Note that in the analysis of QNMs, by assuming small perturbations, one just considers quadratic fluctuations of the bulk fields in the disturbed action, which gives linearized equations of motion for these perturbations.}; it may be also used to provide an independent check of the accuracy of the numerical solutions at late times; finally, it can unveil some of the main differences between the early and the late time equilibration dynamics of the system.

Part of the spectra of QNMs of the 1RCBH plasma were recently obtained in Ref. \cite{Finazzo:2016psx}. By following the classification given in Ref. \cite{DeWolfe:2011ts} for the gauge and diffeomorphism invariant perturbations of the EMD fields in the homogeneous zero wavenumber limit, in which case the system has a rotational $SO(3)$ symmetry, one has three different channels to analyze corresponding to the three lowest dimensional representations of $SO(3)$: the tensor/quintuplet channel, the vector/triplet channel, and the scalar/singlet channel. By working with the non-normalizable modes\footnote{These are the solutions of the linearized equations of motion for the perturbations with the conditions that these perturbations are regular at the horizon and normalized to unity at the boundary.} of each channel, one obtains the associated transport coefficients through the use of Kubo formulas, namely, the shear viscosity from the $SO(3)$ quintuplet channel, the charge conductivity and diffusion from the $SO(3)$ triplet channel, and the bulk viscosity from the $SO(3)$ singlet channel \cite{DeWolfe:2011ts}. On the other hand, by working with the normalizable modes\footnote{These are the solutions of the linearized equations of motions which are regular at the horizon and subjected to the Dirichlet condition that these perturbations vanish at the boundary.}, one obtains the QNMs of each channel. In Ref. \cite{Finazzo:2016psx} we analyzed in details the QNMs of the quintuplet and triplet channels and in Appendix \ref{sec:QNMs} we shall obtain the QNMs of the singlet channel.\footnote{The $SO(3)$ quintuplet channel coincides with the perturbation for a massless external scalar field \cite{DeWolfe:2011ts,Rougemont:2015wca,Finazzo:2016psx}, while the $SO(3)$ singlet channel may be identified more generally with a so-called ``dilaton channel'', as we are going to explain in Appendix \ref{sec:QNMs}.} The lowest non-hydrodynamic QNMs of the $SO(3)$ quintuplet and singlet channels will be compared in what follows with the late time dynamics of the pressure anisotropy and the scalar condensate, respectively.

\begin{figure}[h]
\centering
\includegraphics[width=1.0\textwidth]{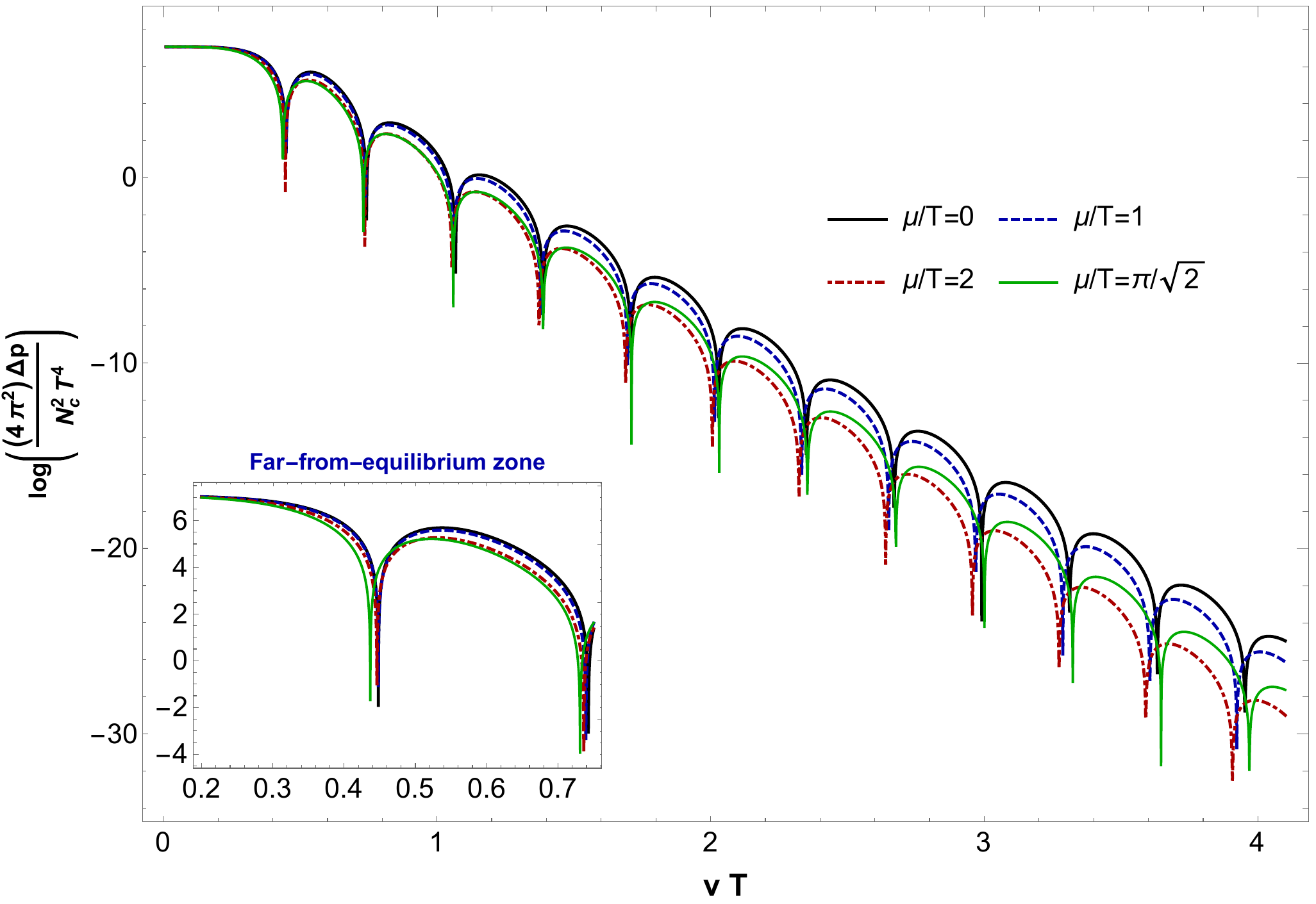}
\caption{(Color online) Time evolution of the pressure anisotropy for different $\mu/T$.}
\label{fig:QNMfig}
\end{figure}

\begin{figure}[h]
\centering
\begin{subfigure}{0.49\textwidth}
\includegraphics[width=\textwidth]{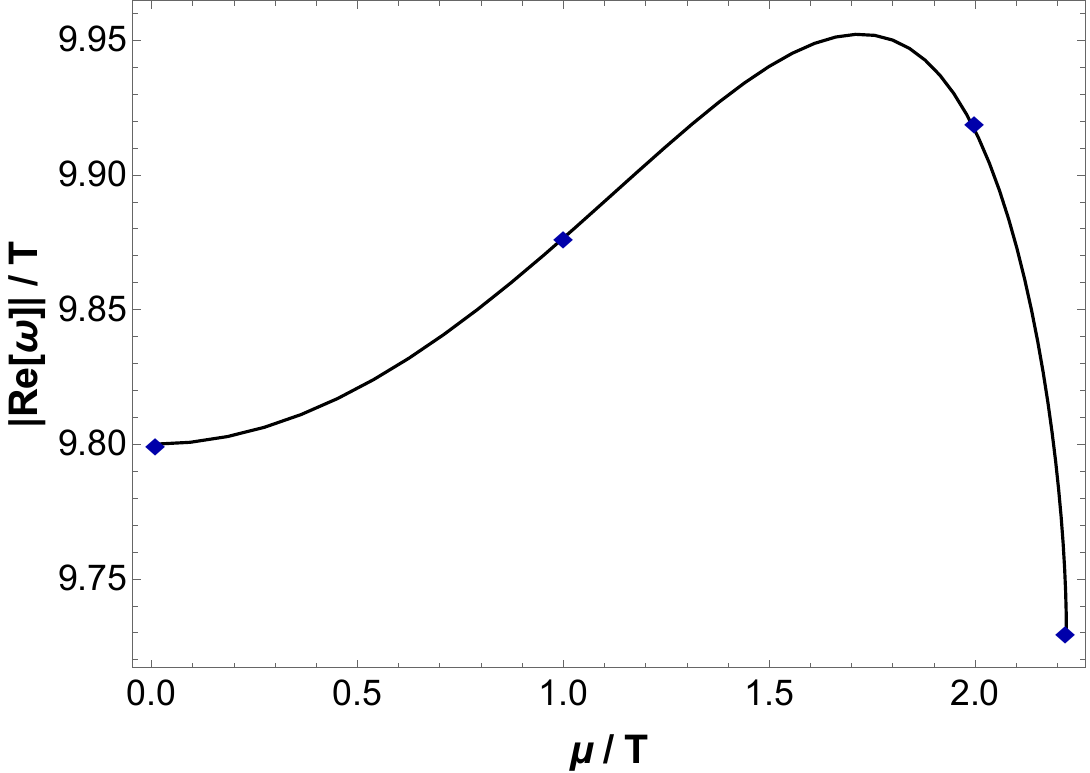}
\caption{}
\end{subfigure}
\begin{subfigure}{0.49\textwidth}
\includegraphics[width=\textwidth]{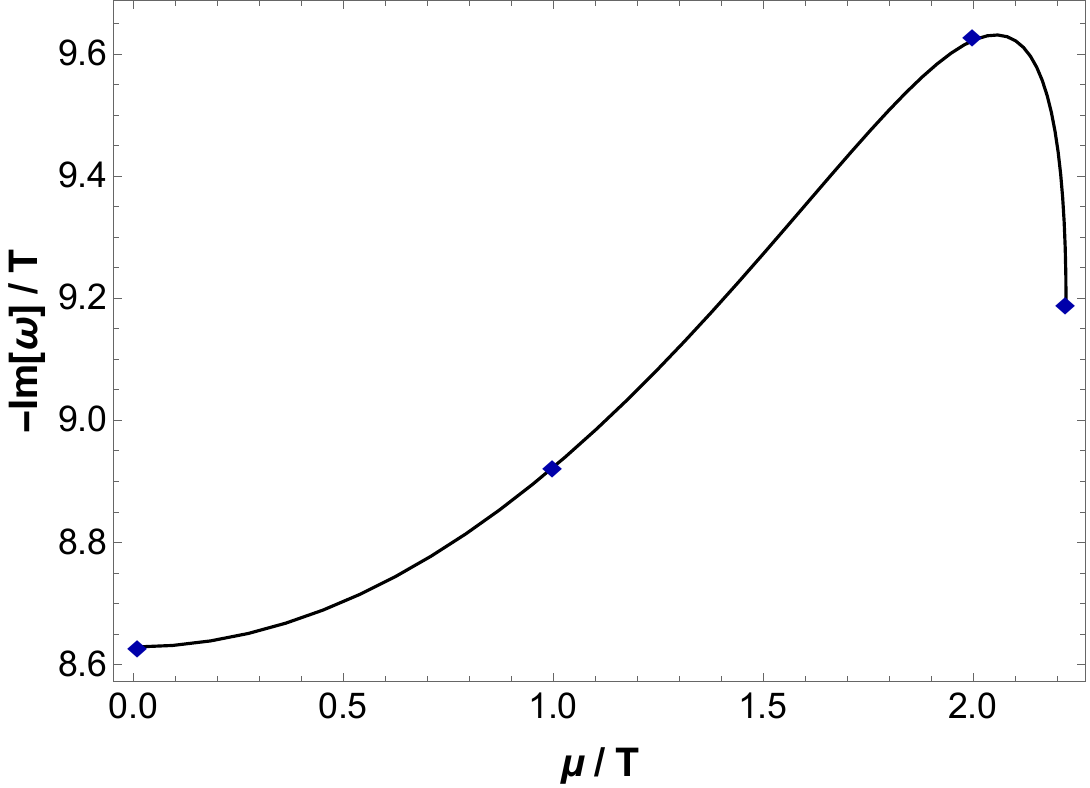}
\caption{}
\end{subfigure}
\caption{(Color online) (a) Real and (b) imaginary parts of the lowest non-hydrodynamic QNM of the $SO(3)$ quintuplet (external scalar) channel (curves) \cite{Finazzo:2016psx} compared with the late time decay of the pressure anisotropy (diamonds) described according to Eq.\ \eqref{eq:LateTime}.}
\label{fig:QNMcomp}
\end{figure}

In Fig. \ref{fig:QNMfig} we display our numerical results for the time evolution of the pressure anisotropy using the initial conditions set in Eq.\ \eqref{eq:BinConCon1} for different values of $\mu/T$.\footnote{We remark that although the early time behavior of this plot changes significantly depending on the initial conditions considered, the qualitative behavior observed at late times is the same for the different initial conditions considered here.} This time we plot the logarithmic of $\Delta p$ which makes it possible to clearly resolve the late time exponential damp of the black hole oscillations. The inset displayed in Fig. \ref{fig:QNMfig} gives a zoom of the behavior of the pressure anisotropy in the far-from-equilibrium zone. To check whether the late time dynamics of the pressure anisotropy can be described by the lowest non-hydrodynamic QNM of the $SO(3)$ quintuple (external scalar) channel, we parametrize $\Delta p$ using the following functional form
\begin{align}
\frac{4\pi^2}{N_c^2}\frac{\Delta p}{T^4} = \mathsf{A}\, e^{\frac{\textrm{Im}[\omega]}{T}\,vT}\cos\left(\frac{\textrm{Re}[\omega]}{T}\,vT + \varphi\right),
\label{eq:LateTime}
\end{align}
where $\textrm{Re}[\omega]/T$, $\textrm{Im}[\omega]/T$, $\mathsf{A}$, and $\varphi$ are fixed by fitting the functional form \eqref{eq:LateTime} to the numerical result for the pressure anisotropy evaluated within the late time interval $vT\in[1.8,4.1]$.\footnote{For the specific value of $\mu/T=\pi/\sqrt{2}$ (critical point), the fit interval used was $vT\in[5.0,7.0]$, since the late time decay of the pressure anisotropy only converges to the critical behavior of the lowest non-hydrodynamic QNM of the external scalar channel at later times when compared to other values of $\mu/T$.} The relevant parameters extracted from this fitting procedure are $\textrm{Re}[\omega]/T$ and $\textrm{Im}[\omega]/T$, which may be then compared with the real and imaginary parts of the non-hydrodynamic QNM with lowest imaginary part (in magnitude), corresponding to the dominant, less damped mode in the external scalar channel.

Such comparison is done in Fig. \ref{fig:QNMcomp}, from which one notes a good agreement between the late time decay of the pressure anisotropy and the lowest non-hydrodynamic QNM of the external scalar channel obtained in Ref. \cite{Finazzo:2016psx}. Such agreement provides an independent check of the accuracy of our numerical routine at late times, as aforementioned. More importantly, we are now able to understand how the isotropization time depends on the presence of a critical point in the phase diagram of the 1RCBH plasma.

We note from Fig.\ \ref{fig:QNMfig} that the early time dynamics of the pressure an\-isot\-ropy is qualitatively different from its late time dynamics. At early times, the pressure anisotropy is always damped as one increases $\mu/T$, however, at late times the pressure anisotropy may decrease or increase with increasing chemical potential depending on whether the system is far or close enough to the critical point. Therefore, the isotropization time of the 1RCBH plasma is affected by the critical point in accordance with the behavior of the ``relaxation time'' extracted from the imaginary part of the lowest non-hydrodynamic QNM of the external scalar channel of the theory \cite{Finazzo:2016psx}. This is explained by the fact that, as shown in Fig. \ref{fig:QNMcomp}, the late time decay of the pressure anisotropy is in fact described by the lowest non-hydrodynamic QNM of the $SO(3)$ quintuplet channel.

\begin{figure}[h]
\centering
\includegraphics[width=1.0\textwidth]{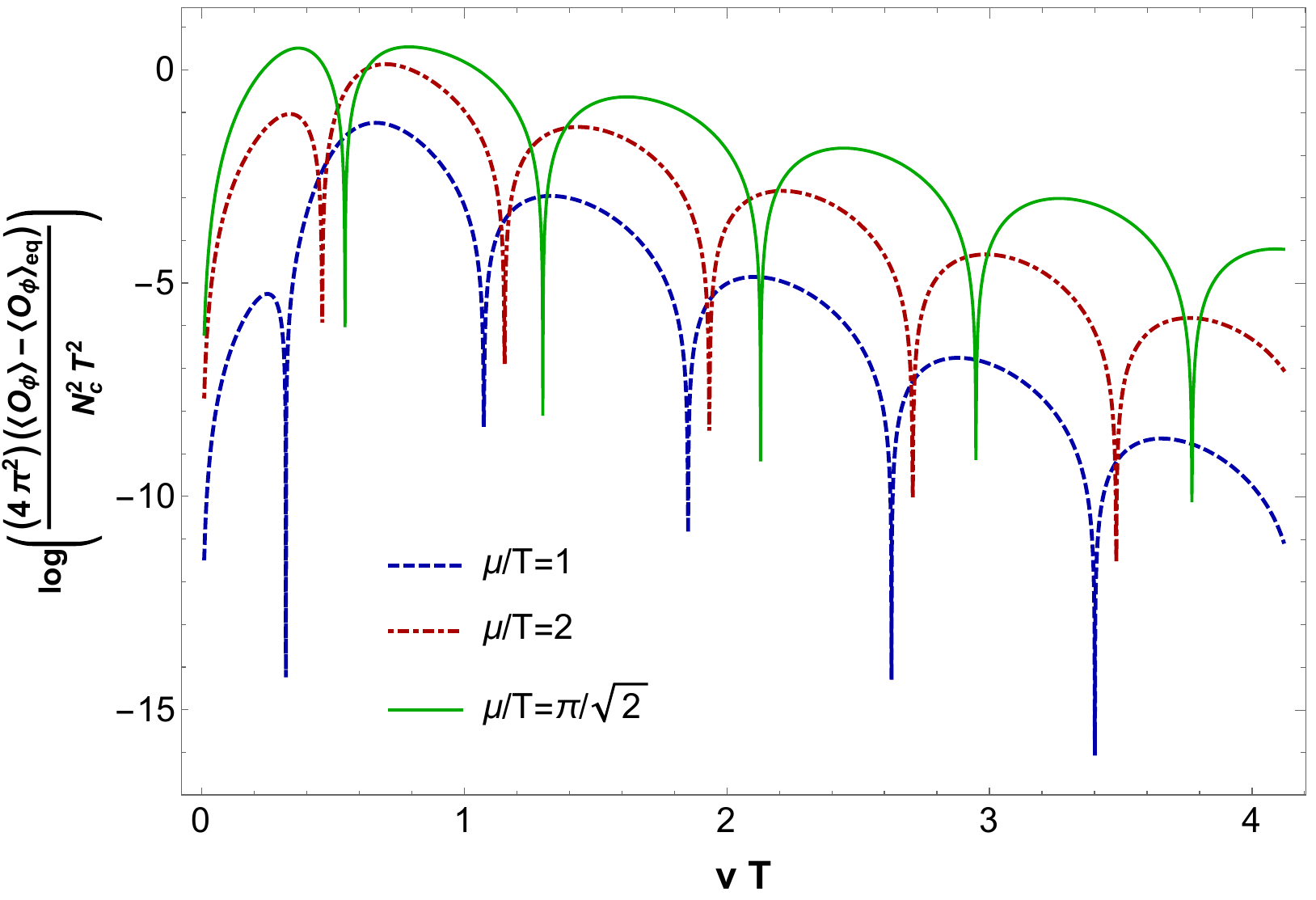}
\caption{(Color online) Time evolution of the difference between the scalar condensate and its equilibrium value for different $\mu/T$.}
\label{fig:QNMfig2}
\end{figure}

This is qualitatively different from the thermalization time discussed before, which is associated with the equilibration of the scalar condensate. This process only sets in after a nearly isotropic state has been already achieved by the system and it always increase with increasing $\mu/T$. In Fig. \ref{fig:QNMfig2} we show our numerical results for the time evolution of the difference between the scalar condensate and its equilibrium value using the initial conditions set in Eq.\ \eqref{eq:BinConCon1} for different values of $\mu/T$. To check whether the late time dynamics of this observable can be described by the lowest non-hydrodynamic QNM of the $SO(3)$ singlet (dilaton) channel, we parametrize it using once more the following functional form
\begin{align}
\frac{4\pi^2}{N_c^2}\frac{\left(\langle\mathcal{O}_\phi\rangle-\langle\mathcal{O}_\phi\rangle_{\textrm{eq}}\right)}{T^2} = \mathsf{A}\, e^{\frac{\textrm{Im}[\omega]}{T}\,vT}\cos\left(\frac{\textrm{Re}[\omega]}{T}\,vT + \varphi\right),
\label{eq:LateTime2}
\end{align}
where $\textrm{Re}[\omega]/T$, $\textrm{Im}[\omega]/T$, $\mathsf{A}$, and $\varphi$ are fixed by fitting the functional form \eqref{eq:LateTime2} to the numerical result for the difference between the scalar condensate and its equilibrium value evaluated within the late time interval $vT\in[1.8,4.1]$.\footnote{At the critical point, however, we performed the fit starting from $vT = 2.2$.}

\begin{figure}[h]
\centering
\begin{subfigure}{0.49\textwidth}
\includegraphics[width=\textwidth]{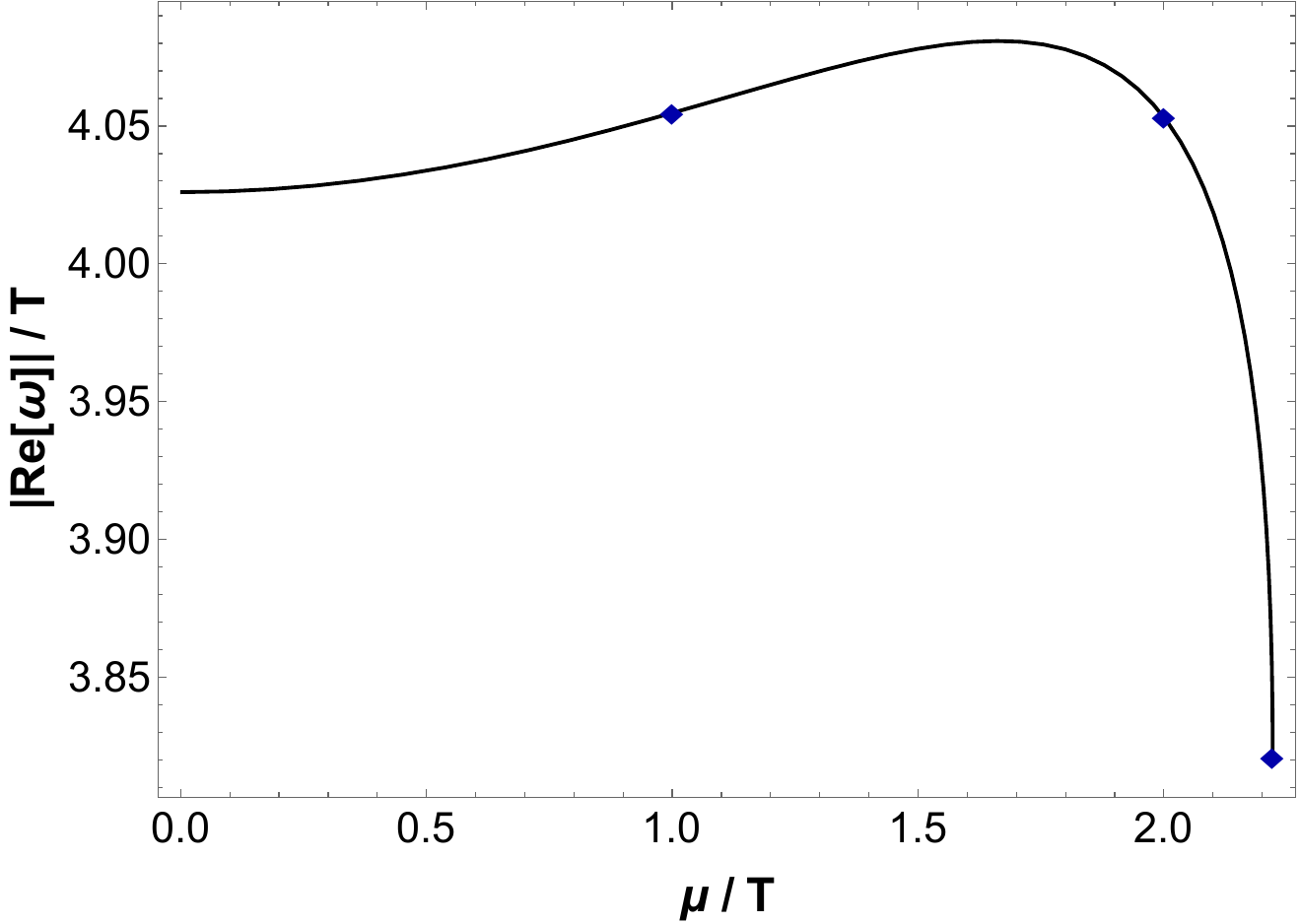}
\caption{}
\end{subfigure}
\begin{subfigure}{0.49\textwidth}
\includegraphics[width=\textwidth]{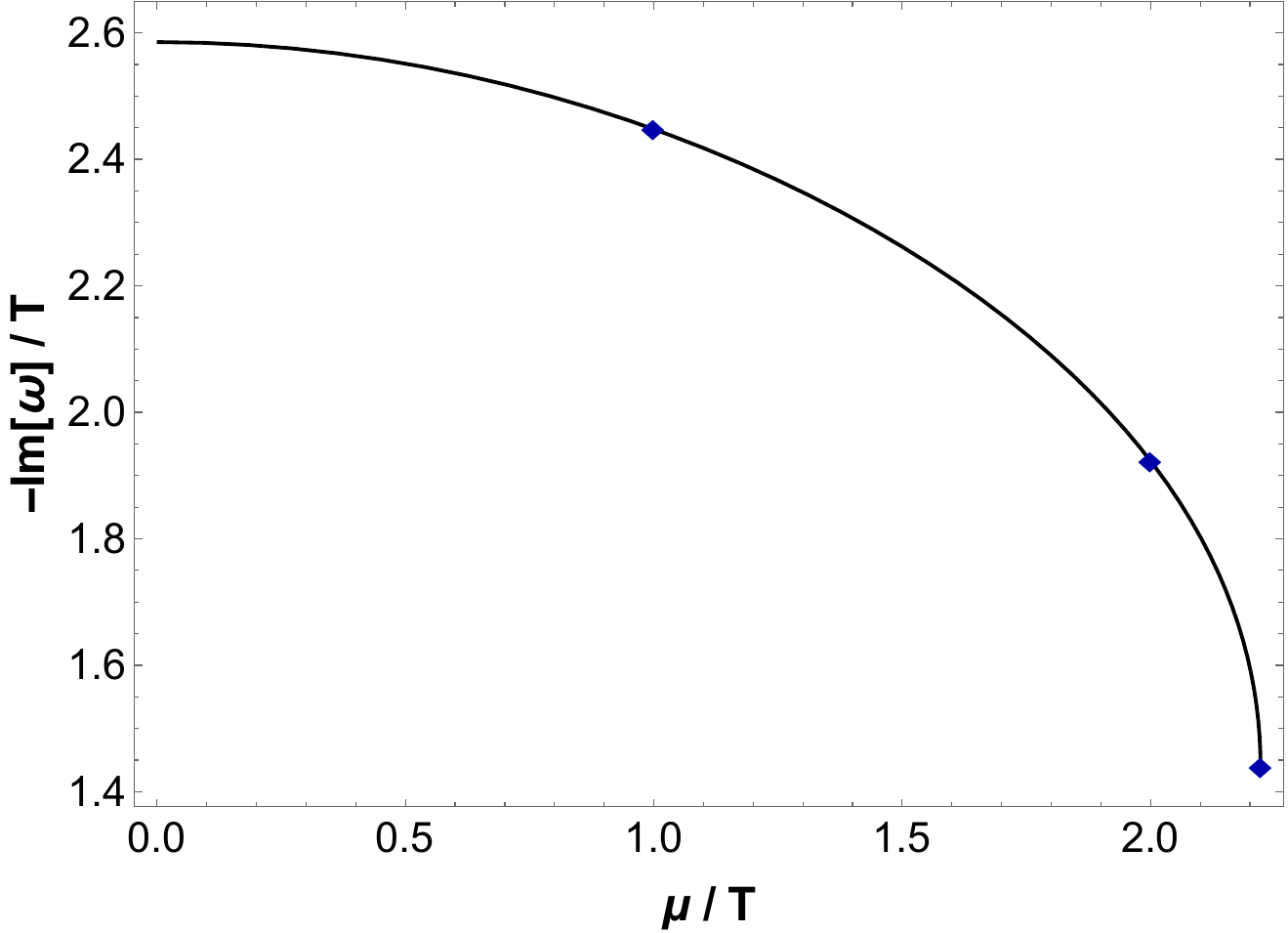}
\caption{}
\end{subfigure}
\caption{(Color online) (a) Real and (b) imaginary parts of the lowest non-hydrodynamic QNM of the $SO(3)$ singlet (dilaton) channel (curves) --- see their derivation in Appendix \ref{sec:QNMs} --- compared with the late time decay of the difference between the scalar condensate and its equilibrium value (diamonds) described according to Eq.\ \eqref{eq:LateTime2}.}
\label{fig:QNMcomp2}
\end{figure}

The comparison between $\textrm{Re}[\omega]/T$ and $\textrm{Im}[\omega]/T$ extracted from these fits to the late time behavior of the full numerical solutions and the lowest non-hydrodynamic QNM of the dilaton channel is displayed in Fig. \ref{fig:QNMcomp2}. One notes a good agreement between both results, which shows that the late dynamics of the scalar condensate is indeed dominated by the lowest non-hydrodynamic QNM of the dilaton channel. This also gives another independent check of the accuracy of our numerical routine at late times.

In summary, we see that the late time dynamics of the pressure anisotropy and the scalar condensate are dominated by the lowest non-hydrodynamic QNMs of the external scalar and dilaton channels, respectively. Consequently, the behavior of the isotropization and thermalization times of the 1RCBH plasma can be correctly inferred from the analysis of the QNMs of the system, even though the early time dynamics of these observables cannot be described by these linearized perturbations.

%%%%%%%%%%%%%%%%%%%%%%%%%
\section{Outlook and final remarks}
\label{sec:outlook}

Now we summarize the main findings of the present work concerning the spatially homogeneous equilibration dynamics of the 1RCBH plasma. The relevant one-point functions of the 1RCBH model correspond to the expectation value of the stress-energy tensor $\langle T^{\mu\nu}\rangle$ dual to the bulk metric $g_{\mu\nu}$, the scalar condensate $\langle\mathcal{O}_\phi\rangle$ dual to the bulk dilaton field $\phi$, and the $U(1)$ R-charge density $\rho_c=\langle J^t\rangle$ associated with the bulk Maxwell field $A_\mu$. The charge density is time-independent in this spatially homogeneous setup and it only depends on the value of the chemical potential of the medium. From $\langle T^{\mu\nu}\rangle$ one extracts the pressure anisotropy $\Delta p$ which describes the isotropization of the system, while the approach of the scalar condensate toward its thermodynamic equilibrium value is associated with the true thermalization of the 1RCBH plasma. This is so because the scalar condensate was found here to always equilibrate only after the system has already reached a nearly isotropic state, being always the last equilibration time scale of the system. In other words, generally the isotropization time is shorter than the thermalization time of the 1RCBH plasma. 

Moreover, while the thermalization time was found to always increase with increasing values of the $U(1)$ R-charge chemical potential, in agreement with the behavior of the lowest non-hydrodynamic QNM of the dilaton channel, the isotropization time shows a non-monotonic dependence on the chemical potential, namely, it decreases or increases with increasing chemical potential depending on how far or close to the critical point the system is, respectively. This behavior of the isotropization time is in consonance with the behavior of the lowest non-hydrodynamic QNM of the external scalar channel of the 1RCBH plasma \cite{Finazzo:2016psx}. These two observations are explained by the fact that the late time dynamics of the pressure anisotropy and the scalar condensate are dominated by the lowest non-hydrodynamic QNMs of the external scalar and dilaton channels, respectively. These qualitative findings are robust for the 1RCBH model, in the sense that they hold for different initial conditions chosen to seed the time evolution of the far-from-equilibrium plasma.

On the other hand, the early time dynamics of the pressure anisotropy and the scalar condensate are strongly dependent on the chosen initial data. More specifically, the backreaction produced in the time evolution of the dilaton field by
changing the initial metric anisotropy was found to be very large, in contrast to what happens with the backreaction produced on the time evolution of the metric anisotropy by changing the initial dilaton profile, which was generally found to be small. This suggests that the pressure anisotropy is robust against the addition of other fields in the gravitational action besides the metric, being only significantly sensitive to the initial profile chosen for the metric anisotropy. In particular, the isotropization time may be significantly modified by changing the initial metric anisotropy. On the other hand, while the early time dynamics of the scalar condensate is strongly dependent on the set of initial data chosen to seed the time evolution of the far-from-equilibrium 1RCBH plasma, its late time dynamics and the associated thermalization time of the medium were found to be remarkably similar for all the different initial conditions considered here.

Some follow-ups of the present work which will appear soon regard the generalization of the analysis of the spatially homogeneous equilibration dynamics of the 1RCBH plasma performed here to consider spatially dependent hydrodynamic flows, such as the Bjorken \cite{Bjorken:1982qr} and the Gubser \cite{Gubser:2010ze,Gubser:2010ui} flows. The latter is particulary interesting since the solution for the hydrodynamic Navier-Stokes approximation in this case displays general inconsistencies, such as regions in the fluid with negative temperature \cite{Gubser:2010ze}, which do not appear after resummation \cite{Marrochio:2013wla,Bemfica:2017wps}. We also intend to investigate the collisions of shock waves in the 1RCBH setup in the near future to study the interplay between critical phenomena and the rapid expansion developed by the system under these conditions.

As discussed at length in the introduction, the 1RCBH plasma (as any conformal plasma) is not suited for direct applications in heavy ion phenomenology. Nonetheless, some of the qualitative features disclosed in the present analysis may be general enough and hold for other strongly coupled holographic fluids. Indeed, the fact that the thermalization time of the 1RCBH plasma increases with increasing chemical potential is in consonance with the suggested delayed equilibration of heavy ion collisions at lower energies or higher densities discussed in the introduction. In order to look for possible signatures of far-from-equilibrium universal dynamics of holographic systems, we also intend to generalize the analysis of the present work to the phenomenologically realistic EMD construction of Ref.\ \cite{Critelli:2017oub}, which provides a quantitative description of QCD thermodynamics at zero and finite baryon chemical potential.

Another perspective for future investigations consists in going beyond the calculation of one-point functions and study also higher order correlation functions \cite{Banks:1998dd,Skenderis:2008dh,Skenderis:2008dg,Keranen:2014lna,Banerjee:2016ray} in far-from-equilibrium holographic settings. In particular, this kind of study may be of relevance for many different ongoing and future low energy heavy ion experiments, such as the BES program \cite{Aggarwal:2010cw} being conducted at RHIC, the future fixed target (FXT) experiments \cite{Meehan:2016qon,Meehan:2017cum} also at RHIC, the ongoing HADES experiment \cite{Agakishiev:2015bwu} at GSI, the future Compressed Baryonic Matter (CBM) experiment at FAIR/GSI \cite{Staszel:2010zza,Ablyazimov:2017guv}, and also experiments in the future NICA facility \cite{NICA}.\footnote{The center of mass collision energies $\sqrt{s}$ (planned to be) reached in these experiments are the following: 7.7 --- 200 GeV (BES), 3.0 --- 7.7 GeV (FXT), 1.0 --- 3.5 GeV (HADES), 2.7 --- 4.9 GeV (CBM), and 4.0 --- 11 GeV (NICA).} This is so because, as recently discussed in Refs.\ \cite{Mukherjee:2015swa,Mukherjee:2016kyu}, the behavior of out-of-equilibrium critical cumulants in the QCD phase diagram may be very different from the equilibrium behavior of these real-time correlation functions \cite{Stephanov:1998dy,Stephanov:1999zu,Rischke:2003mt,Stephanov:2011pb}. Since ratios between some of these cumulants may be compared with ratios between moments of particle multiplicity distributions measured in heavy ion collisions, they are the main smoking gun used in experiments in order to try to identify clear experimental signatures of the QCD critical point in heavy ion collisions. Therefore, an understanding of the behavior of higher order correlation functions in out-of-equilibrium strongly coupled systems may disclose some fundamental insights which may potentially drive the experimental searches for the QCD critical point in the near future.

%%%%%%%%%%%%%%%%%%%%%%%%%
\begin{acknowledgments}
R.C. was supported by Funda\c c\~ao de Amparo \`a Pesquisa do Estado de S\~ao Paulo (FAPESP) grant number 2016/09263-2. R.R. acknowledges financial support by Funda\c c\~ao Norte Riograndense de Pesquisa e Cultura (FUNPEC). J.N. thanks Rutgers University for the hospitality and FAPESP and Conselho Nacional de Desenvolvimento Cient\'ifico e Tecnol\'ogico (CNPq) for financial support.
\end{acknowledgments}

%%%%%%%%%%%%%%%%%%%%%%%%%
\appendix

\section{Holographic renormalization}
\label{sec:HoloRen}

In this Appendix we give the details on how one may obtain the one-point functions $\langle T_{\mu\nu} \rangle$, $\langle J_{\nu} \rangle$, and $\langle \mathcal{O}_{\phi} \rangle$ using the holographic renormalization procedure \cite{Skenderis:2002wp,deHaro:2000vlm,Bianchi:2001kw}.

In what follows, and as it is common in the treatment of holographic renormalization, we adopt the Fefferman-Graham (FG) coordinates in which there is an explicit relation between the renormalization group (RG) flow at the boundary QFT and the bulk radial coordinate
\begin{equation}\label{eq:FGline}
ds_{FG}^{2} = \frac{d\rho^2}{4\rho^2}+\gamma_{\mu\nu}(\rho,x)dx^\mu dx^\nu,
\end{equation}
with the Greek indices running through the coordinates of the dual QFT, $x\in\lbrace t,\vec{x} \rbrace$, and $\rho$ denoting the radial coordinate in the FG chart, where the boundary lies at $\rho=0$.

The first step to renormalize the on-shell action is to identify, in a covariant manner, what are the divergent terms. This analysis is done by integrating out the $\rho$-direction in the on-shell action up to a near-boundary hypersuface $\rho=\epsilon$ that acts as a cutoff, defining then a regulated action, $S_{reg}=(S_{bulk}+S_{GHY})\vert_{\epsilon}$. Once the divergences of the regulated action are identified, the counterterm action $S_{ct}$ is defined as follows \cite{Skenderis:2002wp,deHaro:2000vlm,Bianchi:2001kw}
\begin{equation}
S_{ct} = - (\text{divergent terms of} \ S_{reg}).
\end{equation}
The subtracted action, $S_{sub}$, which is supposed to be evaluated at the cutoff $\rho=\epsilon$, is given by 
\begin{equation} \label{eq:Ssub}
S_{sub}= S_{reg}+S_{ct} = S_{bulk}+S_{GHY}+S_{ct}.
\end{equation}
The renormalized on-shell action is obtained by taking the limit $\rho\rightarrow 0$ on the subtracted action, i.e.
\begin{equation} \label{eq:Sren}
S_{ren} = \lim_{\rho\rightarrow 0} S_{sub}.
\end{equation}

Once the renormalized on-shell action \eqref{eq:Sren} is found, we follow the holographic dictionary and take functional derivatives of the renormalized action with respect to the boundary values of the bulk fields to obtain the corresponding one-point functions in the dual QFT. In particular, for an EMD model, the important one-point functions are\footnote{Note that for a bulk scalar field with dimension $\Delta=2$, as it is the case of the dilaton in the 1RCBH model, we need to introduce an extra $\ln\rho$ term to regulate the expectation value of its dual scalar operator in the boundary QFT.}
\begin{align}
\langle T_{\mu\nu} \rangle &  = - \frac{2}{\sqrt{-g_{(0)}}}\frac{\delta S_{ren}}{\delta g_{(0)}^{\mu\nu}} = -\lim_{\rho\rightarrow 0}\frac{1}{\rho}\frac{2}{\sqrt{-\gamma}}\frac{\delta S_{sub}}{\delta \gamma^{\mu\nu}}, \label{eq:1pt_T} \\
\langle J^{\mu} \rangle & = \frac{1}{\sqrt{-g_{(0)}}}\frac{\delta S_{ren}}{\delta A_{(0)\mu}}= \lim_{\rho\rightarrow 0}\frac{1}{\rho^2}\frac{1}{\sqrt{-\gamma}}\frac{\delta S_{sub}}{\delta A_{\mu}}, \label{eq:1pt_J} \\
\langle \mathcal{O}_{\phi} \rangle & = \frac{1}{\sqrt{-g_{(0)}}}\frac{\delta S_{ren}}{\delta \phi_{(0)}} = \lim_{\rho\rightarrow 0}\frac{\ln\rho}{\rho}\frac{1}{\sqrt{-\gamma}}\frac{\delta S_{sub}}{\delta \phi}, \label{eq:1pt_phi}
\end{align}
where $g_{\mu\nu}=\rho\,\gamma_{\mu\nu}$ is the metric of the boundary QFT, which we shall take to be Minkowski at the end of the calculations. The subscript $(0)$ denotes that these fields are computed at the boundary of the asymptotically AdS space; we will give the precise meaning of it below when we expand the fields near the boundary.

\subsection{Counterterm action}
\label{sec:HoloRen:CT}

The counterterm action for the 1RCBH model, which is an EMD model whose bulk scalar field has dimension $\Delta=2$, is the same counterterm action for the Coulomb branch flow \cite{Bianchi:2001kw}
\begin{align}\label{eq:Sct}
S_{ct} = \frac{1}{\kappa_5^2}\int_{\partial M} d^4x\sqrt{-\gamma}  &\left[ -3-\frac{1}{4}R[\gamma]+\frac{\ln\rho}{16}\left(R^{\mu\nu}[\gamma]R_{\mu\nu}[\gamma]-\frac{1}{3} R[\gamma]^2+f(0)F_{\mu\nu}F^{\mu\nu}\right) \right. \notag\\
  &+ \left. \frac{1}{2}\left(1+\frac{1}{\ln\rho}\right)\phi^2 \right],
\end{align}
where $f(0)=f(\phi=0)$, and $R[\gamma]$, $R_{\mu\nu}[\gamma]$, are the respective Ricci scalar and Ricci tensor of the induced metric at the boundary, $\gamma_{\mu\nu}$. From now on, though, in order to simplify the notation, we will suppress the explicit metric dependence $\gamma$ of the curvature tensors evaluated at the boundary of the bulk space, e.g. $R\equiv R[\gamma]$. Also, from the term multiplying $\ln\rho$ one can already see what is the trace anomaly of the theory, which is zero for the case of the conformal 1RCBH model\footnote{We remark that, although a chemical potential does not induce a trace anomaly, a magnetic field does induce a trace anomaly in the SYM plasma \cite{Fuini:2015hba}.}. Regarding the derivation of the counterterm action, we suggest Ref.\ \cite{Elvang:2016tzz} for very enlightening and clear discussions about it. Also, one may find insightful discussions about the derivation of counterterm actions in the EMD context using the Hamiltonian approach in Ref.\ \cite{Lindgren:2015lia}.

Moreover, it is important to know that, due to the fact that in the 1RCBH model the scalar field and the Abelian gauge field do not break the original conformal symmetry of the SYM plasma, one may add finite contributions to the counterterm action \eqref{eq:Sct}, which unveils the scheme dependence of the holographic renormalization procedure. The finite counterterms that one may add are
\begin{equation}\label{eq:SctFini}
S_{ct}^{finite} = \frac{1}{\kappa_5^2}\int_{\partial M} d^4x\sqrt{-\gamma}\left[c_1 F_{\mu\nu}F^{\mu\nu}+ c_2 \phi^2  \right],
\end{equation}
where $\lbrace c_1, c_2 \rbrace\in \mathbb{R}$. In short, to see why these terms are finite, one just needs to recall that, with the scaling dimensions of the EMD fields for the 1RCBH model, one obtains
\begin{align}
& \sqrt{-\gamma} \sim \rho^{-2},  \ \ \ F_{\mu\nu}F^{\mu\nu}\sim \rho^{2}, \ \ \ \phi^2 \sim \rho^{2}, \notag \\
& \Rightarrow \sqrt{-\gamma}F_{\mu\nu}F^{\mu\nu}\sim \text{constant}, \ \ \text{and} \ \ \sqrt{-\gamma}\phi^2 \sim \text{constant}.
\end{align}
Consequently, one may try to simplify the final expressions for the one-point functions of the dual QFT by including some finite counterterms, i.e.
\begin{equation}
S_{ct} \rightarrow S_{ct}+S_{ct}^{finite}.
\end{equation}
In this work, though, we will not resort to the addition of any extra finite term to the counterterm action \eqref{eq:Sct}.

\subsection{One-point functions}
\label{sec:HoloRen:1PT}

With the counterterm action at hand, we now have the subtracted on-shell action \eqref{eq:Ssub}, which means that we can proceed with the functional derivatives to extract the one-point functions given in Eqs.\ \eqref{eq:1pt_T}---\eqref{eq:1pt_phi}. The analysis for the scalar field carried out here is based on Appendix C of Ref.\ \cite{Elvang:2016tzz}, whilst the vector field analysis is based on Ref.\ \cite{Sahoo:2010sp}.

It is clear from Eqs.\ \eqref{eq:1pt_T}---\eqref{eq:1pt_phi} that we need to expand the fields near the boundary. Thus, we perform the FG expansion of the EMD fields\footnote{Note that in the subscripts $(n,m)$ of the coefficients of these expansions, $n=0$ denotes the leading order term in $\rho$ (as it goes to zero) and $m$ denotes the power of $\ln \rho$. This is reminiscent of the general form of these expansions presented in Eqs.\ \eqref{eq:expA}---\eqref{eq:expPhi}.}
\begin{align}
\gamma_{\mu\nu}(\rho,x) & =\frac{1}{\rho}\gamma_{(0)\mu\nu}(x)+\gamma_{(2)\mu\nu}(x)+ \gamma_{(2,1)\mu\nu}(x)\ln\rho \notag \\
 &+ \rho \left( \gamma_{(4)\mu\nu}(x)+\gamma_{(4,1)\mu\nu}(x)\ln\rho +\gamma_{(4,2)\mu\nu}(x)\ln^2\rho \right)+\mathcal{O}(\rho^2), \label{eq:FGgamma}\\
A_{\mu}(\rho,x) & = A_{(0)\mu}(x)+\rho \left( A_{(2)\mu}(x)+A_{(2,1)\mu}(x)\ln\rho \right)+\mathcal{O}(\rho^2), \label{eq:FGA}\\
\phi(\rho,x) & = \rho\left( \phi_{(0)}(x)+\phi_{(0,1)}(x)\ln\rho \right) + \mathcal{O}(\rho^2). \label{eq:FGphi}
\end{align}
Note that in order to obtain the one-point functions, it suffices to expand the fields up to $\mathcal{O}(\rho)$ since the remaining terms vanish as $\rho\rightarrow 0$. The independent terms of the above expansions are $\lbrace \phi_{(0)},\phi_{(0,1)}, A_{(0)\mu},A_{(2)\mu}, \gamma_{(0)\mu\nu},\gamma_{(4)\mu\nu}\rbrace$ and, thus, any other coefficient may be recast in terms of these independent ones. Furthermore, we are keeping here the analysis fairly general for any EMD model with $\Delta=2$; we shall only specialize to the 1RCBH background at the end of the calculations.

Next, one substitutes the above near-boundary expansions for $\gamma_{\mu\nu}$, $A_{\mu}$, and $\phi$ into Eqs.\ \eqref{eq:1pt_T}---\eqref{eq:1pt_phi} to obtain the explicit formulas for the one-point functions. However, since the required algebra is not so simple, we give some further details below. First, let us provide the formulas for the variation of the regularized on-shell action with respect to the sources\footnote{When we integrated the $\rho$ coordinate by parts in the variation of the integrals we considered that the normal vector $n^{M}=(-2\rho,0,0,0,0)$ at the boundary of the manifold is ``outward-pointing'', i.e. $g_{MN}n^{M}n^{M}=1$.}
\begin{align}
\frac{1}{\rho}\frac{2}{\sqrt{-\gamma}}\frac{\delta S_{sub}}{\delta \gamma^{\mu\nu}} & =\frac{1}{\rho} \left( T_{\mu\nu}^{reg}+T_{\mu\nu}^{ct} \right), \label{eq:deltaSgamma} \\
\frac{1}{\rho^2}\frac{1}{\sqrt{-\gamma}}\frac{\delta S_{sub}}{\delta A_{\mu}} & =  \frac{1}{\rho^2} J^{\mu},  \label{eq:deltaSA} \\ 
\frac{\ln\rho}{\rho}\frac{1}{\sqrt{-\gamma}}\frac{\delta S_{sub}}{\delta \phi} &  =-\frac{1}{\kappa_{5}^{2}}\frac{\ln\rho}{\rho}\left(-\rho\partial_{\rho}\phi+\left(1+\frac{1}{\ln\rho}\right)\phi  \right), \label{eq:deltaSphi}
\end{align}
where
\begin{align}
T_{\mu\nu}^{reg} & = \frac{2}{\sqrt{-\gamma}}\frac{\delta S_{reg}}{\delta \gamma^{\mu\nu}} =  \frac{1}{\kappa_{5}^{2}} \left(K_{\mu\nu}-\gamma_{\mu\nu}K\right), \\
T_{\mu\nu}^{ct} & = \frac{1}{\kappa_{5}^{2}}\left(-2Y_{\mu\nu}+\mathcal{L}_{ct} \gamma_{\mu\nu} \right), \\
J^{\mu} & = \frac{1}{\kappa_{5}^{2}}\left( \rho f(\phi)\gamma^{\mu\nu}\partial_\rho A_{\nu} + \frac{f(\phi)}{4}\nabla_\nu F^{\mu\nu}\ln\rho\right), \label{eq:Jsub}
\end{align}
with $K_{\mu\nu}=\rho\partial_{\rho}\gamma_{\mu\nu}$ denoting the extrinsic curvature of the boundary and $K=\gamma^{\mu\nu}K_{\mu\nu}$ its trace. We also have defined the following objects
\begin{align}
\mathcal{L}_{ct} & = -3-\frac{1}{4}R + \frac{1}{2}\left(1+\frac{1}{\ln\rho}\right)\phi^2 +  \frac{\ln\rho}{16}\left(R^{\mu\nu}R_{\mu\nu}-\frac{1}{3} R^2+f(0)F_{\mu\nu}F^{\mu\nu}\right), \\
Y_{\mu\nu} & = \frac{\delta \mathcal{L}_{ct}}{\delta \gamma^{\mu\nu}} \notag \\
     & =\frac{1}{4} R_{\mu\nu} +  \ln\rho\left[-\frac{f(0)}{8}F_{\mu\sigma}F_{\nu}^{\ \sigma}+\frac{f(0)}{32}F_{\sigma\lambda}F^{\sigma\lambda}\gamma_{\mu\nu}+\frac{1}{32} \gamma_{\mu\nu}R^{\sigma\lambda}R_{\sigma\lambda} \right. \notag\\
    &  \left. + \frac{1}{24}R_{\mu\nu}R -\frac{1}{96} \gamma_{\mu\nu}R -\frac{1}{8}R^{\sigma\lambda}R_{\mu\sigma\nu\lambda}  + \frac{1}{48}(\nabla_\mu\nabla_\nu R)-\frac{1}{16}\square R_{\mu\nu}+\frac{1}{96}\gamma_{\mu\nu}\square R \right].
\end{align}
Substituting Eq.\ \eqref{eq:deltaSphi} into Eq.\ \eqref{eq:1pt_phi}, and performing the asymptotic expansion of the dilaton field \eqref{eq:FGphi}, we obtain the expectation value of the dual scalar operator at the boundary QFT 
\begin{equation}\label{eq:DilVEV}
\boxed{ \langle\mathcal{O}_{\phi}\rangle = -\frac{1}{\kappa_{5}^2}\phi_{(0)}. }
\end{equation}

By the same token, if we substitute Eq.\ \eqref{eq:Jsub} into Eq.\ \eqref{eq:deltaSA}, expand the resulting equation near the boundary and take the limit $\rho\rightarrow 0$, we obtain the renormalized expectation value of the $U(1)$ R-current,
\begin{equation}
\langle J^{\mu} \rangle =  \frac{1}{\kappa_{5}^{2}} \left(  A_{(2)}^{\mu}+A_{(2,1)}^{\mu} \right). 
\end{equation}
Moreover, the leading order solution of Maxwell's equations \eqref{eq:MaxwellEq} under the expansion \eqref{eq:FGgamma}---\eqref{eq:FGphi} give us a relation between $A_{(2,1)\mu}$ and $A_{(0)\mu}$, i.e.
\begin{equation}
A_{(2,1)\mu} = \frac{f(0)}{4}\nabla_\nu F_{(0)\mu}^{ \ \ \ \ \nu},
\end{equation}
which leads us to the final form the $U(1)$ R-current,
\begin{equation}\label{eq:Jmu}
\boxed{ \langle J_{\mu} \rangle =  \frac{1}{\kappa_{5}^{2}} \left(  A_{(2)\mu}+\frac{f(0)}{4}\nabla_\nu F_{(0)\mu}^{ \ \ \ \ \nu} \right). }
\end{equation}
Notice, however, that the last term of the above equation is absent in the 1RCBH background since $\partial_\mu A_{(0)\nu}=0$. 

Regarding the one-point function $\langle T_{\mu\nu} \rangle$, the algebra is a little bit more complicated. Thus, in order to simplify the analysis, we will only focus on the finite contributions of Eq.\ \eqref{eq:deltaSgamma}\footnote{Since all the divergences are mutually canceled out by taking into account the counterterms.}. The finite terms coming from $T_{\mu\nu}^{reg}$ are
\begin{align}\label{eq:Treg}
-\frac{\kappa_{5}^{2}}{\rho} T_{\mu\nu}^{reg} & = -5\gamma_{(4)\mu\nu}-\gamma_{(4,1)\mu\nu}+\gamma_{(2)\mu\nu}\gamma_{(2)\sigma}^{ \ \sigma} + \gamma_{(0)\mu\nu}(2\gamma_{(4)\sigma}^{ \ \sigma}+\gamma_{(4,1)\sigma}^{ \ \sigma} - \gamma_{(2)\sigma\lambda}\gamma_{(2)}^{\sigma\lambda}) \notag \\
& +\text{divergent terms}+\text{vanishing terms as $\rho\rightarrow 0$}.
\end{align}

On the other hand, $T_{\mu\nu}^{ct}$ also contributes to the finite part of the total stress-energy tensor, i.e.
\begin{align}\label{eq:Tct}
-\frac{\kappa_{5}^{2}}{\rho} T_{\mu\nu}^{ct} & = 3\gamma_{(4)\mu\nu}+ \frac{1}{4}\gamma_{(2)\mu\nu}R_{(0)} + \frac{1}{4}\gamma_{(2)}^{\mu\sigma}R_{(0)\mu\sigma\nu\lambda}+ \frac{1}{4}\nabla_{\nu}\nabla_{\mu}\gamma_{(2)\sigma}^{\sigma} - \frac{1}{4}\nabla_\nu\nabla_\sigma\gamma_{(2)\nu}^{\sigma}\notag \\
&  - \frac{1}{4}\nabla_\sigma\nabla_\mu\gamma_{(2)\nu}^{\sigma}  + \frac{1}{4}\square_{(0)} \gamma_{(2)\mu\nu}+ \frac{1}{4}\gamma_{(0)\mu\nu} \left( - R^{\sigma\lambda}_{(0)}\gamma_{(2)\sigma\lambda}  +4\phi_{(0)}^2+8\phi_{(0)}\phi_{(0,1)} \right. \notag \\
& \left.  +\nabla_\sigma\nabla_\lambda\gamma_{(2)}^{\sigma\lambda}- \square_{(0)}\gamma_{(2)\sigma}^{\sigma} \right) +\text{divergent terms}+\text{vanishing terms as $\rho\rightarrow 0$}.
\end{align}
Hence, the counterterms do impact on the finite result for the one-point functions, even though their original purpose was to eliminate the divergences of the on-shell action.

Proceeding with the tensorial algebra to obtain the one-point function of the stress-energy tensor, the next step is to sum Eq.\ \eqref{eq:Treg} with Eq.\ \eqref{eq:Tct}, i.e.
\begin{align}\label{eq:Tsub}
-\kappa_{5}^{2}\langle T_{\mu\nu} \rangle & = -2\gamma_{(4)\mu\nu} -\gamma_{(4,1)\mu\nu}+\gamma_{(2)\mu\nu}\gamma_{(2)\sigma}^{ \ \sigma} + \frac{1}{4}\gamma_{(2)\mu\nu}R_{(0)} + \frac{1}{4}\gamma_{(2)}^{\mu\sigma}R_{(0)\mu\sigma\nu\lambda} \notag \\
& + \frac{1}{4}\nabla_{\nu}\nabla_{\mu}\gamma_{(2)\sigma}^{\sigma} - \frac{1}{4}\nabla_\nu\nabla_\sigma\gamma_{(2)\nu}^{\sigma} - \frac{1}{4}\nabla_\sigma\nabla_\mu\gamma_{(2)\nu}^{\sigma} + \frac{1}{4}\square_{(0)} \gamma_{(2)\mu\nu} \notag \\
 &+ \frac{1}{4}\gamma_{(0)\mu\nu}\left( 8\gamma_{(4)\sigma}^{ \ \sigma}+4\gamma_{(4,1)\sigma}^{ \ \sigma} - 4\gamma_{(2)\sigma\lambda}\gamma_{(2)}^{\sigma\lambda}  - R^{\sigma\lambda}_{(0)}\gamma_{(2)\sigma\lambda}  +2\phi_{(0)}^2+4\phi_{(0)}\phi_{(0,1)} \right. \notag \\
 & \left. +\nabla_\sigma\nabla_\lambda\gamma_{(2)}^{\sigma\lambda}- \square_{(0)}\gamma_{(2)\sigma}^{\sigma} \right).
\end{align}

Now we are close to give the full expression for $\langle T_{\mu\nu} \rangle$. The final step consists in expressing the coefficients of the metric expansion, such as $\gamma_{(2)}$, in terms of the curvature tensors of the boundary metric $\gamma_{(0)\mu\nu}$. This step, though, requires more laborious algebra, and more definitions are needed in order to do it in an simple way.

A convenient way to expand the EMD equations of motion using the FG coordinates is to use the ADM decomposition of general relativity \cite{Arnowitt:1959ah,Arnowitt:1960es} in which one considers spacelike foliations keeping $\rho=\text{constant}$ at each foliation. We suggest at this point Ref.\ \cite{Lindgren:2015lia} for a more complete discussion about this subject. Using the ADM decomposition, from where the Gauss-Codazzi equations are derived, the $(\mu\nu)$-components of Einstein's equations become
\begin{align}
0 & = 2\rho^2\partial_{\rho}^{2}\gamma_{\mu\nu} +\rho^{2}\gamma^{\sigma\lambda}(\partial_{\rho} \gamma_{\mu\sigma})(\partial_{\rho}\gamma_{\nu\lambda})-2\rho^{2}\gamma^{\sigma\lambda}(\partial_{\rho}\gamma_{\sigma\lambda})(\partial_{\rho}\gamma_{\mu\nu})-R_{\mu\nu}  +\frac{1}{2}\partial_{\mu}\phi\partial_{\nu}\phi \notag \\
  & +\frac{1}{3}\gamma_{\mu\nu}V(\phi)+2\rho^2f(\phi)(\partial_{\rho}A_{\mu}) (\partial_{\rho}A_{\nu})-\frac{1}{12}f(\phi)F_{\sigma\lambda}F^{\sigma\lambda}\gamma_{\mu\nu} + \frac{1}{2}f(\phi)F_{\mu\sigma}F_{\nu}^{\ \sigma} \notag \\
 & -\frac{2}{3}f(\phi)\gamma_{\mu\nu}\gamma^{\sigma\lambda}(\partial_{\rho}A_{\sigma}) (\partial_{\rho}A_{\lambda}).
\end{align}  
One can now expand the above equation near the boundary using Eqs.\ \eqref{eq:FGgamma}---\eqref{eq:FGphi}, obtaining
\begin{align}
\gamma_{(2)\mu\nu} & = \frac{1}{12}\left(\gamma_{(0)\mu\nu}R_{(0)}-6R_{(0)\mu\nu}  \right), \label{eq:gamma2} \\
\gamma_{(2,1)\mu\nu} & = 0, \\
\gamma_{(4,2)\mu\nu} & = -\frac{1}{12}\phi_{(0,1)}^2\gamma_{(0)\mu\nu}, \\
\gamma_{(4,1)\mu\nu} & = -\frac{1}{8}f(0)F_{(0)\mu\sigma}F_{(0)\nu}^{\ \ \ \ \sigma}+\frac{1}{8} R_{(0)}^{\sigma\lambda}R_{(0)\mu\sigma\nu\lambda}-\frac{1}{24}R_{(0)\mu\nu}R_{(0)}+ \frac{1}{16}\square_{(0)}R_{(0)\mu\nu} \notag \\
  & -\frac{1}{48}\nabla_{\mu}\nabla_{\nu}R_{(0)} +\gamma_{(0)\mu\nu}\left[\frac{1}{32}f(0)F_{(0)\sigma\lambda}F_{(0)}^{\sigma\lambda} -\frac{1}{32}R_{(0)\sigma\lambda}R_{(0)}^{\sigma\lambda}+\frac{1}{96}R_{(0)}^{2}  \right. \notag \\
  & \left. - \frac{\phi_{(0)}\phi_{(0,1)}}{6} -\frac{1}{96}\square_{(0)}R_{(0)} \right], \label{eq:gamma41} \\
\gamma_{(4)\mu}^{\ \mu} & = \frac{1}{48}f(0)F_{(0)\mu\nu}F_{(0)}^{\mu\nu} +\frac{1}{16}R_{(0)\mu\nu}R^{(0)\mu\nu} -\frac{R_{(0)}^2}{72}-\frac{2}{6}\phi_{(0)}^2-\frac{1}{6}\phi_{(0,1)}^2.\label{eq:gamma4Trace}
\end{align}

Finally, by substituting Eqs.\ \eqref{eq:gamma2}---\eqref{eq:gamma4Trace} into Eq.\ \eqref{eq:Tsub}, we obtain the expectation value of the stress-energy tensor of the dual QFT
\begin{align}\label{eq:GenTmunu}
\kappa_{5}^{2}\langle T_{\mu\nu} \rangle & =  2\gamma_{(4)\mu\nu} - \frac{1}{4}\left(R_{(0)\mu}^{\sigma}R_{(0)\sigma\nu}-\frac{3}{2}R_{(0)}^{\sigma\lambda}R_{(0)\mu\sigma\nu\lambda} + \frac{1}{4}\nabla_\mu\nabla_\nu R_{(0)}-\frac{3}{4}\square_{(0)}R_{(0)\mu\nu} \right) \notag\\
 &  + \frac{1}{8} f(0)F_{(0)\mu\sigma}F_{(0)\nu}^{\ \ \ \ \sigma} -\frac{\gamma_{(0)\mu\nu}}{2} \left(\frac{f(0)}{48}F_{(0)\sigma\lambda}F_{(0)}^{\sigma\lambda}-\frac{\phi_{(0)}^2}{3}+\phi_{(0)}\phi_{(0,1)}-\frac{2\phi_{(0,1)}^2}{3}\right) \notag \\
 & - \frac{1}{32}\gamma_{(0)\mu\nu}\left( R_{(0)\sigma\lambda}R_{(0)}^{\sigma\lambda}+\frac{1}{9}R_{(0)}^2 -\square_{(0)}R_{(0)} \right),
\end{align}
which is valid for any five dimensional EMD theory with a bulk scalar field with dimension $\Delta=2$ and for any kind of boundary.

Specializing the above results for the 1RCBH background will vastly simplify the expression for $\langle T_{\mu\nu} \rangle$. First, the conformal boundary of such theory is flat (i.e., Minkowski), which means that all the curvature tensors are identically zero. Second, if one takes the trace of Eq.\ \eqref{eq:GenTmunu}, the resulting expression reads
\begin{align}
\kappa_{5}^{2}\langle T_{\ \mu}^{\mu} \rangle = - 2\phi_{(0)}\phi_{(0,1)}+\phi_{(0,1)}^2-\frac{1}{8}\left(R_{(0)\mu\nu}R^{\mu\nu}_{(0)} - \frac{1}{3}R_{(0)}^{2} \right) -\frac{f(0)}{8}F_{(0)\mu\nu}F_{(0)}^{\mu\nu}.
\end{align}
Hence, one arrives at a very important result: \emph{in a conformal EMD model with $\Delta=2$ the logarithmic terms on the near-boundary expansions of the bulk fields are absent}. This sentence justifies the prior assumption made in Sec.\ \ref{sec:AsymExp} when we set to zero the logarithmic terms of the near-boundary asymptotic expansions. Furthermore, we remark that the finite counterterm contribution \eqref{eq:SctFini} does not modify the trace anomaly of the theory.

From the previous discussion, the stress-energy tensor for the 1RCBH model reads,
\begin{equation}\label{eq:Tij}
\boxed{ \langle T_{\mu\nu} \rangle = \frac{1}{\kappa_{5}^{2}} \left[ 2\gamma_{(4)\mu\nu}+ \frac{1}{8} f(0)F_{(0)\mu\sigma}F_{(0)\nu}^{\ \ \ \ \sigma} -\gamma_{(0)\mu\nu} \left(\frac{f(0)}{96}F_{(0)\sigma\lambda}F_{(0)}^{\sigma\lambda}-\frac{\phi_{(0)}^2}{6}\right) \right], }
\end{equation}
which is the one-point function adopted throughout this paper. We remark once more that, due to the fact $\partial_\mu A_{(0)\nu}=0$ in our setup, the Maxwell terms in Eq. \eqref{eq:Tij} will vanish.

A minimal internal consistency check of the above results may be done by looking at the trace Ward Identity \cite{Lindgren:2015lia,Papadimitriou:2004ap}, i.e.
\begin{equation}\label{WardId}
\langle T^{\mu}_{\ \mu} \rangle - (4-\Delta)\phi_{(0,1)}\langle\mathcal{O}_{\phi} \rangle = \mathcal{A},
\end{equation} 
where $\mathcal{A}$ denotes the trace anomaly of the theory. For the specific case of the EMD theory with $\Delta=2$, the anomaly is given by
\begin{equation}
\mathcal{A} = \mathcal{A}_{gravity}+\mathcal{A}_{Maxwell}+\mathcal{A}_{dilaton},
\end{equation}
where,
\begin{align}
\mathcal{A}_{gravity} = \frac{1}{8}\left(R_{(0)\mu\nu}R_{(0)}^{\mu\nu}-\frac{1}{3}R_{(0)}^2 \right), \ & \ \ \mathcal{A}_{Maxwell}=-\frac{f(0)}{8}F_{(0)\mu\nu}F_{(0)}^{\mu\nu}, \ \ \ \mathcal{A}_{dilaton}= \phi_{(0,1)}^2.
\end{align}

%%%%%%%%%%%%%%%%%%%%%%%%%
\section{Quasinormal modes for the $SO(3)$ singlet (dilaton) channel}
\label{sec:QNMs}

In this Appendix we obtain the QNMs of the 1RCBH model for the $SO(3)$ singlet (dilaton) channel in the homogeneous, zero wavenumber limit complementing the study carried out in Ref.\ \cite{Finazzo:2016psx} where the $SO(3)$ quintuplet (external scalar) and triplet (vector diffusion) channels have been analyzed in detail.

As discussed in Ref.\ \cite{DeWolfe:2011ts}, at zero wavenumber the EMD system features a rotational $SO(3)$ symmetry under which the gauge and diffeomorphism invariant perturbations of the system are organized into a quintuplet channel corresponding to the five traceless spatial components of the perturbation of the metric field $h_{ij}$, a triplet channel corresponding to the three spatial components of the perturbation of the Maxwell field $a_i$, and a singlet channel corresponding to a combination involving the dilaton perturbation $\varphi$ and the trace of the spatial part of the perturbation of the metric field, namely,
\begin{align}
\mathcal{S}=\varphi-\frac{\phi'}{2A'}\,\frac{1}{3}\left(h_{xx}+h_{yy}+h_{zz}\right),
\label{eq:Spert}
\end{align}
where one sees that the background dilaton field $\phi$ couples the dilaton fluctuation with the spatial trace of the graviton. This $\mathcal{S}$-perturbation is analogous to the $Z_2$-mode of the so-called ``non-conformal channel'' discussed in Ref.\ \cite{Janik:2016btb}.\footnote{Note, however, that in Ref.\ \cite{Janik:2016btb} an $SO(2)$ group was associated with the residual rotational symmetry of the system in the spatial plane orthogonal to the direction of the nontrivial wavenumber of the perturbations.} This was called a non-conformal channel because this mode is intrinsically associated with the background dilaton field, which in the Einstein-dilaton models analyzed in Ref.\ \cite{Janik:2016btb} was responsible for breaking conformal symmetry. However, the dilaton field may also preserve conformal symmetry when in the presence of other fields, as in the case of the 1RCBH model studied here. Therefore, more generally, one could say that this $\mathcal{S}$-perturbation defines the ``dilaton channel''. This nomenclature is also adequate due to the fact that this mode shares the same near-boundary asymptotics of the background dilaton field \cite{DeWolfe:2011ts,Janik:2016btb}.

The linearized equation of motion for the $\mathcal{S}$-perturbation derived in Ref.\ \cite{DeWolfe:2011ts} translates as follows to the modified EF coordinates \eqref{eq:modEF},\footnote{In order to go from the domain-wall coordinates used in Ref. \cite{DeWolfe:2011ts} to the modified EF coordinates \eqref{eq:modEF}, one simply uses that $d/d\tilde{r}=(\partial v/\partial\tilde{r})\partial_v+\partial_{\tilde{r}}$ and $d/dt=(\partial v/\partial t)\partial_v$, where $\partial v/\partial\tilde{r}=\sqrt{-g_{\tilde{r}\tilde{r}}/g_{tt}}=e^{B-A}/h$ and $\partial v/\partial t=1$.}
\begin{align}
&S''+\frac{\left(h \left(4 A'-B'\right)-2 i \omega  e^{B-A}+h'\right)}{h} S' +
\frac{e^{-2 A}}{18 f h \left(A'\right)^2} \left(-18 \left(A'\right)^2 \left(\partial_\phi f\right)^2 \left(\Phi '\right)^2+ \right.\nonumber\\
&\left. f \left(3 \left(A'\right)^2 \left(-6 e^{2 (A+B)} \partial_\phi^2 V+8 e^{2 A} h \left(\phi '\right)^2+3 \partial_\phi^2 f \left(\Phi '\right)^2\right)+6 A' \phi ' \left(e^{2 A} \left(h' \phi '-2 e^{2 B}
  \partial_\phi V\right)+\right.\right.\right.\nonumber\\
&\left.\left.\left. \partial_\phi f \left(\Phi '\right)^2\right)-54 i \omega  e^{A+B} \left(A'\right)^3-e^{2 A} h \left(\phi '\right)^4\right)\right) S=0,
\label{eq:eomS}
\end{align}
where $\omega$ is the frequency of the plane-wave Ansatz for the $\mathcal{S}$-perturbation, which shall give the QNMs of the dilaton channel under appropriate boundary conditions to be discussed below.

In order to solve the eigenvalue problem to be derived next for the quasinormal frequencies of the system we make use of the pseudospectral method as done in Ref.\ \cite{Finazzo:2016psx}. For this, we begin by mapping the radial coordinate $\tilde{r}\in[\tilde{r}_h,\infty)$ into a new radial coordinate $\tilde{r}_h/\tilde{r}=:\tilde{u}\in[0,1]$. By doing so and substituting the equilibrium 1RCBH backgrounds \eqref{eq:modEF} --- \eqref{eq:electromagnetic four-potential} into the equation of motion \eqref{eq:eomS}, one is left with a linear differential equation for the perturbation $\mathcal{S}(\tilde{u})$ on top of the equilibrium 1RCBH backgrounds. In general, the near-boundary, ultraviolet asymptotic behavior of the $\mathcal{S}$-perturbation is given by, $\mathcal{S}(\tilde{u})\sim \tilde{u}^{4-\Delta}G(\tilde{u}) + \tilde{u}^{\Delta}Y(\tilde{u}) + \cdots$, as $\tilde{u}\to 0$. Since in the 1RCBH model $\Delta=2$, we have two degenerate exponents equal to two. As discussed in detail in Ref.\ \cite{Finazzo:2016psx}, the correct eigenvalue problem for the QNMs is obtained by working with the subleading, normalizable mode of the relevant perturbation (in the dilaton channel considered here, the normalizable mode is associated with the expectation value of the boundary operator dual to the dilaton field, $\langle\mathcal{O}_\phi\rangle$), which in the present case corresponds to set the Dirichlet boundary condition $G(0)=0$ with $Y(0)\neq 0$. Then, by substituting $\mathcal{S}(\tilde{u})=: \tilde{u}^2 Y(\tilde{u})$ into the equation of motion for $\mathcal{S}(\tilde{u})$ and defining the dimensionless quasinormal frequency $\bar{\omega}\equiv\omega/T$, one obtains the following differential equation for $Y(\tilde{u})$,
\begin{align}
&\tilde{u} \left(1-\tilde{u}^2\right) \left(Y''\left(\tilde{u}\right)+Y'\left(\tilde{u}\right) \left(-\frac{i \bar{\omega } \tilde{r}_h \left(2 \tilde{r}_h^2+Q^2\right) \sqrt{\frac{Q^2
   \tilde{u}^2}{\tilde{r}_h^2}+1}}{\pi  \left(\tilde{u}^2-1\right) \sqrt{\tilde{r}_h^2+Q^2} \left(\tilde{u}^2 \left(\tilde{r}_h^2+Q^2\right)+\tilde{r}_h^2\right)}+\right.\right.\nonumber\\
   &\left. 2 \tilde{u}
   \left(\frac{\tilde{r}_h^2+Q^2}{\tilde{u}^2 \left(\tilde{r}_h^2+Q^2\right)+\tilde{r}_h^2}+\frac{1}{\tilde{u}^2-1}\right)+\frac{1}{\tilde{u}}\right)+\nonumber\\
   & Y\left(\tilde{u}\right)
   \left(\frac{4 \left(4 Q^4 \left(\tilde{u}^4+6 \tilde{u}^2-2\right) \tilde{u}^2 \tilde{r}_h^2+3 Q^2 \left(8 \tilde{u}^4+3 \tilde{u}^2-2\right) \tilde{r}_h^4+9 \tilde{u}^2
   \tilde{r}_h^6+4 Q^6 \tilde{u}^6\right)}{\left(\tilde{u}^2-1\right) \left(3 \tilde{r}_h^2+2 Q^2 \tilde{u}^2\right){}^2 \left(\tilde{u}^2
   \left(\tilde{r}_h^2+Q^2\right)+\tilde{r}_h^2\right)}-\right.\nonumber\\
   &\left.\left. \frac{i \bar{\omega } \left(2 \tilde{r}_h^2+Q^2\right) \left(\tilde{r}_h^2+2 Q^2 \tilde{u}^2\right)}{2 \pi  \tilde{u}
   \left(\tilde{u}^2-1\right) \tilde{r}_h \sqrt{\tilde{r}_h^2+Q^2} \sqrt{\frac{Q^2 \tilde{u}^2}{\tilde{r}_h^2}+1} \left(\tilde{u}^2
   \left(\tilde{r}_h^2+Q^2\right)+\tilde{r}_h^2\right)}\right)\right) = 0,
\label{eq:eomY}
\end{align}
which defines the relevant generalized eigenvalue problem (GEP) to be solved in order to find the QNMs $\bar{\omega}(\mu/T)$ of the dilaton channel of the 1RCBH model in the zero wavenumber limit.\footnote{Note that in the EF coordinates the infalling wave condition at the horizon is imposed by just requiring regularity of the solutions there. This is one of the main reasons why the EF coordinates are very convenient to deal with the calculation of QNMs. Another reason is that in the domain-wall coordinates one would obtain instead of the GEP \eqref{eq:eomY} a quadratic eigenvalue problem (QEP) for $\bar{\omega}$, which is far more demanding in terms of computational costs. Note also what the QNMs $\bar{\omega}$ are only functions of the dimensionless combination $\mu/T$ controlled by the background parameters $\tilde{r}_h$ and $Q$ (here we restrict our analysis to the QNMs evaluated on the stable branch of black hole solutions).}

\begin{figure}[h]
\centering
\includegraphics[width=0.7\textwidth]{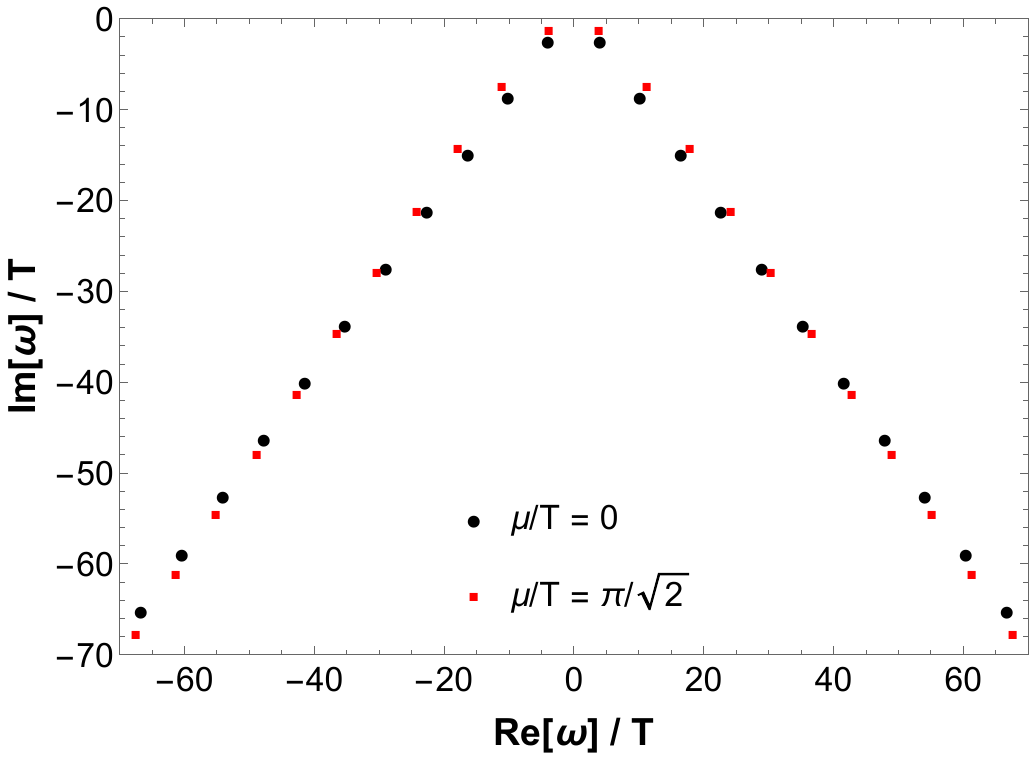}
\caption{(Color online) First 22 non-hydrodynamic QNMs of the dilaton channel evaluated at $\mu/T=0$ and $\mu/T=\pi/\sqrt{2}$ (critical point).}
\label{fig:newQNM}
\end{figure}

\begin{figure}[h]
\centering
\begin{subfigure}{0.49\textwidth}
\includegraphics[width=\textwidth]{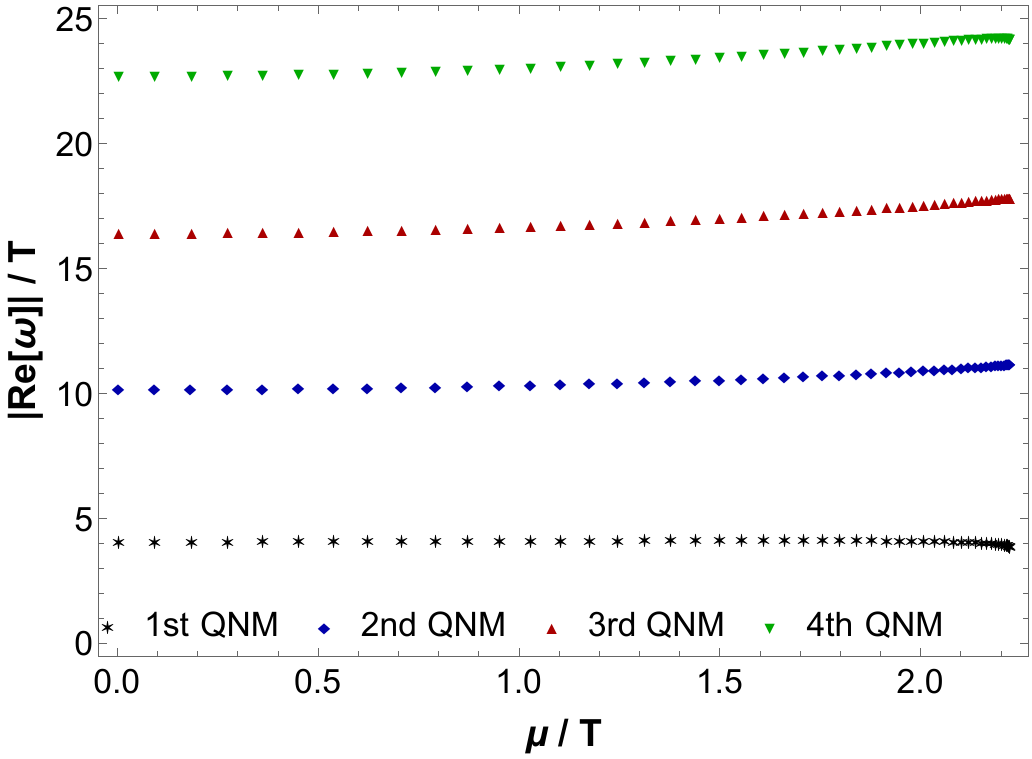}
\caption{}
\end{subfigure}
\begin{subfigure}{0.49\textwidth}
\includegraphics[width=\textwidth]{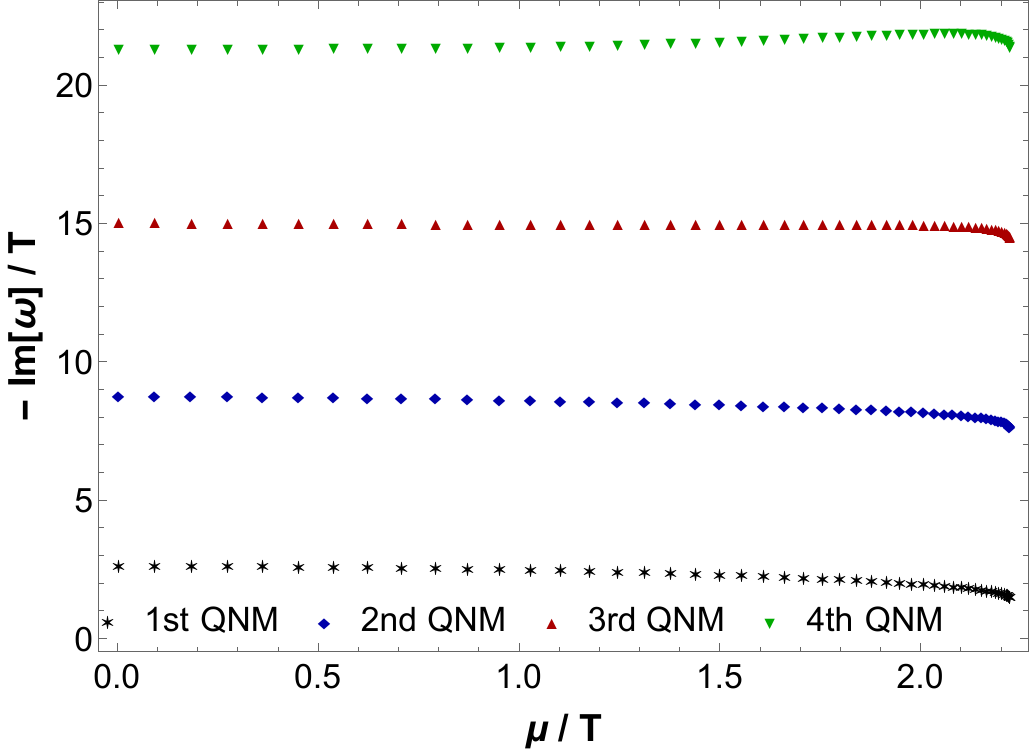}
\caption{}
\end{subfigure}
\caption{(Color online) (a) Real and (b) imaginary parts of the first four non-hydrodynamic QNMs of the dilaton channel as functions of $\mu/T$.}
\label{fig:QNMsDil}
\end{figure}

\begin{figure}[h]
\centering
\includegraphics[width=0.7\textwidth]{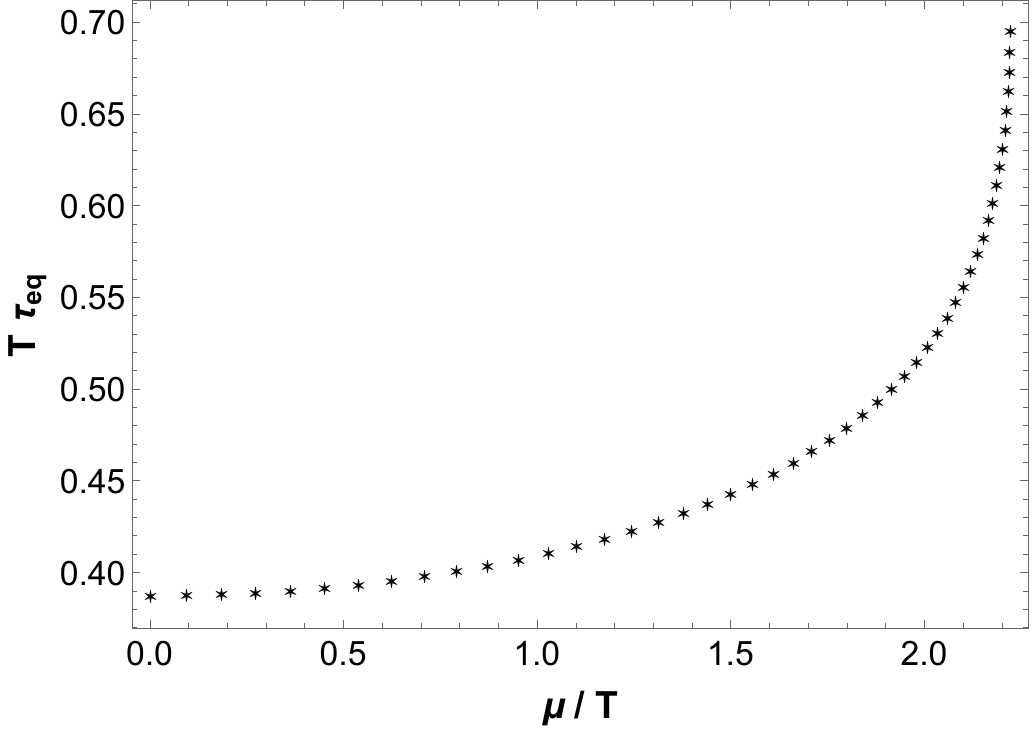}
\caption{Characteristic ``equilibration time'' of the dilaton channel as a function of $\mu/T$.}
\label{fig:EqTimeDil}
\end{figure}

The numerical routine used to solve the GEP equation \eqref{eq:eomY} for the QNMs $\bar{\omega}(\mu/T)$ employed the pseudospectral method and it follows the same general steps discussed in Ref.\ \cite{Finazzo:2016psx} and we refer the interested reader to consult it for technical details. In Figs.\ \ref{fig:newQNM} and \ref{fig:QNMsDil} we show the behavior of the QNMs of the dilaton channel as a function of $\mu/T$. In Fig. \ref{fig:EqTimeDil} we show how the characteristic ``equilibration time'' associated with the inverse of minus the imaginary part of the lowest non-hydrodynamic QNM of the dilaton channel behaves as a function of $\mu/T$. Remarkably, one notes that this characteristic equilibration time is qualitatively different from the equilibration times obtained in Ref. \cite{Finazzo:2016psx} for the external scalar and vector diffusion channels, since the latter are reduced as one increases $\mu/T$ far from the CP while increasing close to the CP. In the dilaton channel, however, the equilibration time always increases with increasing $\mu/T$, in agreement with the late time behavior of the scalar condensate as we have discussed in Subsection \ref{sec:MatQNM}.

%%%%%%%%%%%%%%%%%%%%%%%%%
\bibliographystyle{JHEP}
\bibliography{Bibliography}

\end{document}